%% file: paper.tex
\documentclass[aps,prl,reprint,superscriptaddress,floatfix,longbibliography,nobibnotes]{revtex4-2}

\usepackage[T1]{fontenc}
\usepackage{microtype}
\usepackage{amsmath,amssymb,bm}
\usepackage{upgreek}
\usepackage{graphicx}
\graphicspath{{figs/}}
\usepackage{booktabs}
\usepackage{hyperref}
\usepackage{orcidlink}

\input{setup.tex}

\makeatletter
\renewcommand\section{\@startsection{section}{1}{\parindent}
  {\z@}{-0.1em}{\normalfont\normalsize\itshape}}
\def\@runin@tos@section#1#2{#1#2---}
\def\@runin@to@section#1#2#3{#1#2: #3---}
\renewcommand\paragraph{\@startsection{paragraph}{4}{\parindent}
  {\z@}{-0.1em}{\normalfont\normalsize\itshape}}
\def\@runin@tos@paragraph#1#2{#1#2---}
\makeatother

\hypersetup{hidelinks}

\newcommand{\vect}[1]{\bm{#1}}
\newcommand{\ii}{\mathrm i}
\newcommand{\ee}{\mathrm e}
\newcommand{\dd}{\,\mathrm d}
\newcommand{\Rmat}{\mathcal R}

\newcommand{\unitmu}{\upmu}

\AddToHook{shipout/before}{%
 \pdfpageattr{}%
 \ifnum\value{page}=\getpagerefnumber{sm:fig:ensemble-450}\relax
  \pdfpageattr{/Rotate 90}%
 \fi
 \ifnum\value{page}=\getpagerefnumber{sm:fig:ensemble-520}\relax
  \pdfpageattr{/Rotate 90}%
 \fi
 \ifnum\value{page}=\getpagerefnumber{sm:fig:ensemble-660}\relax
  \pdfpageattr{/Rotate 90}%
 \fi
}

\begin{document}

\title{Emergent Fourfold Symmetry in Polarization Images:
Local Origin and Diffractive Breakdown}

\author{Igor~S.~Lobanov\,\orcidlink{0000-0001-8789-3267}}
\email[Contact author: ]{lobanov@itmo.ru}
\affiliation{ITMO University, Saint Petersburg 197101, Russia}
\author{Tikhon~V.~Reztsov\,\orcidlink{0000-0003-2025-0579}}
\affiliation{ITMO University, Saint Petersburg 197101, Russia}
\author{Artem~N.~Shalev\,\orcidlink{0000-0002-4566-8093}}
\affiliation{ITMO University, Saint Petersburg 197101, Russia}
\author{Tetiana~Orlova\,\orcidlink{0000-0002-1594-291X}}
\affiliation{Yerevan State University, Yerevan 0025, Armenia}
\author{Elena~V.~Aksenova\,\orcidlink{0000-0002-4324-1491}}
\affiliation{Saint Petersburg State University, Saint Petersburg 199034, Russia}
\affiliation{ITMO University, Saint Petersburg 197101, Russia}
\author{Valery~M.~Uzdin\,\orcidlink{0000-0002-9505-0996}}
\affiliation{ITMO University, Saint Petersburg 197101, Russia}
\affiliation{Saint Petersburg State University, Saint Petersburg 199034, Russia}
\author{Nikolay~V.~Petrov\,\orcidlink{0000-0002-8866-7592}}
\affiliation{ITMO University, Saint Petersburg 197101, Russia}
\affiliation{School of Physics, Harbin Institute of Technology, Harbin 150001, China}
\author{Aleksey~V.~Chernykh\,\orcidlink{0000-0003-1297-3267}}
\email[Contact author: ]{chernykh_a@itmo.ru}
\affiliation{ITMO University, Saint Petersburg 197101, Russia}

\date{November 20, 2026}

\begin{abstract}
A polarization image can gain or lose rotational symmetry with no change
in material symmetry.  In axisymmetric anisotropic systems, covariance
restricts the field to azimuthal orders \(m=0,\pm2\), guaranteeing twofold but
not fourfold symmetry; pointwise unitarity adds fourfold symmetry,
which diffraction can destroy.  Holographic
Jones fields of liquid-crystal torons show propagation shifts the
twofold--fourfold balance; for the same director, a marching-wave model
qualitatively reproduces the twofold morphology missed by a local model.
This establishes a symmetry hierarchy for polarization imaging.
\end{abstract}

\maketitle
\raggedbottom

Symmetry is often used to infer microscopic structure, and polarization
microscopy of liquid-crystal solitons is no exception
\cite{smalyukh2010torons,ackerman_smalyukh2017knot,
ackerman_boyle_smalyukh2017squirming,lobanov2022optical_imaging}.
The symmetry of the image, however, is not automatically the symmetry of
the object.
A coherent polarization image combines anisotropic conversion, propagation,
and analyzer projection, so its symmetry belongs to the complete optical map,
not to the material alone.
Focus, wavelength, aperture, and polarizer settings
\cite{ackerman_boyle_smalyukh2017squirming,varanytsia2017topology}
shape the observed pattern; refocusing alone can make
the same soliton appear fourfold or twofold
while its director configuration remains unchanged
\cite{varanytsia2017topology}.

Rotational covariance of the complete optical map is the starting point.  We
show that it fixes the available angular orders and guarantees twofold
intensity symmetry, whereas pointwise unitarity of a local response cancels
the twofold interference and enforces fourfold symmetry.  Spatial nonlocality
can alter interference among the allowed orders without producing
noncovariant field content.  For a covariant response, a change of image
symmetry therefore amounts to a redistribution of interference among the
fixed azimuthal orders \(m=0,\pm2\).
Axisymmetric patterned retarders and unit-charge
\(q\)-plates illustrate the local branch through familiar four-sector images
between fixed linear polarizers
\cite{marrucci2006,rubano2019qplate_review}; diffraction and a finite pupil act
on the resulting vector field
\cite{karimi2009qplate_propagation,jisha2017diffraction_pb,
inoue_kubota1958diffraction_anomaly}.  The same distinction is central to
metasurface polarization optics: local birefringent
meta-atoms can convert circular-polarization angular momentum into structured
wavefronts carrying orbital angular momentum, whereas spatial dispersion and
collective resonances enable deliberately nonlocal control of the wavefront
\cite{arbabi2015dielectric_metasurfaces,devlin2017arbitrary_oam,
kwon2018nonlocal_metasurfaces,song2021nonlocal_metasurfaces}.

This issue is especially acute for liquid-crystal skyrmion tubes, baby
skyrmions \cite{fukuda_zumer2011skyrmions}, and torons, whose polarized
micrographs are used to identify both a localized state and its internal structure
\cite{smalyukh2010torons,
ackerman_boyle_smalyukh2017squirming,tai2020surface_anchoring}.
Varanytsia et al. \cite{varanytsia2017topology} observed toron images
with two bright crescents whose appearance changed with focus.  They attributed
discrepancies between measured and calculated images mainly to toron lensing
omitted from the optical model.
This raises the question addressed here: can light propagation produce a twofold image without breaking material axisymmetry?
A four-sector local-Jones prediction and a two-crescent
measurement therefore need not imply different material symmetries
\cite{tai2020surface_anchoring}.
Torons thus provide an experimental test of the general symmetry hierarchy:
nominal axisymmetry supplies the covariance test, whereas their
three-dimensional director structure and pronounced through-focus
evolution challenge the local
thin-element approximation.

We derive the general hierarchy without invoking a material model, then
propagate the holographically measured toron Jones field through focus
and project it onto the exactly covariant sector, separating symmetry-allowed
changes from covariance-breaking content.  Matched independent-column local-Jones and
marching-wave calculations for the same director test in-layer transverse
coupling, while factorial controls separate amplitude population from
phase-sensitive interference.

\begin{figure*}[t]
  \centering
  \includegraphics[width=\textwidth]{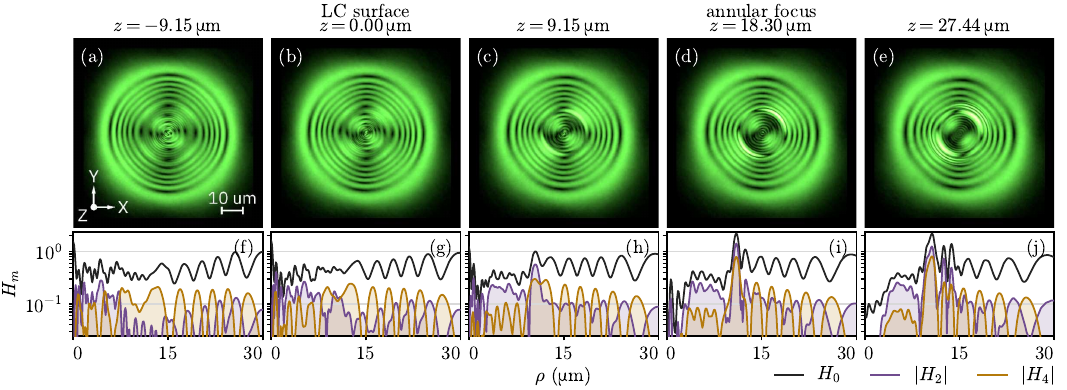}
  \caption{Numerical refocusing of a fixed measured field changes the
  balance of twofold and fourfold image components.  The \(520\)-nm field is
  the coherent average of nine registered Jones reconstructions of distinct
  torons in the same cell, obtained at a single holographic plane.
  (a)--(e) Unprojected crossed
  intensity \(I_\times=|J_{yx}|^2\) at five planes referenced to the estimated
  liquid-crystal exit surface; the second panel shows the estimated exit
  surface and the fourth the annular focus.  (f)--(j) The angular mean
  \(H_0\) and the twofold and fourfold harmonic amplitudes
  \(|H_2|\) and \(|H_4|\) for the same propagated field after
  exact projection onto the rotationally covariant sector, which
  separates symmetry-allowed changes from covariance-breaking content.  The common
  logarithmic ordinate gives the absolute harmonic amplitudes on the
  calibrated incident-intensity scale.}
  \label{fig:through-focus}
\end{figure*}

\section*{Covariance, locality, and image symmetry}

We write \(C_n\) for invariance of the intensity under an in-plane rotation by
\(2\pi/n\).  Let \((\rho,\varphi)\) be polar coordinates in the transverse
plane.  We use \(J(\vect r_\perp)\) to denote either a local Jones matrix or the
effective response of a spatially nonlocal system to two spatially uniform
basis inputs.  Under spatially uniform normal illumination, rotationally
symmetric exterior boundaries, and a centered circular scalar pupil, an
axisymmetric configuration obeys
\begin{equation}
 J(\rho,\varphi+\beta)
 =\Rmat_\beta J(\rho,\varphi)\Rmat_{-\beta},
 \label{eq:rotation-covariance}
\end{equation}
where \(\beta\) is arbitrary and \(\Rmat_\beta\) rotates transverse vectors.
Space and polarization are rotated together; a fixed laboratory analyzer
need not produce a radial image.  In the circular basis
\(\vect e_\pm=(\vect e_x\pm\ii\vect e_y)/\sqrt2\):
\begin{equation}
 J_{\rm c}(\rho,\varphi)=
 \begin{pmatrix}
  A_+(\rho)&B_-(\rho)\ee^{-2\ii\varphi}\\
  B_+(\rho)\ee^{2\ii\varphi}&A_-(\rho)
 \end{pmatrix}.
 \label{eq:circular-jones}
\end{equation}
Covariance therefore permits only \(m=0,\pm2\): \(A_\pm\) preserve circular
polarization, whereas \(B_\pm\) convert between the two circular
polarizations and carry azimuthal orders \(\pm2\).  This selection rule follows
from the representation of in-plane rotations on transverse vectors, not
from a liquid-crystal constitutive law.  Any fixed linear input--analyzer pair
therefore yields a scalar field with complex radial amplitudes
\(u_0,u_+,u_-\):
\begin{align}
 u&=u_0+u_+\ee^{2\ii\varphi}+u_-\ee^{-2\ii\varphi},\nonumber\\
 I=|u|^2&=H_0+2\operatorname{Re}
 \left(H_2\ee^{2\ii\varphi}+H_4\ee^{4\ii\varphi}\right),\label{eq:image-harmonics}\\
 H_0&=|u_0|^2+|u_+|^2+|u_-|^2,\nonumber\\
 H_2&=u_+u_0^*+u_0u_-^*,\qquad H_4=u_+u_-^*.\nonumber
\end{align}
Here \(H_0\) is the angular mean and contains the diagonal contributions of
all three angular orders.  Distinct orders become orthogonal only after
angular integration; locally, all three coexist coherently.  Because
the intensity is quadratic in their sum, \(u_0\) interferes with \(u_\pm\) to
form \(H_2\), while \(u_+\) interferes with \(u_-\) to form \(H_4\).  Covariance
therefore guarantees half-turn symmetry \(C_2\); quarter-turn symmetry
\(C_4\) holds if and only if \(H_2=0\).

\begin{figure*}[t]
  \centering
  \includegraphics[width=\textwidth]{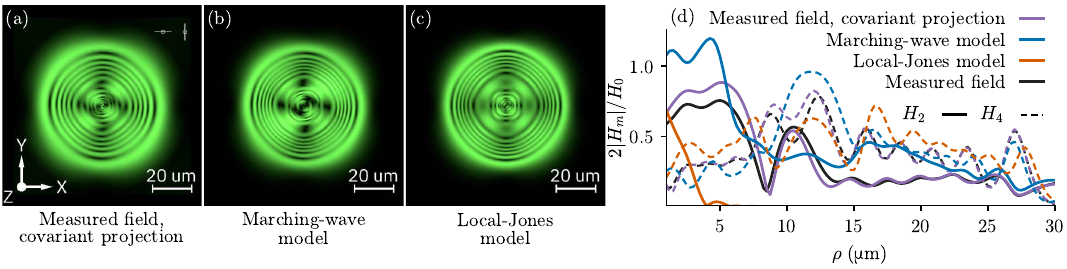}
  \caption{For the same modeled director, the marching-wave model qualitatively
  captures the measured central twofold morphology, whereas the local-Jones
  model remains nearly fourfold.
  Crossed intensity at the model-referenced exit plane at
  \(520\,\mathrm{nm}\) for (a) the averaged measured field after exact covariant
  projection, (b) the marching-wave model, and (c) the local-Jones model.
  The circular-basis projection retains only the angular orders required by
  rotational covariance.  Both models use the same centered circular
  pupil with a cutoff inferred from the measured spectrum.
  (d) Fractional
  twofold \(2|H_2|/H_0\) (solid) and fourfold \(2|H_4|/H_0\) (dashed)
  modulations for the measured field, its covariant projection, and the two
  models.  Projection leaves the interval-averaged measured twofold
  modulation nearly unchanged.}
  \label{fig:optical-comparison}
\end{figure*}

We now specialize the general covariant response to crossed polarizers,
with \(x\)-polarized input and a \(y\)-oriented analyzer.  We write
\(A_\pm=a\mp\ii b\) and \(B_\pm=c\pm\ii d\), where \(a,b,c,d\) are complex
radial amplitudes, not real and imaginary parts.  The crossed field and its
zeroth and second intensity harmonics are
\begin{equation}
 \begin{aligned}
 u_\times&=J_{yx}=b+c\sin2\varphi+d\cos2\varphi,\\
 H_0&=|b|^2+\tfrac12(|c|^2+|d|^2),\\
 H_2&=\operatorname{Re}(b^*d)
 -\ii\operatorname{Re}(b^*c).
 \end{aligned}
 \label{eq:crossed-harmonic}
\end{equation}
Here \(b\) is the matrix-antisymmetric zero-order amplitude, and \(c,d\) are
circular-polarization-conversion amplitudes.  A nonzero \(b\) provides the
zero-order field that can interfere with \(c,d\); their relative phases
determine the observable twofold modulation.

For this crossed-polarizer response, \(H_2\) is the azimuthal-order-two projection
of the pointwise normality defect \(J^\dagger J-JJ^\dagger\)
[Eqs.~\eqref{sm:eq:normality} and \eqref{sm:eq:normality-h2} of the
Supplemental Material \cite{supplemental_material}].  In the reflection-free
local-Jones model considered here, each layer is a unitary retarder in the
orthonormal transverse basis, so \(J^\dagger J=JJ^\dagger=I_2\), forcing
\(H_2=0\).  Thus \(b\) can carry substantial power while its twofold
interference with \(c,d\) cancels.  The quarter-turn identity derived in
Appendix~\ref{app:symmetry} extends this \(C_4\) result to every pair of real
linear input and analyzer states.  Transverse propagation can change
amplitudes and relative phases while preserving covariance, allowing
\(H_2\ne0\) without losing \(C_2\).  Conservation of total power does not
impose pointwise unitarity on \(J(\vect r_\perp)\).  We call the association
between circular-polarization channels and angular orders circular-polarization--order
locking.  \mbox{The resulting symmetry hierarchy is}
\begin{equation}
 \begin{aligned}
 \text{rotational covariance}
 &\;\Rightarrow\; C_2,\\
 \text{covariance + pointwise unitarity}
 &\;\Rightarrow\; C_4.
 \end{aligned}
 \label{eq:symmetry-hierarchy}
\end{equation}
These implications are one-way and do not, by themselves, identify the
physical mechanism.  Phase cancellation can produce accidental \(C_4\); a
local non-normal diattenuating--retarding stack can give \(H_2\ne0\); and a
spatially nonlocal covariant response can retain \(C_2\) while losing
\(C_4\).  Experiments should therefore test nested constraints rather than
compare image morphology alone: circular-polarization--order content tests
covariance, and a nonzero \(H_2\) in the crossed channel within the
covariant sector then rejects
pointwise normality of the effective Jones response.

\section*{Experimental significance in a liquid-crystal toron}

We test the symmetry hierarchy with the Jones
fields of neighboring relaxed torons in a homeotropic
5CB--S811 cell, measured by
polarization-sensitive holography \cite{reztsov2025toron_phm}.  The primary
\(520\)-nm field coherently averages nine registered Jones reconstructions of
distinct torons under the same conditions.  For crossed polarizers,
\(I_\times=|J_{yx}|^2\), and angular Fourier analysis gives the \(H_m\) of
Eq.~\eqref{eq:image-harmonics}.  We quantify the relative strength of
each intensity harmonic by
\(\mathcal V_m=2\lVert H_m\rVert_\rho/\lVert H_0\rVert_\rho\), using radial
\(L^2\) norms over the stated interval (Appendix~\ref{app:experiment}).
The exact estimator, reconstruction, registration, and
consistency tests are given in
Secs.~\ref{sm:sec:experiment}--\ref{sm:sec:ensemble} and
Eq.~\eqref{sm:eq:integrated-visibility} of the Supplemental Material
\cite{supplemental_material}.  Before projection and within
\(\rho\leq28\,\unitmu\mathrm m\), the allowed orders \(m=0,\pm2\) account for
approximately \(99\%\) of the background-subtracted
Jones-matrix power at each of the three nominal wavelengths,
\(450\), \(520\), and \(660\,\mathrm{nm}\)
(Fig.~\ref{sm:fig:experimental-mode-amplitudes}).  The stronger
element-resolved test additionally assesses
circular-polarization--order locking: each order must occupy its covariant
circular-basis Jones element.

Figure~\ref{fig:through-focus} gives a director-model-independent test in
which the propagation distance is the only variable.
Homogeneous propagation of the same measured Jones field changes the radial
balance of \(H_2\) and \(H_4\).  A centered scalar propagator preserves angular
orders, while changing their radial amplitudes and relative phases,
and hence the lobe pattern and
\(C_2\)--\(C_4\) balance without changing material covariance.  Neither image
feature alone identifies the soliton structure.  Full focus maps and the
propagation law are given in
Sec.~\ref{sm:sec:propagation} and Fig.~\ref{sm:fig:visible-c4-focus} of the
Supplemental Material.

\begin{figure*}[!t]
 \centering
 \includegraphics[width=\textwidth]{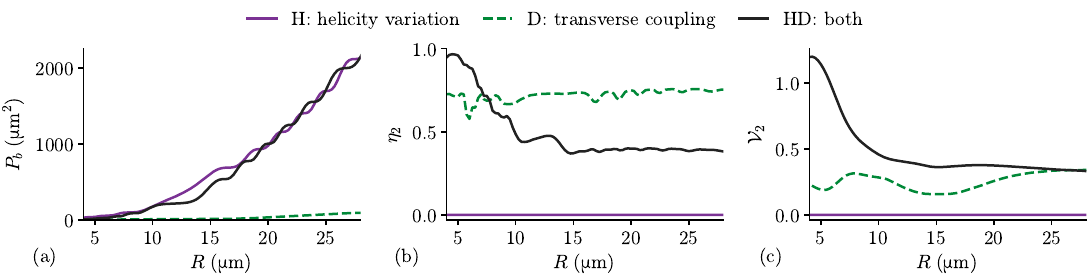}
 \caption{Independent variation of helicity and propagation model
 in axisymmetric controls with a common director polar-tilt profile.
 \(\mathrm H\): depth-varying helicity with local propagation;
 \(\mathrm D\): fixed helicity with transverse coupling;
 \(\mathrm{HD}\): both.  Panels show (a) incident-normalized,
 background-subtracted power \(P_b\) in the azimuth-independent crossed-field
 component \(b\); (b) phase-sensitive interference efficiency \(\eta_2\)
 between \(b\) and the azimuth-dependent field; and (c) relative
 second-harmonic amplitude \(\mathcal V_2\) of the intensity.
 Control \(\mathrm H\) produces large \(b\) power while retaining nearly
 fourfold intensity symmetry; \(\mathrm D\) breaks that symmetry with much
 less \(b\) power.  Quantities are evaluated over the centered disk
 \(\rho\leq R\), at the cell exit before pupil filtering, for
 \(\lambda=520\,\mathrm{nm}\).  The fixed-helicity local control
 \(\mathrm{0}\) is negligible on these scales and is omitted.}
 \label{fig:channel-mechanism}
\end{figure*}

The twofold response also survives removal of the noncovariant field
content.  We project the measured Jones field onto the circular-polarization
and angular-order combinations required by Eq.~\eqref{eq:circular-jones}.
At the model-referenced exit plane, over
\(4\leq\rho\leq28\,\unitmu\mathrm m\), this changes
\(\mathcal V_2\) only from \(0.251\) to \(0.253\).
With the model parameters held fixed, matching the annular focus determines
the axial origin without fitting Jones-component or harmonic amplitudes
(Appendix~\ref{app:experiment}).

An intensity decomposition confirms that the measured noncovariant
remainder is not the dominant source of \(H_2\), and a pointwise normality
test confirms that projection does not create the normality failure
(Secs.~\ref{sm:sec:propagation} and~\ref{sm:sec:experimental-audits} of the
Supplemental Material \cite{supplemental_material}).
Candidate instrumental origins of this remainder, such as residual
centering and registration errors or a polarization-dependent or noncircular
pupil outside the calibration, are discussed in
Sec.~\ref{sm:sec:center-pupil} of the Supplemental Material
\cite{supplemental_material}.

To test a sufficient mechanism, we use an axisymmetric Frank--Oseen toron with
the measured cell thickness and a pitch within its measured range.  The
elastic constants and anchoring strength were selected manually to reproduce
experimental observations.  Its closed skyrmion wall is axisymmetric
and its helicity varies nearly linearly through
the cell.  Local-Jones and marching-wave calculations use the same director,
refractive indices, physical output plane, and spectrum-matched circular
pupil; only the latter includes transverse propagation and anisotropic
longitudinal-field coupling.  Director details are given in
Sec.~\ref{sm:sec:director-model} and
Fig.~\ref{sm:fig:director-quantitative} of the Supplemental Material.

For this director, the marching-wave model qualitatively reproduces the
central twofold-dominated morphology and radial crossover of the covariantly
projected measurement, whereas the local-Jones model is nearly fourfold
(Fig.~\ref{fig:optical-comparison}).  Their respective interval-averaged
\((\mathcal V_2,\mathcal V_4)\)
modulations over the same \(4\leq\rho\leq28\,\unitmu\mathrm m\) interval are
\((0.3387,0.4802)\) and \((0.00472,0.4364)\).
The calculated \(\mathcal V_2\) exceeds the projected
measurement by about a third; we do not interpret this residual gap further.
The model inputs and approximations are collected in
Appendix~\ref{app:numerics} and Sec.~\ref{sm:sec:optical-model} of the
Supplemental Material \cite{supplemental_material}.  A tilted-director positive
control generates odd angular orders forbidden by Eq.~\eqref{eq:circular-jones}
(Fig.~\ref{sm:fig:tilted-control}).
The matched calculations identify transverse propagation as a mechanism
for the observed twofold morphology.  The relaxed director is a comparison texture,
not a unique inverse reconstruction of the experimental one.

\section*{Origin of the observable twofold interference}

The controls in Fig.~\ref{fig:channel-mechanism} produce a strong
twofold image modulation with only a small fraction of the \(b\) power
present in a nearly fourfold control.
Longitudinal rotation of the toron's projected optical axis makes successive
local retarders noncommuting and can populate \(b\).
In a chiral liquid crystal this rotation is supplied by the director
itself \cite{fukuda_zumer2011skyrmions,poy_zumer2020bpm}.
Local lossless propagation nevertheless preserves the phase relation required
by \(H_2=0\).
To separate the generation of \(b\) from its observable interference,
we vary the helicity and the propagation model independently, comparing the
fields at the model-referenced exit surface before pupil filtering.

The synthetic axisymmetric textures share the relaxed toron's radial
director tilt; their helicity is either constant or varies with depth
according to the extracted profile (Supplemental Material,
Sec.~\ref{sm:sec:controls}).  Pairing these textures with local or
marching-wave propagation gives four controls: \(\mathrm{0}\), fixed helicity
and local propagation; \(\mathrm H\), varying helicity and local propagation;
\(\mathrm D\), fixed helicity and marching-wave propagation; and
\(\mathrm{HD}\), both effects.  The synthetic \(\mathrm{HD}\) control is not
the production calculation using the unmodified three-dimensional director.
Figure~\ref{fig:channel-mechanism} reports the cumulative background-subtracted
power \(P_b\) of \(b\), normalized to the incident field; the rms efficiency
\(\eta_2\) with which the phases of \(b,c,d\) generate \(H_2\); and the
observable modulation \(\mathcal V_2[0,R]\).  Thus the three
quantities separate amplitude population, phase-sensitive interference, and
image modulation.  Their exact definitions are given in
Eqs.~\eqref{sm:eq:b-power}, \eqref{sm:eq:phase-unlocking}, and
\eqref{sm:eq:integrated-visibility} of the Supplemental Material
\cite{supplemental_material}.

At \(R=28\,\unitmu\mathrm m\), \(\mathrm H\) carries 99.4\% of the synthetic
\(\mathrm{HD}\) value of \(P_b\), yet gives only
\(\mathcal V_2[0,R]=1.1\times10^{-5}\).  Control
\(\mathrm D\) carries only 4.4\% of the \(P_b\) in \(\mathrm H\), but gives
\(\mathcal V_2[0,R]=0.342\), comparable to 0.332 for
\(\mathrm{HD}\).  Thus longitudinal helicity variation populates \(b\), but
local lossless evolution keeps it almost exactly in quadrature with \(c,d\);
transverse coupling makes the twofold interference observable.  Control
\(\mathrm D\) also shows that transverse coupling can populate \(b\) at fixed
helicity; longitudinal helicity variation is therefore not required for the
twofold response.  The slightly
larger response of \(\mathrm D\) reflects coherent amplitude addition, not
additive intensities.

Control \(\mathrm H\) also makes the imaging statement explicit: nearly
the whole \(b\) population can sit behind a fourfold intensity image.
Polarization-sensitive holography reveals the
phase relations that suppress the twofold modulation
\cite{reztsov2025toron_phm,shalev2025oam_spherulite}.

The diagnostic tests angular-order content rather than resemblance to a
canonical four-lobed image.
The toron's measured twofold response persists within the covariant
sector.  For the same modeled director, the marching-wave model qualitatively
captures the central twofold morphology absent from the local-Jones
approximation.  Transverse propagation can therefore produce a two-lobed
image without changing director symmetry.

Because the derivation concerns a rotationally covariant transverse
\(2\times2\) response rather than a liquid-crystal constitutive law, its scope
extends to other axisymmetric liquid-crystal solitons and unit-charge
\(q\)-plates, as well as to rotationally covariant metasurface optics.
Material covariance constrains the angular orders; the complete
optical map determines which of the allowed interferences appear in the image.

\begin{samepage}
\begin{acknowledgments}
This work was supported by the Russian Science Foundation (Project
No.~26-12-00289).
\end{acknowledgments}
\end{samepage}

\par\addvspace{\baselineskip}
Lobanov and Chernykh conceived the study.  Reztsov and Chernykh performed the
optical experiments.  Orlova prepared the liquid-crystal samples and
topological solitons.  Lobanov and Shalev performed numerical simulations,
developed the software, analyzed the data, and prepared the figures.  Lobanov,
Aksenova, Orlova, Uzdin, Petrov, and Chernykh contributed to the theory,
manuscript preparation, and validation of the results.  Chernykh secured
funding and coordinated the project.

\par\addvspace{\baselineskip}
\paragraph*{Data Availability}
The calculation, analysis, and figure-generation code is
publicly available on GitHub; the matching code and data release is archived
on Zenodo \cite{fourfold_repository}.
A matching experimental-data artifact contains the 27 reconstructed
single-acquisition Jones fields and the five dual-exposure direct-polarimetry
states used here, together with their dark and \texttt{Ground} calibration
frames.  A separate calculated-data artifact provides the two saved
director fields and the derived results.  The registered fields, optical
calculations, analyses, and figures are reproducible from the experimental
inputs and saved directors.

\par\addvspace{\baselineskip}
\paragraph*{Conflicts of Interest}
The authors declare no conflicts of interest.

\par\addvspace{\baselineskip}
\paragraph*{AI Use Disclosure}
The authors used OpenAI Codex and ChatGPT (GPT-5.4 and GPT-6 Astra) to
assist with drafting and revising the manuscript and writing code for
calculations, data analysis, and figure generation.  The authors directed
this work, reviewed the generated text and code, verified the results,
and take full responsibility for the final manuscript.

\nocite{lcsim_git,li2005refractive,topovec_git}
\input{paper.bbl}

\clearpage
\appendix
\setcounter{secnumdepth}{1}
\setlength{\abovedisplayskip}{5pt plus 1pt minus 1pt}
\setlength{\belowdisplayskip}{5pt plus 1pt minus 1pt}
\setlength{\abovedisplayshortskip}{3pt plus 1pt}
\setlength{\belowdisplayshortskip}{3pt plus 1pt minus 1pt}

\section{Experimental reconstruction and material parameters}
\label{app:experiment}

The sample was a homeotropic liquid-crystal (LC) cell (E.H.C., Japan) filled
with a chiral nematic mixture of 5CB (CAS 40817-08-1; Sigma-Aldrich) and the
left-handed chiral dopant S811 (HCCH, China; CAS 87321-20-8) at a concentration
of \(0.27\)\,wt.\%.  Measurements were performed at zero applied voltage.
We measured the LC cell gap to be \(27.4\pm1.3\,\unitmu\mathrm m\) by
interferometry.  The helix pitch was \(29\pm3\,\unitmu\mathrm m\), as determined
from repeated Grandjean--Cano measurements using a KCRS-05 wedge cell and the
manufacturer-specified \(+8\%\) allowance for spacer thickness (E.H.C.,
Japan).  The quoted uncertainty is the standard deviation across the repeated
measurements.  Optical measurements were performed 15 months after
filling the homeotropic LC cell.

The holographic data retain the nominal wavelength
labels \(450\), \(520\), and \(660\,\mathrm{nm}\), whereas the calculations and
propagation of the full Jones matrix use \(453\), \(520\), and
\(662\,\mathrm{nm}\), respectively.  In the nominal \(520\)-nm acquisition,
the horizontal- and vertical-input sources had different central
wavelengths.  The complete reconstructed Jones matrix is propagated at the
common representative wavelength \(520\,\mathrm{nm}\).  Propagating the two
input columns separately at \(518.2\) and \(522\,\mathrm{nm}\), respectively,
changes the projected
exit-plane \(\mathcal V_2\) only from \(0.252982\) to
\(0.252938\), and changes the complete complex \(H_2\) profile by \(0.40\%\)
(Sec.~\ref{sm:sec:experimental-audits} of the Supplemental Material
\cite{supplemental_material}).

Spatially resolved complex Jones fields were reconstructed by
polarization-sensitive digital holography
\cite{reztsov2025toron_phm,shalev2025oam_spherulite}, using the off-axis
Mach--Zehnder implementation described by Reztsov \textit{et al.}
\cite{reztsov2025toron_phm}.  Separate, mutually incoherent diode-laser
modules provided object beams for the two mutually orthogonal linear input
states.  Each object beam was coherent with its corresponding reference
field, and a Wollaston prism imparted distinct tilts to the two reference
fields.  A FLIR BFS-U3-51S5P-C division-of-focal-plane
camera recorded the analyzer angles
\[
 \begin{pmatrix}0^\circ&135^\circ\\45^\circ&90^\circ\end{pmatrix}
\]
in each \(2\times2\) pixel block.  Spatially separated first-order Fourier
sidebands yielded three complex projections
of the output field for each input state, thereby
reconstructing the complete Jones matrix.

At each camera position and wavelength, complex first-order sidebands from 10
sample and 10 empty-field polarization-mosaic frames, acquired with a
\(2\,\mathrm{ms}\) exposure, were averaged separately
to reduce temporal noise.  Empty-field
normalization removed the reference fields and aberrations common to the
sample and empty-field measurements.  Before export, we removed the residual
slowly varying wavefront and the measured complex beam-splitter response.
Nine Jones reconstructions of distinct torons in the same cell at
different camera positions
were centered by minimizing the leading noncovariant angular power, registered
on their physical grids, and combined by drizzle averaging without adjustment
of their global phases.  The resulting coherently registered and cropped
\(96\times96\,\unitmu\mathrm m^2\) field has an output pixel pitch of
\(0.0682\,\unitmu\mathrm m\).  The working native Jones-component pixel pitch,
\(0.1363(63)\,\unitmu\mathrm m\), was obtained from a calibration slide.

We define the system-level coherent detection cutoff
\(\mathrm{NA}_{\mathrm{eff}}=\lambda f_{\perp,c}\) from the common radial
edge \(f_{\perp,c}\) of the four-component Jones spectra.  Across all 27
source fields, the median is \(\mathrm{NA}_{\mathrm{eff}}=0.4990\), with a
robust field-to-field repeatability scatter of 0.0022.  The \(520\)-nm ensemble
used in the Letter has a median of 0.4982 and a scatter of 0.0010.  These values
characterize the repeatability of the complete detection-and-reconstruction
transfer edge rather than that of the nominal objective alone.  The edge has a
typical half-width of approximately 0.02 in numerical aperture.

With the director-model parameters held fixed, matching the measured annular
focus to that of the marching-wave calculation fixes the axial origin.  Only
the axial offset is adjusted; no Jones-component or harmonic amplitude is
fitted during this registration.  Homogeneous propagation by
\(6.23\,\unitmu\mathrm m\) brings the reconstructed field to the model-referenced
exit plane.  There, the noncovariant remainder is
\[
 \frac{\lVert(1-P_{\rm cov})(J-J_\infty)\rVert_2}
 {\lVert J-J_\infty\rVert_2}=11.1\%
\]
over \(0\leq\rho\leq45\,\unitmu\mathrm m\), where \(P_{\rm cov}\) projects onto the
circular-basis component--order combinations in
Eq.~\eqref{eq:circular-jones}, and \(J_\infty\) is the outer background.

For each stated radial interval, we define
\[
 \lVert f\rVert_\rho^2=\int_{\rho_1}^{\rho_2}|f(\rho)|^2\rho\dd\rho,
 \quad \mathcal V_m=2\frac{\lVert H_m\rVert_\rho}{\lVert H_0\rVert_\rho}.
\]

See Supplemental Material, Secs.~\ref{sm:sec:center-pupil}--\ref{sm:sec:ensemble},
\ref{sm:sec:focus-data}, and~\ref{sm:sec:propagation}--\ref{sm:sec:experimental-audits},
for reconstruction, pupil, registration, projection, and independent direct
polarimetric checks \cite{supplemental_material}.

\section{Director and optical calculations}
\label{app:numerics}

The director calculation uses the Frank--Oseen free energy
\begin{equation}
 \begin{aligned}
 F=\frac12\int\!\big[&
 K_1(\nabla\!\cdot\!\vect n)^2
 +K_2(\vect n\!\cdot\!\nabla\!\times\!\vect n+q_0)^2\\
 &+K_3|\vect n\!\times\!\nabla\!\times\!\vect n|^2\big]\dd^3\vect r.
 \end{aligned}
 \label{eq:frank}
\end{equation}
Here \(\vect n\) is the unit director field, \(K_1,K_2,K_3\) are the splay,
twist, and bend elastic constants, and \(q_0\) is the preferred chiral
wavenumber in the stated sign convention.
The archived toron was numerically relaxed for \(1000\,\mathrm{s}\) of
simulated time using the \texttt{lcsim} code of Ref.~\cite{lcsim_git} on a
\(200\times200\times51\) grid with an isotropic node spacing of
\(0.548\,\unitmu\mathrm m\) and a thickness of \(27.4\,\unitmu\mathrm m\).  At the
lateral boundaries, the director is fixed along the cell normal,
\(\vect n=\vect e_z\).  Both substrates use the finite-strength homeotropic
\texttt{SoftAxis} condition with anchoring strength
\(W_{\rm s}=5.48\times10^{-6}\,\mathrm{J\,m^{-2}}\); its surface functional is
given in Sec.~\ref{sm:sec:director-model} of the Supplemental Material
\cite{supplemental_material}.  We use \(K_1=6.4\), \(K_2=3.0\),
\(K_3=9.1\,\mathrm{pN}\), and \(q_0=-0.2285\,\unitmu\mathrm m^{-1}\),
corresponding to a \(27.5\,\unitmu\mathrm m\) pitch.  The archived metastable
state serves as the common input to both optical models.  It is a comparison
texture, not an inverse reconstruction of the experimental director.

The elastic constants and anchoring strength are model parameters selected
manually to reproduce experimental observations.  This selection did not
constitute a systematic multiparameter fit, and the comparison
includes no sensitivity analysis of these constants.  The resulting director is held
fixed in the comparison of local-Jones and marching-wave propagation.

For the primary \(520\)-nm calculation, the pure-5CB dispersion reported in
Ref.~\cite{li2005refractive} gives the ordinary and extraordinary indices
\(n_o=1.5441\) and \(n_e=1.7336\), respectively, at
\(25\,{}^\circ\mathrm C\).  Dispersion data for the precise 5CB--S811 mixture
were unavailable, so the mixture is approximated using these pure-5CB values.
Measurements in the additional nominal \(450\)- and \(660\)-nm data sets and
calculations at \(453\) and \(662\,\mathrm{nm}\), respectively, give the same
qualitative distinction: the local-Jones patterns are closest to
fourfold symmetry, while marching-wave propagation changes the radial phase,
fringe positions, and relative lobe intensities.  The corresponding simulation
wavelengths, refractive indices, and comparisons are given in
Secs.~\ref{sm:sec:optical-model} and~\ref{sm:sec:linear-comparisons} of the
Supplemental Material \cite{supplemental_material}.

The forward, lossless uniaxial model omits absorption, substrate reflections,
backward waves, and multiple reflections.  Centered isotropic planar
interfaces preserve the rotational selection rule, although the
omitted effects can shift radial phases and amplitudes.  The local-Jones model
computes
\[
 J_{\rm loc}=\mathcal P\exp\left[\ii\int_0^L G(z)\dd z\right],
\]
independently in each transverse column.  Here \(L\) is the cell thickness,
\(G\) is the local Jones generator, and \(\mathcal P\) denotes ordering along
the propagation direction.  The published \texttt{topovec} 0.1.17
marching-wave solver of Ref.~\cite{topovec_git} propagates the forward
electric--magnetic branch, including anisotropic longitudinal-field coupling
and transverse Fourier propagation.
Poy and \v{Z}umer separate local phase evolution from diffraction and
walk-off in their beam-propagation formulation
\cite{poy_zumer2020bpm}.
The full electric--magnetic formulation
and its one-way radial reduction are given in
Secs.~\ref{sm:sec:maxwell} and~\ref{sm:sec:radial-evolution} of the
Supplemental Material \cite{supplemental_material}.  The director grid is
interpolated by a factor of two in the transverse directions, giving
\(0.274\,\unitmu\mathrm m\) optical sampling.

The maximum internal phase increment for the primary \(520\,\mathrm{nm}\)
marching-wave field is \(0.025\) rad.  For the same finite-anchoring director
at \(520\,\mathrm{nm}\), before application of the matched pupil, reducing
this increment to \(0.0125\) rad decreases the normalized amplitude of \(b\)
by \(0.246\%\) and the integrated crossed-intensity twofold amplitude by
\(3.09\%\).  After one optimal global complex scaling, the
background-subtracted Jones fields differ by \(0.677\%\) rms within
\(\rho\leq28\,\unitmu\mathrm m\).  These differences characterize phase-step
sensitivity for the reported field; they do not bound uncertainties due to the
director model, the one-way approximation, or the finite transverse domain.
All calculations retain only the forward branch.

The published scripts \cite{fourfold_repository} generate the figures from
measured and calculated data.

\section{Strict symmetry identities}
\label{app:symmetry}

For a rotationally covariant field, the radial coefficients can be extracted
directly from the Cartesian matrix elements.  With
\(X=(J_{xx}-J_{yy})/2\) and \(Y=(J_{xy}+J_{yx})/2\),
\begin{equation}
 \begin{aligned}
 a&=\tfrac12(J_{xx}+J_{yy}),&
 b&=\tfrac12(J_{yx}-J_{xy}),\\
 c&=X\cos2\varphi+Y\sin2\varphi,&
 d&=-X\sin2\varphi+Y\cos2\varphi.
 \end{aligned}
 \label{eq:coefficient-extraction}
\end{equation}
These coefficients are independent of \(\varphi\) in the exact symmetry
sector.  Any residual \(\varphi\)-dependence of the extracted coefficients
signals noncovariance, including circular-polarization--order mismatches that
remain within the aggregate order set \(m=0,\pm2\).

For any invertible \(2\times2\) matrix \(M\),
\(\Rmat_{\pi/2}M\Rmat_{-\pi/2}=\det(M)M^{-\mathsf T}\).  Applying this identity
to Eq.~\eqref{eq:rotation-covariance} at \(\beta=\pi/2\) for a pointwise
unitary Jones matrix gives
\begin{equation}
 J(\rho,\varphi+\pi/2)=\det J(\rho)J(\rho,\varphi)^*.
 \label{eq:quarter-turn}
\end{equation}
Pointwise unitarity gives \(|\det J|=1\).  For real linear input and analyzer
states, Eq.~\eqref{eq:quarter-turn} therefore implies
\(I(\varphi+\pi/2)=I(\varphi)\).  This proves the emergent local \(C_4\)
symmetry without requiring \(b\), \(c\), or \(d\) to vanish.

For transverse input and output coordinates \(\vect r_i\) and \(\vect r_o\),
a nonlocal system with transfer kernel \(\mathcal T\) relates the transverse
electric fields by
\[
 \vect E_{\rm out}(\vect r_o)=
 \int \mathcal T(\vect r_o,\vect r_i)
 \vect E_{\rm in}(\vect r_i)\dd^2\vect r_i.
\]
Losslessness constrains the complete kernel as an operator on the full
transverse field and hence conserves total output power; it does not make the
reduced response \(J(\vect r_o)\) pointwise unitary or normal.  For a
rotationally covariant system, the angular selection rule and \(C_2\) remain
exact, whereas Eq.~\eqref{eq:quarter-turn} and \(C_4\) need not.  The full
kernel covariance and electric--magnetic radial Maxwell state are derived in
Sec.~\ref{sm:sec:maxwell}; the one-way reduction, substrate-exchange
reciprocity, and controlled helicity/diffraction tests are given in
Secs.~\ref{sm:sec:radial-evolution},~\ref{sm:sec:surface-exchange},
and~\ref{sm:sec:controls} of the Supplemental Material
\cite{supplemental_material}.

\arxivStartSupplement{Supplemental Material for
\textquotedblleft Emergent Fourfold Symmetry in Polarization Images:
Local Origin and Diffractive Breakdown\textquotedblright}
\input{supplement.tex}

\end{document}

%% file: setup.tex
\usepackage{overpic}
\DeclareMathSizes{11}{10.95}{8}{6}
\makeatletter
\let\arxiv@section\section
\let\arxiv@paragraph\paragraph
\let\arxiv@hangfrom@section\@hangfrom@section
\let\arxiv@sectioncntformat\@sectioncntformat
\let\arxiv@runin@tos@section\@runin@tos@section
\let\arxiv@runin@to@section\@runin@to@section
\let\arxiv@runin@tos@paragraph\@runin@tos@paragraph

\newcommand{\arxivSupplementContents}{}
\newcommand{\arxivTableOfContents}{%
  \begingroup
  \let\arxiv@contentsline\contentsline
  \long\def\contentsline##1##2##3##4{}%
  \let\appendix\relax
  \def\arxivSupplementContents{\let\contentsline\arxiv@contentsline}%
  \tableofcontents
  \endgroup
}

\newcommand{\arxivStartSupplement}[1]{%
  \clearpage
  \onecolumngrid
  \renewcommand{\floatpagefraction}{0.6}%
  \renewcommand{\dblfloatpagefraction}{0.6}%
  \let\section\arxiv@section
  \let\paragraph\arxiv@paragraph
  \let\@hangfrom@section\arxiv@hangfrom@section
  \let\@sectioncntformat\arxiv@sectioncntformat
  \let\@runin@tos@section\arxiv@runin@tos@section
  \let\@runin@to@section\arxiv@runin@to@section
  \let\@runin@tos@paragraph\arxiv@runin@tos@paragraph
  \@removefromreset{equation}{section}%
  \def\theequation@prefix{}%
  \setcounter{section}{0}%
  \setcounter{subsection}{0}%
  \setcounter{subsubsection}{0}%
  \setcounter{equation}{0}%
  \setcounter{figure}{0}%
  \setcounter{table}{0}%
  \setcounter{secnumdepth}{2}%
  \setcounter{tocdepth}{2}%
  \renewcommand{\thesection}{\Roman{section}}%
  \renewcommand{\thesubsection}{\Alph{subsection}}%
  \renewcommand{\thesubsubsection}{\arabic{subsubsection}}%
  \renewcommand{\theequation}{S\arabic{equation}}%
  \renewcommand{\thefigure}{S\arabic{figure}}%
  \renewcommand{\thetable}{S\arabic{table}}%
  \renewcommand{\theHsection}{S.\arabic{section}}%
  \renewcommand{\theHsubsection}{\theHsection.\arabic{subsection}}%
  \renewcommand{\theHsubsubsection}{\theHsubsection.\arabic{subsubsection}}%
  \renewcommand{\theHequation}{S.\arabic{equation}}%
  \renewcommand{\theHfigure}{S.\arabic{figure}}%
  \renewcommand{\theHtable}{S.\arabic{table}}%
  \renewcommand\normalsize{%
    \@setfontsize\normalsize{11}{14}%
    \abovedisplayskip 8pt plus 2pt minus 2pt
    \belowdisplayskip 8pt plus 2pt minus 2pt
    \abovedisplayshortskip 3pt plus 1pt
    \belowdisplayshortskip 3pt plus 1pt minus 1pt
  }%
  \normalsize
  \phantomsection
  \label{arxiv:supplement}%
  \addtocontents{toc}{\protect\arxivSupplementContents}%
  \begin{center}\large\bfseries #1\end{center}
}
\makeatother

%% file: supplement.tex
This Supplemental Material presents the experimental data and calibration,
the symmetry derivations, the two forward models, and their numerical
validation.  All quantitative experimental statistics are derived from
holographic measurements of nine distinct neighboring torons at each
wavelength and an independently acquired
through-focus sequence of the same sample.  The holographic statistics
quantify consistency among relaxed torons in the same cell under common
conditions.

\begingroup
\makeatletter
\def\toc@@font{%
 \def\baselinestretch{1}%
 \small
 \setlength{\parskip}{0pt}%
}
\def\l@f@section{\addvspace{0.35em plus 0.1em}\bfseries}
\makeatother
\arxivTableOfContents
\endgroup
\clearpage

\section{Experimental data, calibration, and optical comparisons}
\label{sm:sec:data}

\subsection{Sample and polarization-sensitive holography}
\label{sm:sec:experiment}

The sample was a homeotropic LC cell (E.H.C., Japan) filled with a chiral
nematic mixture of 5CB (CAS 40817-08-1; Sigma-Aldrich) and the left-handed
chiral dopant S811 (HCCH, China; CAS 87321-20-8) at a concentration of
\(0.27\)\,wt.\%.  Measurements were performed at zero applied voltage.
We measured the LC cell gap to be \(27.4\pm1.3\,\unitmu\mathrm m\) by
interferometry.  The helix pitch was \(29\pm3\,\unitmu\mathrm m\), as determined
by repeated Grandjean--Cano measurements using a KCRS-05 wedge cell with the
manufacturer-specified \(+8\%\) allowance for spacer thickness (E.H.C.,
Japan).  The quoted uncertainty is the standard deviation over the repeated
measurements.  Optical measurements were performed 15 months after
filling the homeotropic LC cell.

The holographic archives retain the nominal wavelength labels
450, 520, and 660~nm;
calculations and homogeneous propagation of the complete Jones fields use
453, 520, and 662~nm, respectively. In the nominal 520-nm acquisition,
spectra of the beams supplying the horizontal and vertical input states were
measured with an Avesta ASP100 spectrometer and centered at
\(518.2\pm0.5\,\mathrm{nm}\) and
\(522.5\pm0.5\,\mathrm{nm}\), respectively.
The complete reconstructed Jones matrix is propagated at the common
representative wavelength
\(520\,\mathrm{nm}\).  A split-column sensitivity control instead propagates
the horizontal and vertical input columns at \(518.2\) and
\(522\,\mathrm{nm}\), respectively, before the circular-basis projection and
harmonic analysis; its results are given in
Sec.~\ref{sm:sec:experimental-audits}.

The spatial Jones matrix is defined by
\begin{equation}
 \vect E_{\rm out}(x,y)=J(x,y)\vect E_{\rm in},\qquad
 J=\begin{pmatrix}J_{xx}&J_{xy}\\J_{yx}&J_{yy}\end{pmatrix},
 \label{sm:eq:jones-convention}
\end{equation}
where the first index labels the analyzed output component and the second the
input component.  Spatially resolved complex Jones fields were reconstructed
by polarization-sensitive digital holography
\cite{reztsov2025toron_phm,shalev2025oam_spherulite}, using the off-axis
Mach--Zehnder implementation described by Reztsov \textit{et al.}
\cite{reztsov2025toron_phm}.  Separate, mutually incoherent diode-laser
modules supplied object beams for the two mutually orthogonal linear input
states.  Each object field was coherent with its corresponding reference
field, and a Wollaston prism gave the two reference fields distinct tilts.  A FLIR
BFS-U3-51S5P-C division-of-focal-plane camera recorded four analyzer
projections in each \(2\times2\) pixel block,
\begin{equation}
 \begin{pmatrix}0^\circ&135^\circ\\45^\circ&90^\circ\end{pmatrix}.
 \label{sm:eq:camera-mosaic}
\end{equation}
The two reference-wave tilts placed the first-order Fourier sidebands in
spatially separated spectral regions.
Their complex amplitudes yielded three projections
of the output field for each input state and
therefore all four elements of Eq.~\eqref{sm:eq:jones-convention}.

At each camera position and wavelength, complex first-order sidebands from 10
sample and 10 empty-field polarization-mosaic frames, acquired with a
\(2\,\mathrm{ms}\) exposure, were averaged separately
to reduce temporal noise.  Empty-field
normalization removed the reference fields and aberrations common to the
sample and empty-field measurements.  Before export, we removed the remaining
slowly varying wavefront and the measured complex response of the
post-objective beam splitter.  An independent sideband reconstruction
confirmed the component ordering and phase convention.  The exported Jones
fields preserve the
complex scale established by the empty-field calibration, which sets the
incident-field normalization for all subsequent propagation and comparisons.
The working native Jones-component pixel pitch in the reconstruction was
\(0.1363(63)\,\unitmu\mathrm m\), as determined from a calibration slide.

For a unit linearly polarized input
\(\vect e_\beta=(\cos\beta,\sin\beta)^{\mathsf T}\), the analyzer-resolved
intensity is
\begin{equation}
 I_{\alpha\leftarrow\beta}(x,y)
 =\left|\vect e_\alpha^{\mathsf T}J(x,y)\vect e_\beta\right|^2.
 \label{sm:eq:linear-channel-definition}
\end{equation}
Unless stated otherwise, the figures use \(\beta=0\) and abbreviate this
quantity as \(I_\alpha\).  The corresponding total output intensity is
\(S_0=\lVert J\vect e_0\rVert^2=I_0+I_{90}\).  For each analysis,
\(J_\infty\) denotes the stated estimate of the complex response in its outer
background annulus; the relevant analysis specifies whether an arithmetic or
robust mean is used.  Jones amplitudes and derived intensities
therefore remain on the incident-field scale fixed by the empty-field
calibration; no panel-dependent radiometric adjustment is applied.

With \((\rho,\varphi)\) denoting polar coordinates about the fitted toron
center, the crossed-polarizer intensity is
\(I_\times=I_{90\leftarrow0}\).  Its angular coefficients are
\begin{equation}
 H_m(\rho)=\frac{1}{2\pi}\int_0^{2\pi}
 I_\times(\rho,\varphi)\ee^{-\ii m\varphi}\dd\varphi.
 \label{sm:eq:experimental-harmonics}
\end{equation}
For the stated radial interval, let
\(\lVert f\rVert_\rho^2=\int_{\rho_1}^{\rho_2}|f(\rho)|^2\rho\dd\rho\).
We quantify the local and integrated harmonic amplitudes by
\begin{align}
 v_m(\rho)
 &=\frac{2|H_m(\rho)|}{H_0(\rho)},
 \label{sm:eq:local-harmonic-visibility}\\
 \mathcal V_m[\rho_1,\rho_2]
 &=2\frac{\lVert H_m\rVert_\rho}{\lVert H_0\rVert_\rho},
 \qquad m=2,4.
 \label{sm:eq:integrated-visibility}
\end{align}

Two axial coordinates are used.  In equations for propagation inside the
liquid crystal, \(z\) denotes material depth.  Homogeneous propagation from a
measured or calculated output plane is initially parameterized by \(\zeta\).
After the single axial registration defined in Sec.~\ref{sm:sec:focus-data}, the
propagation figures reuse \(z=\zeta-\zeta_{\rm MW}\) as a model-referenced
observation coordinate, with the origin at the estimated liquid-crystal exit
plane.  The relevant convention is stated in each context.

\subsection{Centering, coherent registration, and pupil calibration}
\label{sm:sec:center-pupil}

At each wavelength, nine Jones fields of distinct neighboring torons
in the same cell were recorded at
different camera positions.  For each trial center, the field is transformed as
\(\widehat J=\Rmat_{-\varphi}J\Rmat_\varphi\) and decomposed into angular
harmonics.  An exactly rotationally covariant response makes
\(\widehat J\) independent of \(\varphi\); equivalently, in the circular basis
its same-helicity Jones components contain only \(m=0\), while its
helicity-converting Jones components contain only \(m=\pm2\).  Miscentering
generates leading \(m=\pm1\) components of \(\widehat J\).  We locate the center
by minimizing their power relative to that of the radially varying Jones field
after subtraction of its annular mean.  The objective is integrated over
\(4\leq\rho\leq30\,\unitmu\mathrm m\) in radial steps of
\(0.75\,\unitmu\mathrm m\).  The search uses three nested grids: a broad
\(\pm16\)-pixel grid sampled every 2 pixels, a \(\pm2\)-pixel grid sampled
every 0.25 pixel, and a final \(\pm0.5\)-pixel grid sampled every 0.05 pixel.
The three stages use 90, 180, and 360 azimuthal samples, respectively.
Evaluation over each complete grid confirms that the reported minimum does
not lie on a grid boundary.  Joint registration then uses pairwise
complex-field correlations.  Across the 27 source fields, the median pairwise
residuals are 0.019, 0.025, and 0.092 native pixels at 450, 520, and 660 nm,
respectively; the corresponding median closure errors around registration
triangles are 0.050, 0.050, and 0.158 pixels.  These quantities assess internal
registration consistency, not absolute position accuracy in the laboratory
frame.  A radius inferred independently from the radial
polarization-conversion signal is retained as a cross-check but is not used to
locate the center.

The relaxed torons were measured under the same conditions and give
consistent registered optical fields (Sec.~\ref{sm:sec:ensemble}).  Coherent
averaging suppresses background interference fringes caused by instrumental
wavefront distortions: their pattern varies between acquisitions, whereas the
toron structure remains aligned.  The nine complex fields at each wavelength
are registered on
a common physical grid.  Pairwise complex correlations determine the
relative translations, and drizzle resampling onto a grid with half the native
pixel pitch preserves the measured subpixel offsets.  No global phase offset is
fitted because the empty-field calibration already fixes the background
phase.  We verify phase consistency in the outer annulus before averaging and
otherwise leave each source-field phase unchanged.  The resulting
coherently registered and cropped field spans
\(96\times96\,\unitmu\mathrm m^2\) at a pixel pitch of
\(0.0682\,\unitmu\mathrm m\).  Its manifest records the source centers,
phase-coherence diagnostics, resampling rule, and physical scale.

We define the system-level effective coherent detection cutoff as
\(\mathrm{NA}_{\rm eff}=\lambda f_{\perp,c}\), where \(f_{\perp,c}\) is the
common radial edge in the spatial-frequency spectra of all 27 source Jones
fields.  For each field, we subtract the mean Jones matrix over pixels
satisfying
\(\rho_{\rm px}\geq0.44\min(N_x,N_y)\), where \(\rho_{\rm px}\) is the radial
distance from the array center in pixels and \(N_x,N_y\) are the array
dimensions.  We then apply a two-dimensional Tukey window with taper fraction
0.2, sum the Fourier power of all four complex Jones components, and average it
in
radial bins of width \(0.005\,\unitmu\mathrm m^{-1}\).  A cubic
Savitzky--Golay filter of width \(0.15\,\unitmu\mathrm m^{-1}\) smooths the
logarithmic power.  The cutoff is the quadratically interpolated minimum of
its derivative over \(0.40\leq\mathrm{NA}\leq0.56\), where
\(\mathrm{NA}=\lambda f_\perp\) and \(f_\perp\) is the radial spatial
frequency.
The median effective numerical aperture is
\(\mathrm{NA}_{\rm eff}=0.4990\), with a robust field-to-field repeatability
scatter of 0.0022, defined as 1.4826 times the median absolute deviation.
This value characterizes the complete detection-and-reconstruction transfer
rather than the nominal \(50\times/0.55\) objective alone.
Individual edge fits have a typical half-width of about 0.02 in numerical
aperture.  Because the measured Jones fields already include this coherent
transfer function, exit-plane instrument-matched comparisons apply an
equivalent centered, circular, scalar pupil once, component by component, to
the calculated fields only.  No second pupil is applied to the measured field
in the exit-plane symmetry analysis.  Separately, the through-focus recovery
below uses the same cutoff as an explicit numerical propagation band limit in
Eq.~\eqref{sm:eq:focus-forward}.  The calibration does not include a
polarization-dependent or noncircular pupil; either would constitute a
distinct instrumental mechanism.

\subsection{Consistency across nine holographic reconstructions}
\label{sm:sec:ensemble}

Figures~\ref{sm:fig:ensemble-450}--\ref{sm:fig:ensemble-660} show every Jones field
included in the coherent averages.  The numbered rows are the individual
reconstructions; the columns show crossed-polarizer intensity, total output
intensity for horizontal input, a phase-sensitive field representation, and
the converted circular-polarization Jones component used to verify
registration.
The final row shows the coherent average on the common physical grid.  Each
source row carries its own scale bar.

Each 520-nm source field is also propagated independently to the common
model-referenced exit plane, where the harmonic and covariance analyses are
repeated before coherent averaging.  Here
\(\mathcal V_m\) is the radial-\(L^2\) crossed-intensity
harmonic amplitude defined in Eq.~\eqref{sm:eq:integrated-visibility} over
\(4\leq\rho\leq28\,\unitmu\mathrm m\).  To obtain the covariant field
\(J_{\rm cov}\), we apply the exact orthogonal projector
\(P_{\rm cov}\), which retains the circular-basis components and angular
orders specified in Sec.~\ref{sm:sec:selection}.  The background response is
estimated with a robust complex Jones average over
\(41\leq\rho\leq45\,\unitmu\mathrm m\), and the
covariance-residual norm is integrated over
\(0\leq\rho\leq45\,\unitmu\mathrm m\).  Across the nine acquisitions, the
medians and robust acquisition-to-acquisition scatters, defined as
\(1.4826\,\operatorname{median}|x-\operatorname{median}x|\), are
\[
 (\mathcal V_2,\mathcal V_4)
 =(0.260\pm0.013,\,0.401\pm0.006),
 \qquad
 \epsilon_{\rm cov}\equiv
 \frac{\lVert(1-P_{\rm cov})(J-J_\infty)\rVert_2}
 {\lVert J-J_\infty\rVert_2}
 =0.148\pm0.005 .
\]
Projecting each field separately onto its rotationally covariant component
gives \((0.254\pm0.005,\,0.409\pm0.002)\).  Before projection, the corresponding
ranges are \(0.241\)--\(0.273\) and \(0.394\)--\(0.415\); after projection they
are \(0.246\)--\(0.263\) and \(0.405\)--\(0.416\).  These values quantify
the observed variation among nine distinct torons under common
conditions, including measurement noise and any differences between the
torons.  The coherently averaged surface field gives
\(\mathcal V_2=0.2512\), inside the individual-acquisition
range and below its median \(0.260\); coherent averaging therefore does not
inflate the reported twofold amplitude.

\clearpage
\begingroup
\addtolength{\textwidth}{-15pt}
\begin{turnpage}
\begin{figure*}[p]
 \centering
 \includegraphics[width=0.95\linewidth,height=0.82\textwidth,
 keepaspectratio]{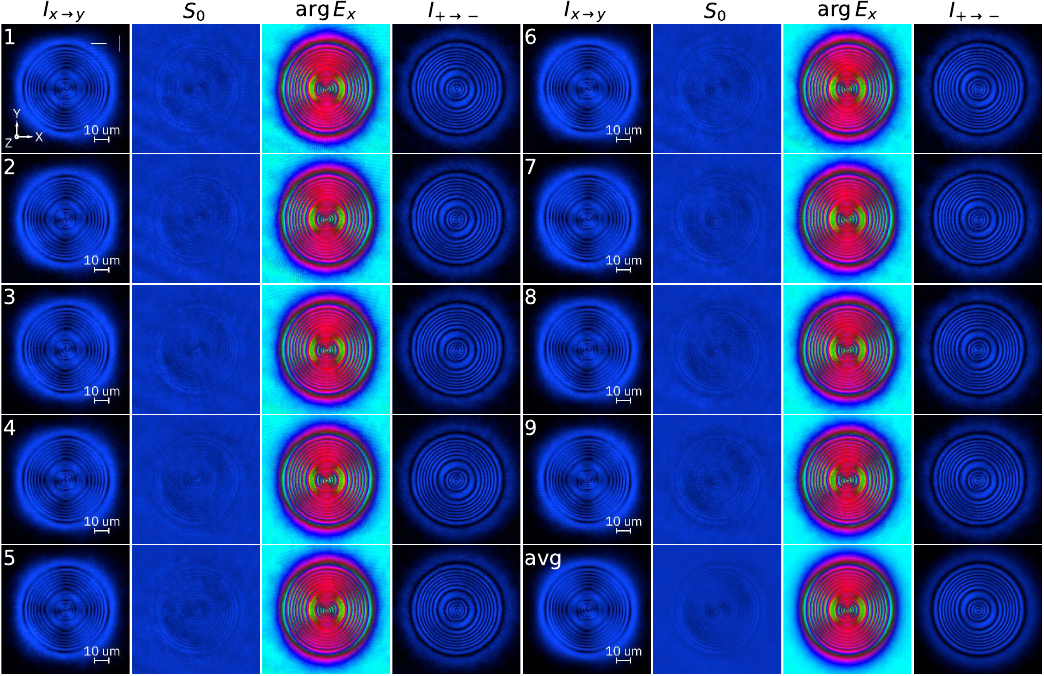}
 \caption{Nine registered Jones measurements of distinct torons at nominal
 \(450\,\mathrm{nm}\) and their coherent complex average.  Each source row
 carries its own scale bar.  Their agreement in toron position, ring order,
 and phase-sensitive structure after registration provides a direct
 consistency check for the subsequent analysis.}
 \label{sm:fig:ensemble-450}
\end{figure*}
\end{turnpage}
\clearpage
\endgroup

\begingroup
\addtolength{\textwidth}{-15pt}
\begin{turnpage}
\begin{figure*}[p]
 \centering
 \includegraphics[width=0.95\linewidth,height=0.82\textwidth,
 keepaspectratio]{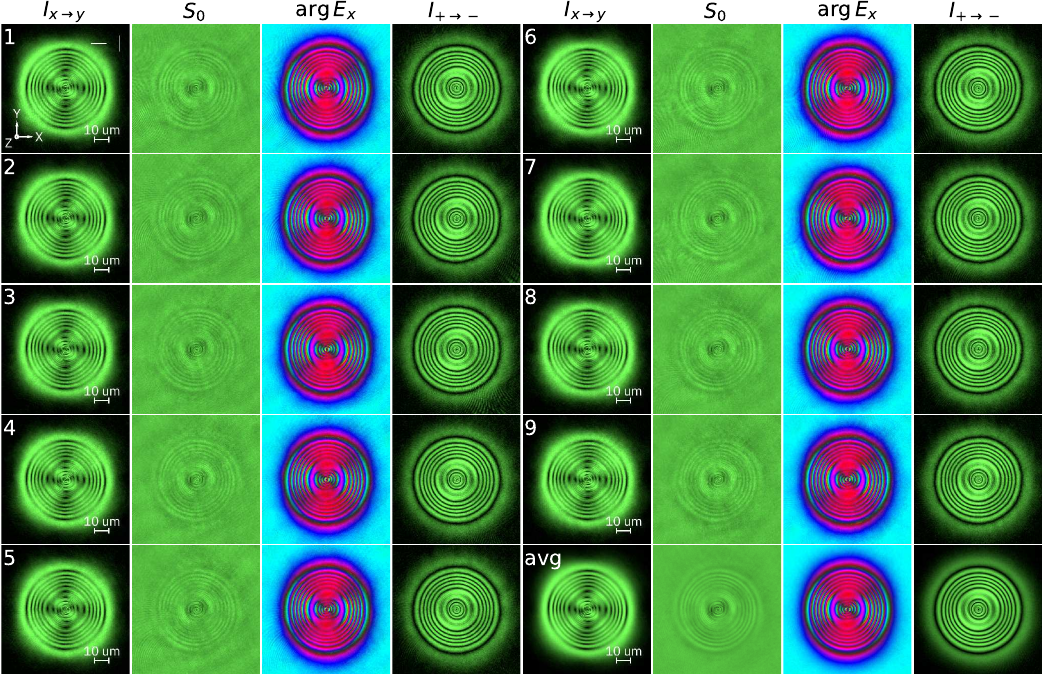}
 \caption{Nine registered Jones measurements at nominal
 \(520\,\mathrm{nm}\) and their coherent complex average, arranged as in
 Fig.~\ref{sm:fig:ensemble-450}.  This averaged field is used for the primary
 \(\lambda=520\,\mathrm{nm}\) comparison below.}
 \label{sm:fig:ensemble-520}
\end{figure*}
\end{turnpage}
\clearpage
\endgroup

\begingroup
\addtolength{\textwidth}{-15pt}
\begin{turnpage}
\begin{figure*}[p]
 \centering
 \includegraphics[width=0.95\linewidth,height=0.82\textwidth,
 keepaspectratio]{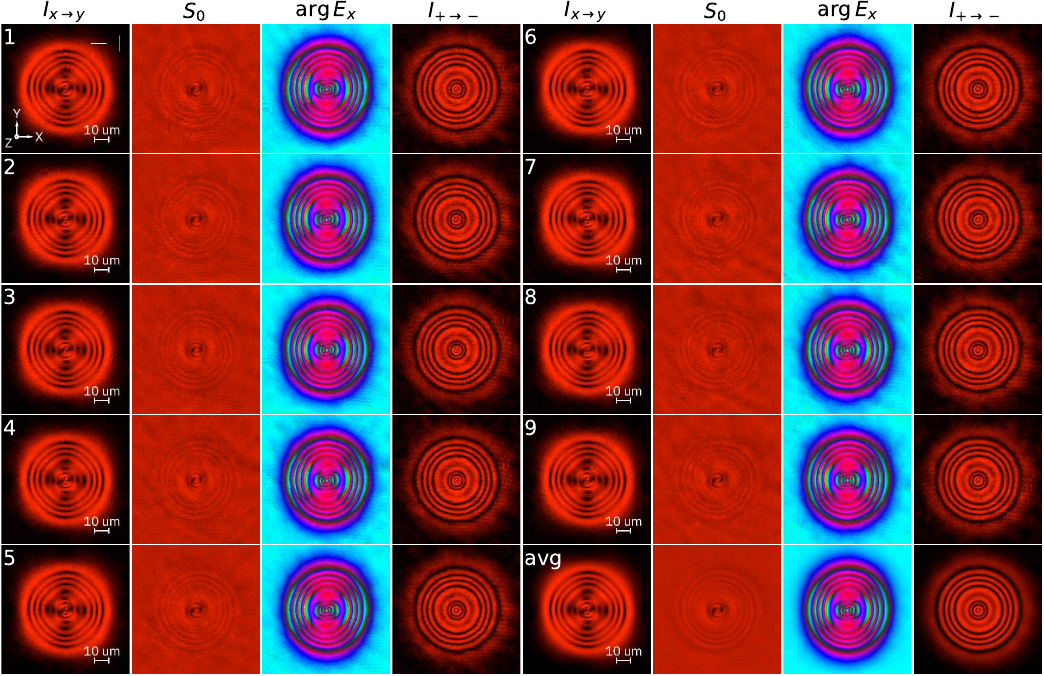}
 \caption{Nine registered Jones measurements at nominal
 \(660\,\mathrm{nm}\) and their coherent complex average, arranged as in
 Fig.~\ref{sm:fig:ensemble-450}.}
 \label{sm:fig:ensemble-660}
\end{figure*}
\end{turnpage}
\clearpage
\endgroup

\subsection{Direct experimental angular-order test}
\label{sm:sec:experimental-modes}

We test the field-order selection rule before imposing rotational covariance.
In the fixed laboratory polarization basis, let
\begin{equation}
 \Delta J_m(\rho)=\frac{1}{2\pi}\int_0^{2\pi}
 [J(\rho,\varphi)-J_\infty]\ee^{-\ii m\varphi}\dd\varphi .
 \label{sm:eq:jones-angular-mode}
\end{equation}
The normalized, radially integrated amplitude of the complete matrix mode is
\begin{equation}
 q_m=
 \left[
 \frac{2\pi\int_0^R\|\Delta J_m(\rho)\|_F^2\rho\dd\rho}
 {\int_{\rho\leq R}\|J-J_\infty\|_F^2\dd^2r}
 \right]^{1/2},
 \qquad p_m=q_m^2 .
 \label{sm:eq:jones-mode-amplitude}
\end{equation}
This amplitude includes all four complex Jones elements and is invariant
under any spatially uniform unitary change of input and output polarization
bases.  The aggregate amplitude outside the allowed order set is
\begin{equation}
 \epsilon_{\rm orders}=
 \left[1-q_{-2}^2-q_0^2-q_{+2}^2\right]^{1/2}.
 \label{sm:eq:allowed-order-leakage}
\end{equation}
This out-of-set angular-order leakage is not \(\epsilon_{\rm cov}\), the
distance to the covariant sector: covariance also locks each order to a
particular circular-basis Jones element.

\begin{figure*}[t]
 \centering
 \includegraphics[width=\textwidth]{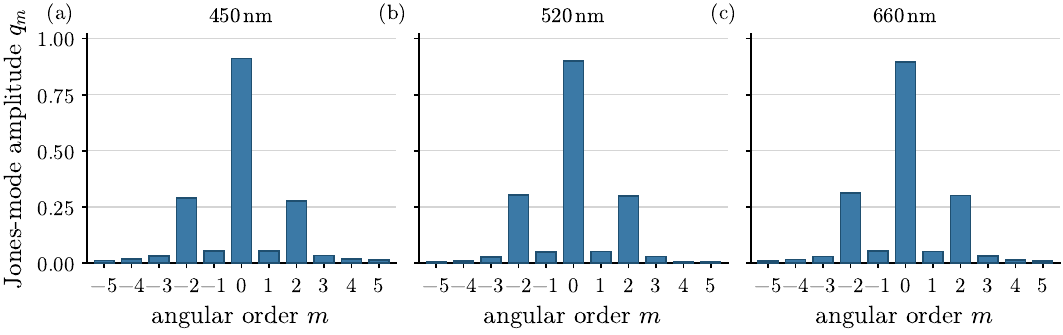}
 \caption{Amplitudes \(q_m\) of the signed angular orders \(m\) in the complete
 background-subtracted, unprojected coherent averages of nine registered
 experimental Jones fields at nominal wavelengths (a) \(450\), (b) \(520\),
 and (c) \(660\,\mathrm{nm}\).
 All panels use \(R=28\,\unitmu\mathrm m\), the fixed laboratory polarization
 basis, and a common linear scale.  The displayed interval
 \(-5\leq m\leq5\) contains more than \(99.8\%\) of the total
 background-subtracted Jones power at
 every wavelength.  Including unplotted orders, the out-of-set angular-order
 leakage
 amplitudes \(\epsilon_{\rm orders}\) are \(10.08\%\), \(8.75\%\), and
 \(9.98\%\), corresponding to power fractions \(1.02\%\), \(0.77\%\), and
 \(1.00\%\), respectively.  The stronger exact circular-basis test, evaluated over
 \(R=45\,\unitmu\mathrm m\), gives
 \(\epsilon_{\rm cov}=15.46\%\), \(11.21\%\), and \(10.84\%\);
 unlike \(\epsilon_{\rm orders}\), it also detects an allowed angular order
 occurring in the wrong circular-polarization Jones component.  Because the
 two diagnostics use different outer radii, their numerical difference is
 not attributable to circular-polarization--order locking alone.  Over
 \(4\leq\rho\leq28\,\unitmu\mathrm m\), the corresponding measured twofold
 amplitudes \(\mathcal V_2\) are \(0.245\),
 \(0.206\), and \(0.193\), respectively.}
\label{sm:fig:experimental-mode-amplitudes}
\end{figure*}

Figure~\ref{sm:fig:experimental-mode-amplitudes} applies these definitions
directly to the coherently averaged measured fields at their holographic
planes.  No projection onto the covariant sector is applied before
calculating the displayed \(q_m\) spectra or \(\epsilon_{\rm orders}\); the
separately quoted \(\epsilon_{\rm cov}\) is the norm rejected by the exact
covariant projector.  Neither diagnostic uses homogeneous propagation or a
calculated field.  Each spectrum is calculated from the coherent average of
the nine registered complex Jones fields; it is
neither a single-acquisition spectrum nor an average of nine spectra.  The
origin is fixed by the nine-position centering and registration procedure
described above rather than refitted for the displayed spectra.
Because suppression of the leading co-rotating \(m=\pm1\) residual contributes
to locating that origin, those orders alone are not an independent test of
covariance.  The simultaneous suppression of the remaining forbidden orders
is not imposed by the two fitted translation parameters.

At \(520\,\mathrm{nm}\), shifting the fixed registered center by
\(0.10\,\unitmu\mathrm m\) along a Cartesian axis changes
\(\epsilon_{\rm orders}\) from \(8.75\%\) to \(10.1\)--\(11.5\%\), and a
\(0.25\,\unitmu\mathrm m\) shift gives \(16.8\)--\(18.6\%\).  The corresponding
\(\epsilon_{\rm cov}\) ranges are \(12.28\)--\(12.64\%\) and
\(17.13\)--\(17.79\%\), versus \(11.21\%\) without an offset.  Replacing the
robust background by the arithmetic annular mean changes either nominal
amplitude by less than \(10^{-8}\); the registered center, rather than the
choice of background estimator, controls this sensitivity.

\subsection{Linear-polarization comparisons at three wavelengths}
\label{sm:sec:linear-comparisons}

\begin{figure}[p]
 \centering
 \includegraphics[height=0.408\textheight,trim=21 0 21 11.5,clip]
 {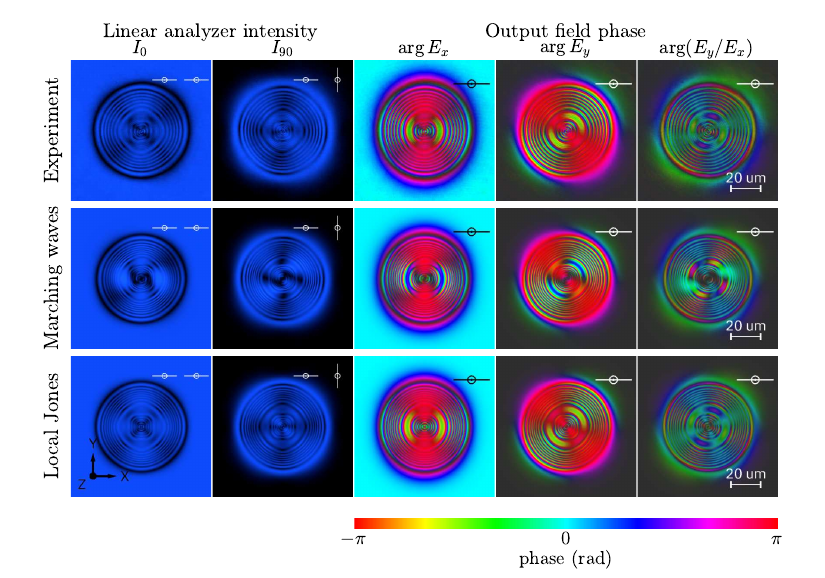}\par\vspace{-3pt}
 \includegraphics[height=0.408\textheight,trim=19.5 0 19.5 11.5,clip]
 {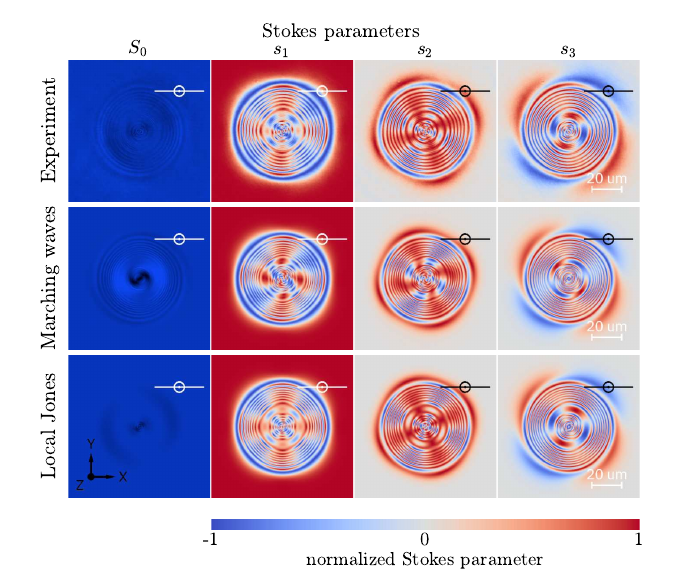}
 \caption{Linear-polarization observables at the nominal wavelength
 \(450\,\mathrm{nm}\) for \(x\)-polarized input.  Top: linear-analyzer
 intensities and output-field phases; columns show \(I_0\), \(I_{90}\),
 \(\arg E_x\), \(\arg E_y\), and \(\arg(E_y/E_x)\).  Bottom: total intensity
 \(S_0\) and normalized Stokes parameters \(s_i=S_i/S_0\), \(i=1,2,3\).
 In both blocks, rows show experiment, marching-wave propagation, and
 independent-column Jones propagation.  Identical panel dimensions give the
 same physical image scale in both blocks.}
 \label{sm:fig:linear-450}
\end{figure}

\begin{figure}[p]
 \centering
 \includegraphics[height=0.408\textheight,trim=21 0 21 11.5,clip]
 {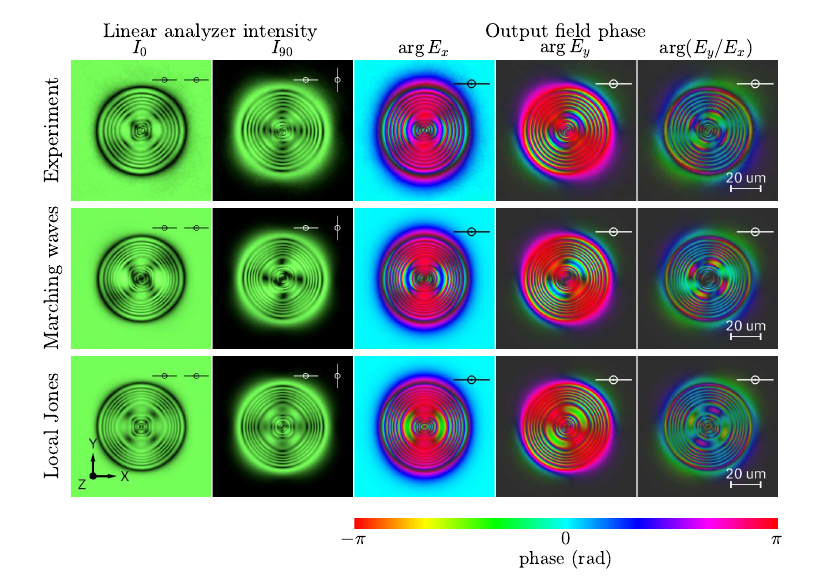}\par\vspace{-3pt}
 \includegraphics[height=0.408\textheight,trim=19.5 0 19.5 11.5,clip]
 {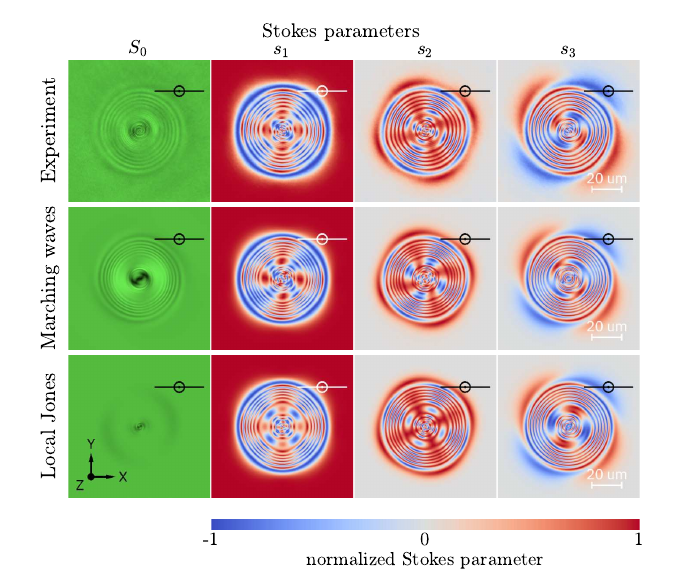}
 \caption{Linear-polarization observables at the nominal wavelength
 \(520\,\mathrm{nm}\), arranged and scaled as in
 Fig.~\ref{sm:fig:linear-450}.  The upper block shows linear intensities and
 output-field phases; the lower block shows \(S_0\) and normalized Stokes
 parameters.}
 \label{sm:fig:linear-520}
\end{figure}

\begin{figure}[p]
 \centering
 \includegraphics[height=0.408\textheight,trim=21 0 21 11.5,clip]
 {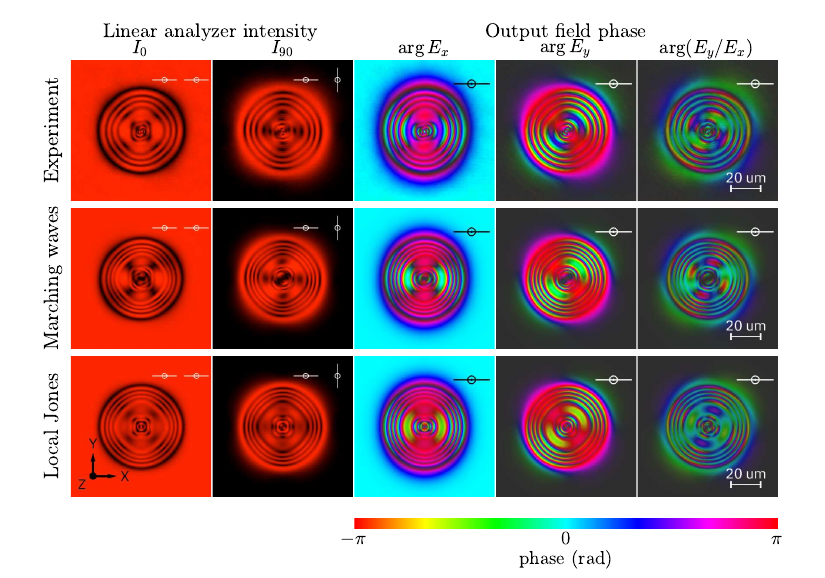}\par\vspace{-3pt}
 \includegraphics[height=0.408\textheight,trim=19.5 0 19.5 11.5,clip]
 {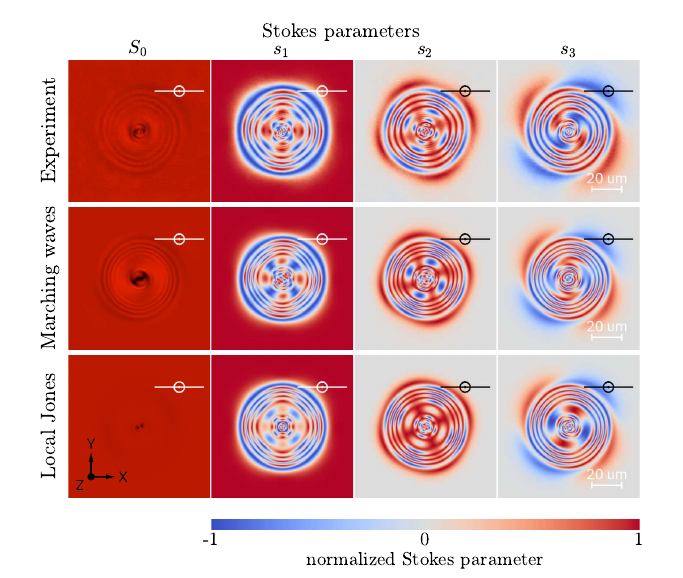}
 \caption{Linear-polarization observables at the nominal wavelength
 \(660\,\mathrm{nm}\), arranged and scaled as in
 Fig.~\ref{sm:fig:linear-450}.  The upper block shows linear intensities and
 output-field phases; the lower block shows \(S_0\) and normalized Stokes
 parameters.}
 \label{sm:fig:linear-660}
\end{figure}

Figures~\ref{sm:fig:linear-450}--\ref{sm:fig:linear-660} compare the averaged
experimental fields with the two forward models specified in
Sec.~\ref{sm:sec:methods}.  Both models use the same relaxed director with finite
surface anchoring.  In each figure, the rows show the experiment,
marching-wave propagation, and independent-column Jones propagation.  Before
comparison, the calculated fields are filtered by an ideal centered circular
pupil whose cutoff is inferred from the measured spatial-frequency spectra;
their Jones amplitudes are not otherwise adjusted.

At all three wavelengths, the independent-column model produces the patterns
closest to fourfold symmetry.  Marching-wave propagation changes the radial
phase, fringe positions, and relative lobe intensities while preserving the
symmetry-allowed angular orders up to the small numerical leakage quantified
below.  Residual experiment--calculation differences may arise from the
reconstructed incident wavefront, the transverse scale and detailed structure
of the modeled director, and optical effects omitted by the forward-only
marching-wave approximation.

\subsection{Circular-basis structure at
\texorpdfstring{\(520\,\mathrm{nm}\)}{520 nm}}
\label{sm:sec:circular-data}

\begin{figure}[p]
 \centering
 \includegraphics[scale=1.03,trim=19.5 26 20.5 11.5,clip]
 {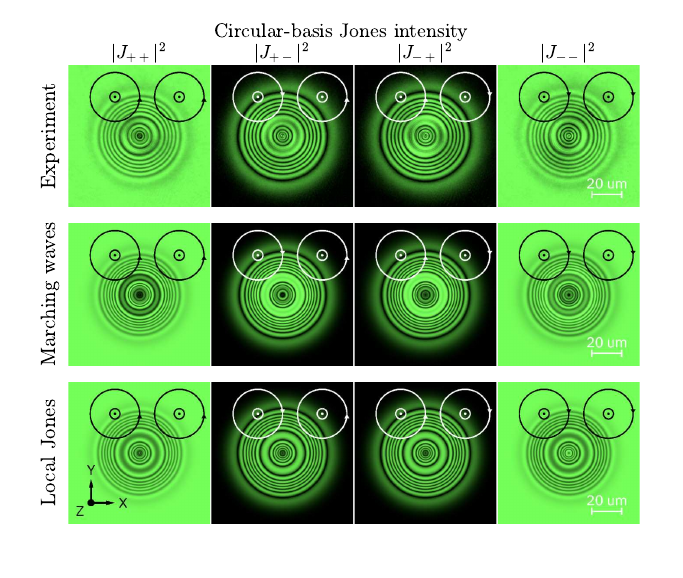}\par\vspace{-3pt}
 \includegraphics[scale=1.03,trim=19.5 0 19 11.5,clip]
 {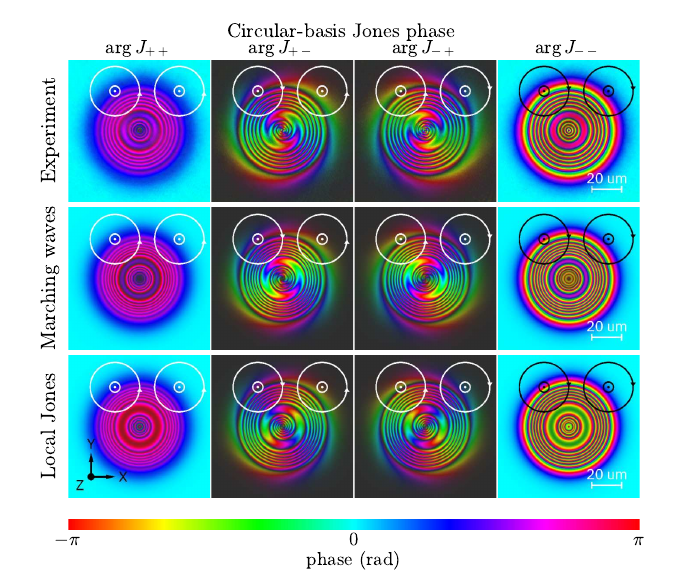}
 \caption{Circular-basis Jones response at
 \(\lambda=520\,\mathrm{nm}\).  Top: intensities
 \(|J_{++}|^2,|J_{+-}|^2,|J_{-+}|^2,|J_{--}|^2\).  Bottom: phases of the
 same elements, wrapped to \([-\pi,\pi]\) and masked where the corresponding
 amplitude is below \(10^{-3}\).  Rows show experiment, marching-wave
 propagation, and independent-column Jones
 propagation.  The dominant diagonal response is nearly radial, and the
 dominant off-diagonal components carry opposite azimuthal phase windings;
 departures from exact covariance are quantified below.}
 \label{sm:fig:circular-intensity}
\end{figure}

The circular-polarization basis separates the symmetry-allowed angular orders
without projection onto a linear analyzer.  We use
\(\vect e_\pm=(\vect e_x\pm\ii\vect e_y)/\sqrt2\), with the output index
written first.  Figure~\ref{sm:fig:circular-intensity} shows the
\(520\,\mathrm{nm}\) response used in the primary symmetry analysis.

\subsection{Independent through-focus imaging}
\label{sm:sec:focus-data}

For the direct comparison, we retain five adjacent axial positions (source
frames 45--49) acquired with the same \(50\times\) optical system at exposure
times of 12 and 42 ms.  Each raw
\(2448\times2048\) frame contains four \(1224\times1024\) analyzer images.
Their physical sampling uses the same calibration-slide scale as the
holographic Jones components.
The nominal stage increment is \(10\,\unitmu\mathrm m\).  Because the focal
surface traverses glass, liquid crystal, and glass while the cell--objective
air gap changes, the mechanical stage coordinate is not a homogeneous
propagation distance in any single medium.  This horizontally illuminated
sequence used the same nominal 520-nm module as the horizontal-input
holographic measurement.  Its measured central wavelength,
\(518.2\pm0.5\,\mathrm{nm}\), is used for this direct-POM comparison, while
figure labels retain the nominal 520-nm dataset designation.

A dark frame is subtracted separately for each analyzer channel and exposure
time.  Each 12-ms frame is registered to its 42-ms counterpart, and their
relative exposure factor is estimated from unsaturated pixels.  The short
exposure contributes only where the measured saturation taper downweights the
long exposure; pixels invalid in both exposures remain masked.  At each axial
position, this procedure yields a radiometrically linear,
saturation-corrected set of four analyzer images and an associated confidence
map.  Pixels saturated in both exposures retain zero confidence and do not
enter the radial fits.
The detector-fixed, low-spatial-frequency component of the archived
\texttt{Ground} reference frames provides a single incident level
\(I_{\rm inc}\) per frame.  Division of all four channels by this scalar puts
them on a common incident-intensity scale.  The spatial pattern in the
\texttt{Ground} frames is neither subtracted from the object images nor used
to divide them.

The direct frames are reflected once about the project \(x\) axis to match
the imported holography.  Under this parity operation,
\(I_{45}\leftrightarrow I_{135}\), whereas \(I_0\) and \(I_{90}\) are
unchanged.  The directly imaged field still passes through the post-objective
beam splitter, while the holographic Jones matrices have already been
corrected for it.  Before comparison, the propagated holographic field is
therefore transformed at the complex-amplitude level:
\begin{equation}
 \vect E_{\rm det}=\mathsf G_{\rm det}\vect E_{\rm hol},\qquad
 \mathsf G_{\rm det}=\operatorname{diag}(1,\sqrt2).
 \label{sm:eq:bs2}
\end{equation}
An independent comparison in the stable \(I_{90}\) channel supports this
amplitude correction.  Fitted 1--3\% scalar analyzer corrections improve
Stokes-parameter closure only marginally, by less than the spatially
structured residual; no such correction is applied.

\begin{figure}[p]
 \centering
 \includegraphics[width=\linewidth]{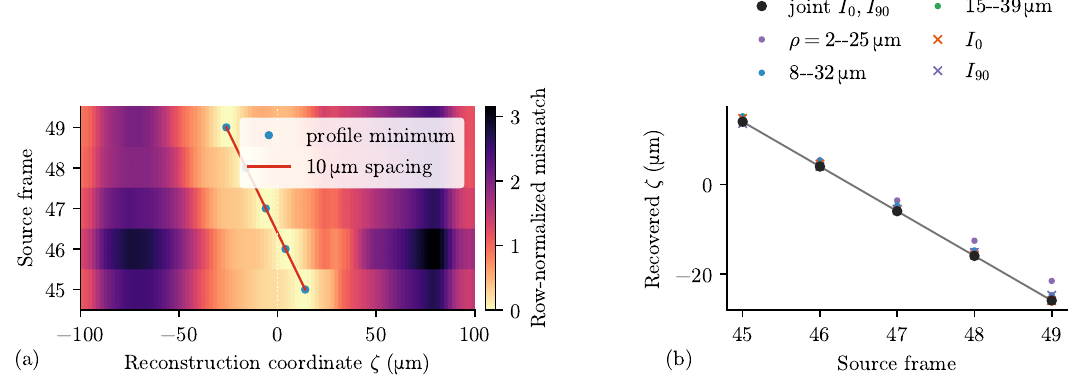}
 \caption{Focus recovery and registration for the direct analyzer-resolved
 five-state subset.  (a) Radial-profile mismatch, normalized separately for
 each displayed row.  Blue points mark the framewise minima and the red line
 their common \(10\,\unitmu\mathrm m\) spacing.  Repeated diagonal valleys are
 ring-order aliases; the dotted line marks the holographic plane,
 \(\zeta=0\).  (b) Minima obtained jointly from \(I_0,I_{90}\), from three
 radial intervals, and from either analyzer channel alone.}
 \label{sm:fig:focus-cost}
\end{figure}

With \(J_\infty\) the robust outer response, the holographic field is first
propagated from its reconstruction plane by an internal coordinate \(\zeta\):
\begin{equation}
 J_\zeta=J_\infty+\mathcal F^{-1}\!\left[
  \Theta(k\,{\rm NA}_{\rm eff}-k_\perp)
  \ee^{\ii \zeta(\sqrt{k^2-k_\perp^2}-k)}
  \mathcal F[J-J_\infty]\right].
 \label{sm:eq:focus-forward}
\end{equation}
Here \(\mathcal F\) is the two-dimensional Fourier transform,
\(\Theta\) is the unit-step function, \(k=2\pi/\lambda\) is the image-space
wavenumber for refractive index \(n=1\), and
\(k_\perp=2\pi f_\perp\) is the radial spatial angular frequency.  The
coordinate \(\zeta\) is therefore an air-equivalent image-space distance from
the holographic reconstruction plane.  Each frame is fitted entirely in this
coordinate; only afterward is the liquid-crystal exit plane located by matching
an annular-focus landmark to the model.  The fit uses only the radial
profiles of \(I_0\) and \(I_{90}\), reserving the diagonal-analyzer channels
for validation.  The calibrated intensity scale remains fixed throughout the
axial search.  Each radial profile is a confidence-weighted azimuthal mean
about a center estimated independently for that frame.  Denoting the measured
and propagated profiles for frame \(i\) and analyzer channel \(c\) by
\(p_{ic}(\rho)\) and \(\widehat p_c(\rho;\zeta)\), respectively, the
framewise objective is
\begin{equation}
 \chi_i^2(\zeta)=\frac12\sum_{c\in\{0,90\}}
 \frac{\sum_\rho w_{ic}(\rho)
 [p_{ic}(\rho)-\widehat p_c(\rho;\zeta)]^2}
 {\sum_\rho w_{ic}(\rho)p_{ic}^2(\rho)} .
 \label{sm:eq:focus-cost}
\end{equation}
The sums cover \(2\leq\rho\leq39\,\unitmu\mathrm m\), with
\(w_{ic}=\max(\rho,0.5\,\unitmu\mathrm m)f_{ic}^2\), where \(f_{ic}\) is the
radial confidence coverage.  The two analyzer costs have equal weight, and no
intensity scale or offset is fitted.  Equation~\eqref{sm:eq:focus-forward} is
evaluated over \(-100\leq\zeta\leq100\,\unitmu\mathrm m\) in
\(2\,\unitmu\mathrm m\) increments with twofold transverse zero padding.  Each
row in Fig.~\ref{sm:fig:focus-cost} is normalized solely to display the shape of
its cost function; this operation does not shift the minimum.  Positive
\(\zeta\) denotes the propagation direction fixed by the phase factor in
Eq.~\eqref{sm:eq:focus-forward}, not the signed mechanical displacement of any
single glass or liquid-crystal interface.
For frames 45--49 the profile minima are respectively
\(\zeta=(14,4,-6,-16,-26)\,\unitmu\mathrm m\), exactly following the nominal
\(-10\,\unitmu\mathrm m\) spacing at the \(2\,\unitmu\mathrm m\) search resolution.
The minima obtained from three radial bands span at most
\(4\,\unitmu\mathrm m\); the independently fitted \(I_0\) and \(I_{90}\) minima
differ by at most \(2\,\unitmu\mathrm m\); and the separation from the nearest
competing ring-order alias is 0.041--0.088 in the normalized objective.  The
liquid-crystal exit plane is inferred separately.
Aligning the polarization-independent annular focus with that predicted by
the marching-wave model, whose exit plane is known, gives
\(\zeta_{\rm MW}=6.23\,\unitmu\mathrm m\).  The propagation maps use
\(z=\zeta-\zeta_{\rm MW}\), so \(z=0\) is the common comparison plane and the
model-based estimate of the liquid-crystal exit surface.  This one-parameter
procedure defines a model-referenced axial origin; it is not an independent
measurement of the interface position.  The independent-column model
predicts a different annular-focus position, and the inner and outer radial
profiles retain distinct residuals after the common shift.

\begin{figure}[p]
 \centering
 \includegraphics[width=\linewidth]{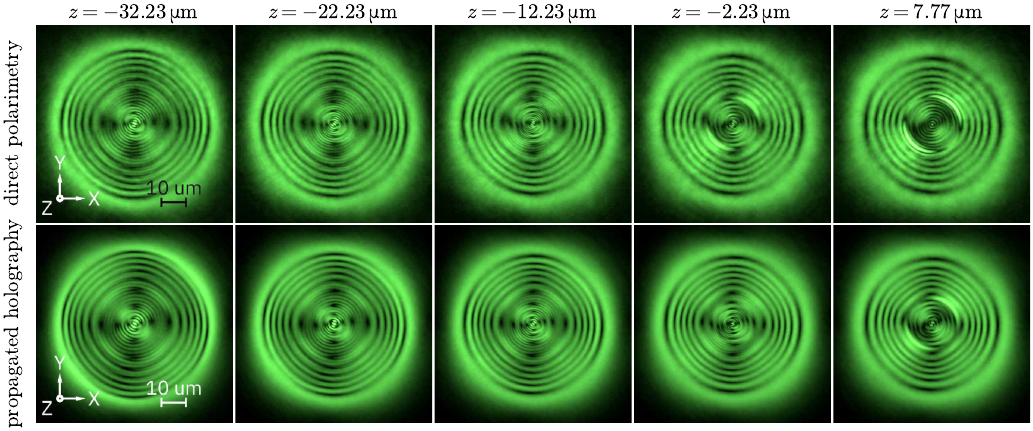}
 \caption{Matched-focus comparison in the calibrated \(I_{90}\) channel at
 the nominal wavelength \(520\,\mathrm{nm}\).  Columns are the five acquired
 planes nearest the holographic plane; titles give the model-referenced
 coordinate \(z\).  The upper row shows direct analyzer-resolved measurements,
 and the lower row shows the averaged holographic Jones field propagated to
 the fitted coordinate and transformed by Eq.~\eqref{sm:eq:bs2}.  Each image is
 recentered on its independently estimated toron center.  All panels share the
 same physical field of view and calibrated intensity scale.}
 \label{sm:fig:focus-comparison}
\end{figure}

For display in Fig.~\ref{sm:fig:focus-comparison} only, narrow
double-saturated gaps (at most \(2.83\) pixels from valid data) are filled by
local continuation so that missing values are not rendered as black.

The median radial-profile correlations are 0.819 for \(I_0\) and 0.841 for
\(I_{90}\); the held-out \(I_{45}\) and \(I_{135}\) channels give 0.682 and
0.770.  Their median relative root-mean-square discrepancies are 27.3\%,
27.0\%, 23.8\%, and 36.7\%, respectively.  Agreement in ring order and in the
direction of radial motion supports the use of homogeneous propagation for
qualitative axial registration, but not for quantitative reconstruction of
every radial amplitude.  The remaining discrepancy includes direct-image
calibration errors and the replacement of the actual stratified
glass--liquid-crystal optical path by scalar \(n=1\) propagation.

\section{Axisymmetric Maxwell theory and optical mechanisms}
\label{sm:sec:theory}

The derivation is organized by level of approximation.  Rotational covariance
about the optical axis and the associated angular-momentum selection rule
follow exactly from Maxwell's equations under the stated illumination and
boundary symmetries.  An eight-function radial electric--magnetic system
retains both propagation directions.  Its four radial electric-response
functions form a closed first-order state only after projection onto the
forward branch; their exact decomposition is defined within the full
electric--magnetic system.  We use the paraxial electric-only equations to
identify physical mechanisms, while the numerical marching-wave calculation
propagates the forward electric--magnetic state.  Finally, the
independent-column Jones model removes all transverse derivatives and serves
as the local reference.  These three levels are kept distinct throughout the
comparisons.

\subsection{Symmetry assumptions and implications}
\label{sm:sec:symmetry-hierarchy}

We write \(C_n\) for invariance of a measured intensity under an in-plane
rotation by \(2\pi/n\).  Material symmetry, symmetry of the
polarization-transfer operator, and symmetry of a fixed-polarizer image are
distinct statements \cite{armitage2014jones_symmetry}.  In particular,
rotating an axisymmetric anisotropic object while holding the laboratory
polarizers fixed is not equivalent to rotating space and polarization
together.

Let \(J(\rho,\varphi)\) be the \(2\times2\) matrix whose columns are the
electric output fields generated by two spatially uniform, orthogonal
transverse inputs.  This definition covers both a pointwise Jones element and
a spatially nonlocal optical system.  In the latter case,
\(J\) is the response of the complete system to the two uniform inputs, not a
local constitutive tensor.  For normal illumination, rotationally symmetric
exterior boundaries, and a centered circular pupil, material axisymmetry
implies joint space--polarization covariance of \(J\).  The Maxwell
derivation below shows that this covariance locks \(m=0\) to the
helicity-preserving Jones components and \(m=\pm2\) to the
helicity-converting Jones components, so any intensity measured with fixed
linear input and analyzer polarizations contains only \(m=0,\pm2,\pm4\).  The
resulting half-turn symmetry \(C_2\) is required by rotational covariance;
quarter-turn symmetry \(C_4\) is not.

The additional \(C_4\) symmetry follows when the reduced response is unitary
at every point, as in lossless independent-column propagation.  An
untruncated lossless system can preserve both rotational covariance and the
global unitarity of its propagation operator while diffraction makes the
response to a uniform input nonunitary at an individual output point.  A
centered circular pupil preserves covariance but, when finite, need not
preserve global unitarity.  Loss of \(C_4\) therefore does not by itself imply
loss of material axisymmetry.  We call the association between circular-
polarization channels and angular orders circular-polarization--order
locking.  The implications used below are
\begin{align}
 \text{material axisymmetry}
 &\ \Longrightarrow\ \text{order locking at }m=0,\pm2
 \ \Longrightarrow\ C_2,\nonumber\\
 \text{covariant response + pointwise unitarity}
 &\ \Longrightarrow\ C_4.
 \label{sm:eq:symmetry-hierarchy}
\end{align}
For crossed input and analyzer polarizations, pointwise normality is already
sufficient to force \(H_2=0\), whereas pointwise unitarity gives
\(C_4\) for every fixed real linear input--analyzer pair;
Sec.~\ref{sm:sec:c4} derives both statements.
Jones-field orders outside \(m=0,\pm2\),
circular-polarization--order mismatches within that set, and loss of \(C_2\)
are forbidden by the axisymmetric model.  Their presence signals a breakdown
of rotational covariance in the object, illumination, pupil, or another part
of the instrument.

\subsection{From Maxwell equations to radial response functions}
\label{sm:sec:maxwell}

We adopt the time dependence \(\exp(-\ii\omega t)\), where \(\omega\) is the
angular frequency.  In a nonmagnetic dielectric with relative permittivity
tensor \(\varepsilon\),
\begin{equation}
 \nabla\times\vect E=\ii\omega\mu_0\vect H,\qquad
 \nabla\times\vect H=-\ii\omega\varepsilon_0\varepsilon\vect E,
 \qquad
 \nabla\cdot(\varepsilon\vect E)=0,\quad \nabla\cdot\vect H=0.
 \label{sm:eq:maxwell}
\end{equation}
Here \(\varepsilon_0\) and \(\mu_0\) are the vacuum permittivity and
permeability, and \(k_0=\omega/c\) is the vacuum wavenumber.
On a plane of constant \(z\), the tangential state
\(\Phi=(E_x,E_y,H_x,H_y)^{\mathsf T}\) retains both forward- and
backward-propagating waves.  To write its evolution without implicit
longitudinal components, we partition the relative permittivity tensor and
define the transverse scalar curl by
\begin{equation}
 \varepsilon=
 \begin{pmatrix}\varepsilon_\perp&\vect g\\
 \vect g^{\mathsf T}&\varepsilon_{zz}\end{pmatrix},\qquad
 \operatorname{curl}_{\!\perp}\vect F_\perp
 =\partial_xF_y-\partial_yF_x,\qquad
 \mathsf R=\begin{pmatrix}0&1\\-1&0\end{pmatrix}.
 \label{sm:eq:epsilon-blocks}
\end{equation}
The \(z\) components of the two curl equations give the exact algebraic
relations
\begin{equation}
 H_z=\frac{\operatorname{curl}_{\!\perp}\vect E_\perp}
 {\ii\omega\mu_0},\qquad
 E_z=-\frac{\vect g^{\mathsf T}\vect E_\perp}
               {\varepsilon_{zz}}
     -\frac{\operatorname{curl}_{\!\perp}\vect H_\perp}
               {\ii\omega\varepsilon_0\varepsilon_{zz}}.
 \label{sm:eq:longitudinal-fields}
\end{equation}
Substitution into the transverse curl equations produces the closed system
\begin{align}
 \partial_z\vect E_\perp={}&
 \ii\omega\mu_0\mathsf R\vect H_\perp
 -\nabla_\perp(\vect\eta^{\mathsf T}\vect E_\perp)
 -\nabla_\perp\!\left[
   \frac{\operatorname{curl}_{\!\perp}\vect H_\perp}
        {\ii\omega\varepsilon_0\varepsilon_{zz}}\right],
 \label{sm:eq:maxwell-transverse-e}\\
 \partial_z\vect H_\perp={}&
 -\ii\omega\varepsilon_0\mathsf R
     \bar\varepsilon_\perp\vect E_\perp
 +\mathsf R\vect\eta\,
   \operatorname{curl}_{\!\perp}\vect H_\perp
 +\nabla_\perp\!\left[
   \frac{\operatorname{curl}_{\!\perp}\vect E_\perp}
        {\ii\omega\mu_0}\right],
 \label{sm:eq:maxwell-transverse-h}
\end{align}
where
\begin{equation}
 \vect\eta=\frac{\vect g}{\varepsilon_{zz}},\qquad
 \bar\varepsilon_\perp=
 \varepsilon_\perp-\frac{\vect g\vect g^{\mathsf T}}
                           {\varepsilon_{zz}}.
 \label{sm:eq:schur-epsilon}
\end{equation}
Equations~\eqref{sm:eq:maxwell-transverse-e} and
\eqref{sm:eq:maxwell-transverse-h} explicitly define
\(\partial_z\Phi=\mathcal M[\varepsilon,\nabla_\perp]\Phi\); no longitudinal
field or unspecified vector operator remains.  The system is local in \(z\),
differential in the transverse coordinates, and retains both propagation
directions.  Because \(\Phi\) is continuous across a planar interface in the
absence of surface currents or charges, this form is also a natural starting
point for reflection- and multiple-pass calculations
\cite{berreman1972,yeh1980,yeh_gu1999}.

Let \(R_\beta^{(3)}\) rotate three-dimensional vectors about \(z\), and let
\(\Rmat_\beta\) be its transverse restriction.  Axisymmetry of the material
is expressed by
\begin{equation}
 \varepsilon(\Rmat_\beta\vect r_\perp,z)
 =R_\beta^{(3)}\varepsilon(\vect r_\perp,z)
 R_\beta^{(3)\mathsf T}.
 \label{sm:eq:epsilon-covariance}
\end{equation}
Uniqueness of the Maxwell boundary-value problem then implies the exact
covariance of the transmission kernel \(\mathcal T(\vect r_o,\vect r_i)\),
where \(\vect r_i\) and \(\vect r_o\) are transverse input and output
coordinates:
\begin{equation}
 \mathcal T(\Rmat_\beta\vect r_o,\Rmat_\beta\vect r_i)
 =\Rmat_\beta \mathcal T(\vect r_o,\vect r_i)\Rmat_{-\beta}.
 \label{sm:eq:kernel-covariance}
\end{equation}
For two uniform transverse basis inputs, the electric output fields form the
Jones response
\begin{equation}
 J_E(\rho,\varphi+\beta,z)
 =\Rmat_\beta J_E(\rho,\varphi,z)\Rmat_{-\beta}.
 \label{sm:eq:jones-covariance}
\end{equation}
The analogous magnetic response \(J_H\) obeys the same relation.
Equation~\eqref{sm:eq:jones-covariance} does not imply that a fixed laboratory
polarization produces a circularly symmetric image.  Rather, it states that
the boundary-value problem is unchanged when the observation point and the
polarization basis are rotated together.  An axisymmetric object can
therefore produce a twofold or fourfold image between fixed polarizers.

Define
\begin{equation}
 K=\begin{pmatrix}0&-1\\1&0\end{pmatrix},\quad
 \Sigma_1=\begin{pmatrix}\cos2\varphi&\sin2\varphi\\
 \sin2\varphi&-\cos2\varphi\end{pmatrix},\quad
 \Sigma_2=\begin{pmatrix}-\sin2\varphi&\cos2\varphi\\
 \cos2\varphi&\sin2\varphi\end{pmatrix}.
 \label{sm:eq:generator-basis}
\end{equation}
Thus the in-plane rotation matrix introduced in
Eq.~\eqref{sm:eq:epsilon-blocks} is
\(\mathsf R=-K\).
With \(I_2\) denoting the \(2\times2\) identity, every solution of
Eq.~\eqref{sm:eq:jones-covariance} has the exact form
\begin{equation}
 J_F=a_F I_2+b_FK+c_F\Sigma_1+d_F\Sigma_2,\qquad F=E,H,
 \label{sm:eq:eh-abcd}
\end{equation}
with eight complex functions of \((\rho,z)\).  For the electric response,
\(a_E\) is the scalar amplitude, \(b_E\) is the antisymmetric amplitude of
angular order zero, and \(c_E,d_E\) are the two quadratures of the
circular-polarization-conversion response.  Exact covariance makes all four
functions independent of \(\varphi\).  The curl relations fix the angular
orders of the longitudinal responses to \(m=\pm1\).  To write the radial
equations explicitly, we transform to the co-rotating polar basis,
\begin{equation}
 \widehat J_F=\Rmat_{-\varphi}J_F\Rmat_\varphi
 =\begin{pmatrix}
 a_F+c_F&-b_F+d_F\\ b_F+d_F&a_F-c_F
 \end{pmatrix},\qquad
 \vect q_F=(a_F,b_F,c_F,d_F)^{\mathsf T}.
 \label{sm:eq:polar-jones-matrix}
\end{equation}
For a matrix \(M\) written in this basis, define its coefficient projection
\begin{equation}
 \Pi_4(M)=\frac12
 \begin{pmatrix}
 M_{11}+M_{22}\\ M_{21}-M_{12}\\
 M_{11}-M_{22}\\ M_{12}+M_{21}
 \end{pmatrix}.
 \label{sm:eq:radial-matrix-projection}
\end{equation}
The transverse curl of either response is a row vector that maps the uniform
input polarization to the longitudinal field.  Its components in the polar
basis are
\begin{equation}
 \vect t_F^{\mathsf T}=\mathcal C(\vect q_F)^{\mathsf T}
 =\left(
 \partial_\rho b_F+\partial_\rho d_F+\frac{2d_F}{\rho},\;
 \partial_\rho a_F-\partial_\rho c_F-\frac{2c_F}{\rho}
 \right).
 \label{sm:eq:radial-curl-map}
\end{equation}
The corresponding transverse divergence is
\begin{equation}
 \vect v_F^{\mathsf T}=\mathcal D(\vect q_F)^{\mathsf T}
 =\left(
 \partial_\rho a_F+\partial_\rho c_F+\frac{2c_F}{\rho},\;
 -\partial_\rho b_F+\partial_\rho d_F+\frac{2d_F}{\rho}
 \right).
 \label{sm:eq:radial-divergence-map}
\end{equation}
Likewise, if \(\vect s^{\mathsf T}=(s_\rho,s_\varphi)\) is a covariant row
vector, the matrix \(\nabla_\perp(\vect s^{\mathsf T}\vect p)\), for a fixed
Cartesian input \(\vect p\), has radial coefficients
\begin{equation}
 \mathcal G(\vect s)=\frac12
 \begin{pmatrix}
 \partial_\rho s_\rho+s_\rho/\rho\\
 -\partial_\rho s_\varphi-s_\varphi/\rho\\
 \partial_\rho s_\rho-s_\rho/\rho\\
 \partial_\rho s_\varphi-s_\varphi/\rho
 \end{pmatrix}.
 \label{sm:eq:radial-gradient-map}
\end{equation}
The \(1/\rho\) terms arise from differentiating the rotating polar basis and
are essential for a complete radial equation.

For the uniaxial director
\(\vect n=\sin\theta\cos\tau\,\vect e_\rho+
\sin\theta\sin\tau\,\vect e_\varphi+\cos\theta\,\vect e_z\), where
\(\theta\) is the polar tilt and \(\tau\) the in-plane angle from
\(\vect e_\rho\), set \(\Delta\varepsilon=n_e^2-n_o^2\), with \(n_o,n_e\)
the ordinary and extraordinary refractive indices, and
\(\vect u_\tau=(\cos\tau,\sin\tau)^{\mathsf T}\).  The material quantities
in Eqs.~\eqref{sm:eq:maxwell-transverse-e}--\eqref{sm:eq:maxwell-transverse-h} are
then, in the polar basis,
\begin{equation}
 \begin{aligned}
 \varepsilon_{zz}&=n_o^2+\Delta\varepsilon\cos^2\theta,\\
 \vect g&=\Delta\varepsilon\sin\theta\cos\theta\,\vect u_\tau,\\
 \varepsilon_\perp&=n_o^2I_2+
 \Delta\varepsilon\sin^2\theta\,\vect u_\tau\vect u_\tau^{\mathsf T},\\
 \vect\eta&=\frac{\Delta\varepsilon\sin\theta\cos\theta}
                  {\varepsilon_{zz}}\vect u_\tau,\\
 \bar\varepsilon_\perp&=n_o^2I_2+
 \frac{\Delta\varepsilon n_o^2\sin^2\theta}{\varepsilon_{zz}}
 \vect u_\tau\vect u_\tau^{\mathsf T}.
 \end{aligned}
 \label{sm:eq:radial-material-blocks}
\end{equation}
Define the row
\begin{equation}
 \vect s_E^{\mathsf T}=\vect\eta^{\mathsf T}\widehat J_E+
 \frac{\vect t_H^{\mathsf T}}
      {\ii\omega\varepsilon_0\varepsilon_{zz}}.
 \label{sm:eq:radial-longitudinal-row}
\end{equation}
Equations~\eqref{sm:eq:maxwell-transverse-e} and
\eqref{sm:eq:maxwell-transverse-h} then reduce to the following closed radial
Maxwell system:
\begin{align}
 \partial_z\vect q_E={}&
 \ii\omega\mu_0
 \begin{pmatrix}b_H\\-a_H\\d_H\\-c_H\end{pmatrix}
 -\mathcal G(\vect s_E),
 \label{sm:eq:exact-radial-e}\\
 \partial_z\vect q_H={}&
 \Pi_4\!\left[
 -\ii\omega\varepsilon_0\mathsf R\bar\varepsilon_\perp\widehat J_E
 +\mathsf R\vect\eta\vect t_H^{\mathsf T}\right]
 +\mathcal G\!\left(\frac{\vect t_E}{\ii\omega\mu_0}\right).
 \label{sm:eq:exact-radial-h}
\end{align}
All quantities on the right-hand sides are defined component by component in
Eqs.~\eqref{sm:eq:polar-jones-matrix}--\eqref{sm:eq:radial-longitudinal-row}; thus
Eqs.~\eqref{sm:eq:exact-radial-e} and \eqref{sm:eq:exact-radial-h} contain no
unspecified operator.  For example, the two equations for the antisymmetric
amplitudes contain the longitudinal-gradient terms
\begin{align}
 \partial_z b_E={}&-\ii\omega\mu_0a_H
 +\frac12\left(\partial_\rho+\frac1\rho\right)s_{E,\varphi},
 \label{sm:eq:exact-b-e}\\
 \partial_z b_H={}&
 \left\{\Pi_4\!\left[
 -\ii\omega\varepsilon_0\mathsf R\bar\varepsilon_\perp\widehat J_E
 +\mathsf R\vect\eta\vect t_H^{\mathsf T}\right]\right\}_b
 -\frac12\left(\partial_\rho+\frac1\rho\right)
 \frac{t_{E,\varphi}}{\ii\omega\mu_0}.
 \label{sm:eq:exact-b-h}
\end{align}
At \(\rho=0\), these expressions are evaluated through the regular limits
imposed by the \(m=0\) and \(|m|=2\) parities.  The eight-function reduction
is exact and retains reflected waves; the boundary conditions select its
forward- and backward-propagating parts.

The calculations reported here retain only the forward branch.  Reflections
at the liquid-crystal--glass interfaces are weak and cannot break the angular
selection rule while the interfaces remain axisymmetric.  Moreover, the
forward calculation alone produces a twofold response of the observed scale.
In each homogeneous Fourier-space substep, the numerical propagator applies
the forward transverse-electric/transverse-magnetic (TE/TM) projector and then
integrates the anisotropic residual in Eqs.~\eqref{sm:eq:exact-radial-e}--
\eqref{sm:eq:exact-radial-h}.  The four electric functions
\(\vect q=(a,b,c,d)^{\mathsf T}\) form a closed first-order state only after
the magnetic field has been restricted to this forward branch.  The
Jones-like radial evolution introduced below is therefore an approximation,
not an exact replacement for Maxwell's equations.

For later use, define the radial Laplacian by
\begin{equation}
 \mathcal L\vect q_F=
 \left(\mathcal L_0a_F,\mathcal L_0b_F,
 \mathcal L_2c_F,\mathcal L_2d_F\right)^{\mathsf T},
 \qquad
 \mathcal L_m=\partial_\rho^2+\rho^{-1}\partial_\rho-\frac{m^2}{\rho^2}.
 \label{sm:eq:radial-laplacian-map}
\end{equation}

\paragraph{Electric-only closure used for the analytical discussion.}
The magnetic amplitudes can be eliminated before making the paraxial
approximation.  The resulting four-function model exposes the mechanisms
discussed below, whereas the numerical implementation advances the first-order
\(E\)--\(H\) state above.  Let the longitudinal electric response be
\(J_z=\vect j_z^{\mathsf T}\) in the co-rotating input basis.  The
\(z\) and transverse components of
\(\nabla\times\nabla\times\vect E=k_0^2\varepsilon\vect E\) give the exact
electric-only pair
\begin{align}
 (k_0^2\varepsilon_{zz}+\mathcal L_1)\vect j_z^{\mathsf T}
 ={}&\partial_z\mathcal D(\vect q_E)^{\mathsf T}
 -k_0^2\vect g^{\mathsf T}\widehat J_E,
 \label{sm:eq:exact-electric-longitudinal}\\
 \partial_z^2\vect q_E+\mathcal L\vect q_E
 +k_0^2\Pi_4[
 \varepsilon_\perp\widehat J_E+\vect g\vect j_z^{\mathsf T}]
 -\mathcal G[\mathcal D(\vect q_E)]
 -\mathcal G(\partial_z\vect j_z)={}&0.
 \label{sm:eq:exact-electric-radial}
\end{align}
Here \(\mathcal L\) is the diagonal \(m=0,2\) radial Laplacian in
Eq.~\eqref{sm:eq:radial-laplacian-map}, and \(\mathcal L_1\) is the same scalar
operator with \(m=1\), acting on each component of the longitudinal row.
Eliminating \(H\) is exact at this stage, but it converts the first-order
Maxwell system into a second-order electric-field equation that still contains
both propagation directions.

For the analytical one-way reduction, choose an isotropic reference index
\(n_b\), set the carrier wavenumber \(k_b=k_0n_b\), and write
\[
 J_E=\ee^{\ii k_bz}U,\qquad
 J_z=\ee^{\ii k_bz}\vect\ell^{\mathsf T},\qquad
 \widehat U=\Rmat_{-\varphi}U\Rmat_\varphi,\qquad
 \widetilde{\vect q}=(\widetilde a,\widetilde b,
 \widetilde c,\widetilde d)^{\mathsf T}.
\]
To leading one-way order, the longitudinal electric response is fixed by the
transverse field:
\begin{equation}
 \vect\ell^{\mathsf T}=
 -\vect\eta^{\mathsf T}\widehat U+
 \frac{\ii k_b}{k_0^2\varepsilon_{zz}}\,
 \mathcal D(\widetilde{\vect q})^{\mathsf T}.
 \label{sm:eq:paraxial-longitudinal-electric}
\end{equation}
Equation~\eqref{sm:eq:paraxial-longitudinal-electric} follows from the \(z\)
component of
\(\nabla\times\nabla\times\vect E=k_0^2\varepsilon\vect E\), after neglecting
\(\mathcal L_1\ell\) relative to
\(k_0^2\varepsilon_{zz}\ell\) and replacing the rapidly varying part of
\(\partial_z\) by \(\ii k_b\).  Projecting the transverse components and
neglecting \(\partial_z^2U\) and \(\partial_z\ell\) gives
\begin{equation}
 \begin{split}
 \partial_z\widetilde{\vect q}
 =\frac{1}{2\ii k_b}\bigg\{&
 k_b^2\widetilde{\vect q}
 -\mathcal L\widetilde{\vect q}
 -k_0^2\Pi_4\!\left[
   \varepsilon_\perp\widehat U+\vect g\vect\ell^{\mathsf T}\right]\\
 &+\mathcal G\!\left[\mathcal D(\widetilde{\vect q})\right]
 +\ii k_b\mathcal G(\vect\ell)\bigg\},
 \end{split}
 \label{sm:eq:electric-only-radial-system}
\end{equation}
Together with the component maps \(\mathcal D\), \(\mathcal G\), and
\(\Pi_4\), Eqs.~\eqref{sm:eq:paraxial-longitudinal-electric} and
\eqref{sm:eq:electric-only-radial-system} form a closed system for
\(\widetilde a,\widetilde b,\widetilde c,\widetilde d\).  In an isotropic
medium, \(\vect g=\vect\eta=0\) and
\(\varepsilon_\perp=\varepsilon_{zz}=n_b^2I_2\), so the two
gradient--divergence terms cancel.
Equation~\eqref{sm:eq:electric-only-radial-system} then reduces to
\(\partial_z\widetilde{\vect q}=\ii\mathcal L\widetilde{\vect q}/(2k_b)\),
confirming the sign and normalization of the homogeneous paraxial limit.

\subsection{Circular-polarization basis, allowed modes, and direct extraction}
\label{sm:sec:selection}

In the circular basis
\(\vect e_\pm=(\vect e_x\pm\ii\vect e_y)/\sqrt2\),
Eq.~\eqref{sm:eq:eh-abcd} becomes
\begin{equation}
 J_{\rm c}=
 \begin{pmatrix}
  A_+&B_-\ee^{-2\ii\varphi}\\
  B_+\ee^{2\ii\varphi}&A_-
 \end{pmatrix},\qquad
 \begin{matrix}
  A_+=a-\ii b,&A_-=a+\ii b,\\
  B_-=c-\ii d,&B_+=c+\ii d.
 \end{matrix}
 \label{sm:eq:circular}
\end{equation}
Thus rotational covariance about \(z\) permits only \(m=0,\pm2\): \(A_\pm\)
preserve circular polarization, whereas \(B_+\) and \(B_-\) convert between
opposite circular polarizations and carry azimuthal orders \(+2\) and \(-2\),
respectively.  This selection rule follows from covariance alone and is
independent of locality, unitarity, director helicity, and the radial profile.

With unit \(x\)-polarized input and a crossed \(y\) analyzer, the detected
field is
\begin{align}
 u_\times=\vect e_y^\dagger J\vect e_x=J_{yx}
 &=b+c\sin2\varphi+d\cos2\varphi\nonumber\\
 &=b-\frac{\ii}{2}B_+\ee^{2\ii\varphi}
     +\frac{\ii}{2}B_-\ee^{-2\ii\varphi}.
 \label{sm:eq:crossed-field}
\end{align}
Thus \(b\) is the angular-order-zero amplitude.  The real quadratures of its
cross products with the oppositely wound conversion amplitudes determine the
amplitude and orientation of the twofold intensity modulation.

The coefficients are obtained pointwise from a Cartesian Jones map.  With
\(X=(J_{xx}-J_{yy})/2\) and \(Y=(J_{xy}+J_{yx})/2\),
\begin{equation}
 \begin{aligned}
 a&=\tfrac12(J_{xx}+J_{yy}),&
 b&=\tfrac12(J_{yx}-J_{xy}),\\
 c&=X\cos2\varphi+Y\sin2\varphi,&
 d&=-X\sin2\varphi+Y\cos2\varphi.
 \end{aligned}
 \label{sm:eq:coefficient-extraction}
\end{equation}
For an exactly rotationally covariant response, these coefficients are
independent of \(\varphi\).  For a general measured map, the exact orthogonal
projector \(P_{\rm cov}\) can therefore be implemented either by
angularly averaging \(a,b,c,d\) at fixed \(\rho\), or equivalently in the
circular basis by retaining \(m=0\) in \(J_{++},J_{--}\), \(m=+2\) in
\(J_{-+}\), and \(m=-2\) in \(J_{+-}\).  All reported projections use this
circular-basis operation; the rejected field is called the noncovariant
residual.  One-dimensional profiles use smoothed annular averages, whereas
nonlinear products are always evaluated pixel by pixel before averaging.

\subsection{\texorpdfstring{Linear-polarization images, \(H_2,H_4\), mirror
axes, and \(S_0\)}{Linear-polarization images, H2, H4, mirror axes, and S0}}
\label{sm:sec:observables}

More generally, fixed real unit Jones vectors \(\vect e_{\rm in}\) (input) and
\(\vect e_{\rm a}\) (analyzer) select the scalar field
\(u(\rho,\varphi)=\vect e_{\rm a}^\dagger J(\rho,\varphi)\vect e_{\rm in}\).
Denote its complex radial amplitudes of orders \(0,+2,-2\) by
\(u_0,u_+,u_-\).  Then
\begin{align}
 u&=u_0+u_+\ee^{2\ii\varphi}+u_-\ee^{-2\ii\varphi},\nonumber\\
 I=|u|^2&=H_0+2\operatorname{Re}
 \left(H_2\ee^{2\ii\varphi}+H_4\ee^{4\ii\varphi}\right),\label{sm:eq:i-harmonics}\\
 H_0&=|u_0|^2+|u_+|^2+|u_-|^2,\nonumber\\
 H_2&=u_+u_0^*+u_0u_-^*,\qquad H_4=u_+u_-^*.\nonumber
\end{align}
The angular mean \(H_0\) is the sum of the three order-resolved intensities.
The three field orders coexist coherently at each image point.  Because
intensity is quadratic in their sum, \(u_0\) interferes with \(u_\pm\) to form
\(H_2\), while \(u_+\) interferes with \(u_-\) to form \(H_4\).  These cross
products vanish only upon integration over \(\varphi\), when distinct angular
orders become orthogonal.  The intensity is exactly invariant under a half
turn, so \(C_2\) is the generic symmetry of this angular sector.  Fourfold rotational
symmetry is equivalent to \(H_2=0\).

For \(x\)-polarized input and crossed \(y\) analysis,
\begin{align}
 u_\times&=b+c\sin2\varphi+d\cos2\varphi,\nonumber\\
 H_0&=|b|^2+\tfrac12\left(|c|^2+|d|^2\right),\nonumber\\
 H_2&=\operatorname{Re}(b^*d)
 -\ii\operatorname{Re}(b^*c),\label{sm:eq:crossed-h2}\\
 H_4&=-\tfrac14(c+\ii d)(c-\ii d)^*.\nonumber
\end{align}
Thus \(H_0\) combines the order-zero contribution \(|b|^2\) with the
summed order-\(\pm2\) contribution
\(\tfrac12(|c|^2+|d|^2)\).
Hence \(b=0\) is sufficient for a fourfold crossed-polarizer image, but it is
not necessary: when \(b\ne0\), both real interference quadratures may still
vanish, leaving \(H_2=0\).

For any fixed linear input--analyzer pair, the phase of a nonzero
harmonic determines the family of angular extrema,
\begin{equation}
 \psi_m(\rho)=-\frac1m\arg H_m(\rho)
 \pmod{\frac{\pi}{m}},\qquad m=2,4.
 \label{sm:eq:axis-family}
\end{equation}
Thus \(\arg H_4\) directly determines the orientation of the four-lobed
intensity pattern.  Inferring the director helicity from this orientation
would require an additional weak-retardance inverse model and is not part of
the present symmetry analysis.
When both harmonics are present, their extrema share an exact mirror family
only when
\begin{equation}
 \arg H_4-2\arg H_2=0\ \text{or}\ \pi\pmod{2\pi}.
 \label{sm:eq:axis-compatibility}
\end{equation}
Thus the presence of both harmonics does not by itself imply a mirror axis.
One family of intensity maxima for each harmonic is
\begin{equation}
 \psi_m^{(+)}=-\frac1m\arg H_m
 \pmod{\frac{2\pi}{m}},\qquad
 \Theta_m=\operatorname{wrap}_{[-\pi,\pi)}(-\arg H_m)
 =m\psi_m^{(+)}.
 \label{sm:eq:reduced-harmonic-orientation}
\end{equation}
A complete \(2\pi\) cycle of the reduced orientation \(\Theta_m\) therefore
corresponds to a physical rotation of \(2\pi/m\): \(\pi\) for \(C_2\) and
\(\pi/2\) for \(C_4\).  These are harmonic orientations, not tests of the
rotational or mirror symmetry of the complete image.

A mirror axis of the complete image is defined more directly by reflection
than by comparing the phases of the two harmonics separately.  Let
\(\mathcal S_\psi I(\varphi)=I(2\psi-\varphi)\),
\(\delta I=I-H_0\), and
\(H_m=|H_m|\exp(\ii\alpha_m)\).  The normalized correlation with a reflected
copy follows analytically from Eq.~\eqref{sm:eq:i-harmonics}:
\begin{align}
 \Gamma(\psi)
 &=\frac{\langle\delta I(\varphi)
 \mathcal S_\psi\delta I(\varphi)\rangle_\varphi}
 {\langle\delta I^2(\varphi)\rangle_\varphi}\nonumber\\
 &=\frac{
 \mathcal H_2\cos[2(\alpha_2+2\psi)]
 +\mathcal H_4\cos[2(\alpha_4+4\psi)]}
 {\mathcal H_2+\mathcal H_4},
 \qquad
 \mathcal H_2=|H_2|^2,\quad \mathcal H_4=|H_4|^2.
 \label{sm:eq:reflection-correlation}
\end{align}
Equivalently,
\begin{equation}
 1-\Gamma(\psi)=
 \frac{\langle[I(\varphi)-I(2\psi-\varphi)]^2\rangle_\varphi}
 {2\langle\delta I^2(\varphi)\rangle_\varphi}.
 \label{sm:eq:reflection-rms}
\end{equation}
Thus \(\Gamma(\psi)=1\) is equivalent to zero normalized root-mean-square
(rms) reflection error.  When
\(\mathcal H_2+\mathcal H_4=0\), the angular modulation vanishes and its axis
is undefined; such points are suppressed by the visibility factor below.

The quality and orientation of the best common mirror line are
\begin{equation}
 \Gamma_M=\max_\psi\Gamma(\psi),\qquad
 \psi_M\mathrel{\in}\operatorname*{arg\,max}_\psi\Gamma(\psi).
 \label{sm:eq:best-reflection-axis}
\end{equation}
With \(x=\alpha_2+2\psi\), \(\Delta_\alpha=\alpha_4-2\alpha_2\),
the stationary points of \(\Gamma\) obey
\begin{equation}
 \mathcal H_2\sin2x+2\mathcal H_4\sin(4x+2\Delta_\alpha)=0.
 \label{sm:eq:axis-stationary}
\end{equation}
After \(\xi=\exp(2\ii x)\), Eq.~\eqref{sm:eq:axis-stationary} is the quartic
\begin{equation}
 2\mathcal H_4\ee^{2\ii\Delta_\alpha}\xi^4
 +\mathcal H_2\xi^3-\mathcal H_2\xi
 -2\mathcal H_4\ee^{-2\ii\Delta_\alpha}=0.
 \label{sm:eq:axis-quartic}
\end{equation}
Evaluating Eq.~\eqref{sm:eq:reflection-correlation} at the stationary points
corresponding to the unit-circle roots selects the global maximum without a
scan over trial axes.  For nonzero
\(\mathcal H_2,\mathcal H_4\), \(\Gamma_M=1\) precisely when
\(\Delta_\alpha=0\) or \(\pi\) modulo \(2\pi\); otherwise the two harmonics
have no exact common mirror line.  Because the image already has rotational
\(C_2\), the equivalent lines \(\psi_M\) and \(\psi_M+\pi/2\) form one
geometric family.  We therefore plot the branch-independent family phase
\begin{equation}
 \Phi_M=\operatorname{wrap}_{[-\pi,\pi)}(4\psi_M).
 \label{sm:eq:mirror-family-phase}
\end{equation}
If distinct axis families attain the same maximum on a ring, the
plotted phase represents one of them.

Equation~\eqref{sm:eq:reflection-rms} gives a normalized rms reflection defect.
For visualization, we map it to the bounded fidelity
\begin{equation}
 \varepsilon_M=\sqrt{2(1-\Gamma_M)},\qquad
 F_M=[1-\varepsilon_M]_+,
 \label{sm:eq:mirror-defect}
\end{equation}
where \([x]_+=\max(x,0)\).  Thus \(\varepsilon_M=0\) identifies an exact
mirror family, whereas a large phase incompatibility between \(H_2\) and \(H_4\) suppresses \(F_M\).

Rotational \(C_2\) follows from the allowed even harmonics and must not be
confused with mirror symmetry.  Rotational \(C_4\) is tested by a quarter-turn
comparison, whose normalized rms defect reduces to
\begin{align}
 \varepsilon_4
 &=\frac12\sqrt{\frac{
 \langle[I(\varphi+\pi/2)-I(\varphi)]^2\rangle_\varphi}
 {\langle\delta I^2(\varphi)\rangle_\varphi}}
 =\sqrt{\frac{\mathcal H_2}{\mathcal H_2+\mathcal H_4}},\nonumber\\
 \Gamma_4&=1-\varepsilon_4^2
 =\frac{\mathcal H_4}{\mathcal H_2+\mathcal H_4}.
 \label{sm:eq:c4-rotation-defect}
\end{align}
Hence exact rotational \(C_4\) is equivalent to \(H_2=0\), independently of
the phase of \(H_4\).  In that limit the remaining harmonic has mirror lines
\begin{equation}
 \psi_4=-\frac14\arg H_4\pmod{\frac{\pi}{4}},\qquad
 \Phi_{4M}=\operatorname{wrap}_{[-\pi,\pi)}(8\psi_4).
 \label{sm:eq:c4-mirror-family}
\end{equation}
The nested \(C_4\)-compatible mirror fidelity is
\begin{equation}
 F_{4M}=\min\!\left(F_M,[1-\varepsilon_4]_+\right),
 \qquad 0\le F_{4M}\le F_M\le1.
 \label{sm:eq:nested-mirror-fidelity}
\end{equation}
It is high precisely when the complete pattern has a resolved mirror family
and is also close to rotational \(C_4\); by construction,
\(F_{4M}\leq F_M\).

Finally, a mathematically exact mirror line is not experimentally meaningful
when the angular modulation is vanishingly weak.  We therefore define the
pointwise angular visibility
\begin{equation}
 V=\min\!\left(1,
 \frac{\sqrt{2(\mathcal H_2+\mathcal H_4)}}{H_0+\epsilon}\right),
 \qquad W_M=VF_M,\quad W_{4M}=VF_{4M}.
 \label{sm:eq:axis-visibility}
\end{equation}
Here \(\epsilon>0\) only regularizes regions of vanishing mean intensity and
is fixed once for every set of maps being compared; it is never adjusted
pointwise.  The plotted fields are evaluated pointwise at every \((z,\rho)\) and satisfy
\(W_{4M}\le W_M\).

To quantify local power redistribution, normalize by the constant outer
response:
\begin{equation}
 T=JJ_\infty^{-1},\quad Q=T^\dagger T,\quad
 I_{\rm av}=\tfrac12\operatorname{tr}Q.
 \label{sm:eq:gram}
\end{equation}
We define
\begin{equation}
 D_I=(I_{\rm av}-1)^2,\qquad
 D_P=\tfrac12\|Q-I_{\rm av}I_2\|_F^2,\qquad
 \tfrac12\|Q-I_2\|_F^2=D_I+D_P.
 \label{sm:eq:defects}
\end{equation}
Here \(D_I\) measures the polarization-independent part of the local intensity
defect, whereas \(D_P\) measures its traceless, polarization-dependent part.
For incident \(x\)-polarization, \(\vect e_x=(1,0)^{\mathsf T}\),
\begin{align}
 S_0^{(x)}
 =&\,|a|^2+|b|^2+|c|^2+|d|^2\nonumber\\
 &+2[\operatorname{Re}(a^*c)+\operatorname{Re}(b^*d)]\cos2\varphi\nonumber\\
 &+2[-\operatorname{Re}(a^*d)+\operatorname{Re}(b^*c)]\sin2\varphi.
 \label{sm:eq:s0-abcd}
\end{align}
For a linear input at angle \(\alpha\), the total intensity is
\(S_0^{(\alpha)}=\lVert J\vect e_\alpha\rVert^2\).  Write
\(Q_J=J^\dagger J\).  In the circular basis, covariance gives
\((Q_J)_{+-}=q_2(\rho)\ee^{-2\ii\varphi}\), and therefore
\begin{equation}
 S_0^{(\alpha)}
 =\tfrac12\operatorname{tr}Q_J+\operatorname{Re}
 [q_2(\rho)\ee^{-2\ii(\varphi-\alpha)}].
 \label{sm:eq:s0}
\end{equation}
The total output intensity is radially symmetric for every linear input if
and only if \(q_2=0\).  This condition is weaker than pointwise unitarity and
provides an analyzer-independent observable of polarization-dependent
focusing.

\subsection{Emergent fourfold symmetry and the normality test}
\label{sm:sec:c4}

For any invertible \(2\times2\) matrix,
\begin{equation}
 KMK^{-1}=\det(M)M^{-\mathsf T}.
 \label{sm:eq:two-by-two}
\end{equation}
Combining this identity with Eq.~\eqref{sm:eq:jones-covariance} at
\(\beta=\pi/2\) gives, for pointwise unitary \(J\),
\begin{equation}
 J(\rho,\varphi+\pi/2)=\det J(\rho)J(\rho,\varphi)^*.
 \label{sm:eq:quarter-turn}
\end{equation}
For real linear input and analyzer vectors, a quarter turn therefore
complex-conjugates the measured amplitude up to a unit-modulus scalar and
leaves the intensity invariant.  Material axisymmetry supplies
Eq.~\eqref{sm:eq:jones-covariance} and \(C_2\), while pointwise unitarity supplies
the quarter-turn identity and hence \(C_4\).  A pointwise local description is
approached for optically thin or transversely broad textures and for weak
in-sample diffraction.  Commuting local retarders form the still more
restrictive limit in which the locally generated antisymmetric amplitude
vanishes.

The exact condition for the crossed-polarizer response is more general.  Let
\begin{equation}
 N=J^\dagger J-JJ^\dagger.
 \label{sm:eq:normality}
\end{equation}
Direct substitution yields
\begin{equation}
 H_2=\frac18\operatorname{tr}
 [N(\Sigma_1+\ii\Sigma_2)].
 \label{sm:eq:normality-h2}
\end{equation}
Thus the measured second harmonic is the angular-order-two projection of the
pointwise normality defect.  Unitarity is sufficient but not necessary for
\(N=0\): a lossy normal matrix also has \(N=0\), and a non-normal matrix may
have a vanishing angular-order-two projection.  Conversely, once rotational
covariance and instrumental polarization have been established independently,
reproducible nonzero \(H_2\) excludes the entire class of
pointwise-normal effective Jones matrices.
Equation~\eqref{sm:eq:normality-h2} is therefore a strict null test for pointwise
normality of that matrix.  A local diattenuating--retarding stack with
noncommuting eigenaxes remains a local alternative to a nonlocal kernel.

The null test applies to the complete
matter--propagation--measurement map up to the observation plane.  A
rotationally covariant vector pupil can preserve the covariant
\(m=0,\pm2\) sector while changing its radial coefficients and pointwise
normality.  Projection onto the covariant \(a,b,c,d\) sector separately
quantifies the noncovariant content of the experimental map.  After
independently constraining contributions from local polarization-dependent
loss combined with noncommuting retardance and from any polarization-dependent
pupil, a nonzero \(H_2\) provides evidence of transverse nonlocality.
The calibrated response in the present experiment fails the null test for
pointwise normality of its effective Jones matrix.  For the selected
director, the comparison of independent-column and marching-wave calculations
further shows that in-layer transverse coupling is sufficient to produce a
twofold modulation comparable in magnitude to the measured one.

With nonlocal propagation,
\begin{equation}
 \vect E_{\rm out}(\vect r_o)=
 \int \mathcal T(\vect r_o,\vect r_i)\vect E_{\rm in}(\vect r_i)
 \dd^2\vect r_i.
 \label{sm:eq:nonlocal}
\end{equation}
Losslessness constrains the complete operator and integrated output power, not
\(J(\vect r_o)^\dagger J(\vect r_o)\) at each output point.  Consequently,
the \(m=0,\pm2\) field selection and \(C_2\) remain exact, while
Eq.~\eqref{sm:eq:quarter-turn} may fail.

\subsection{Axisymmetric director geometry: skyrmion tubes, baby skyrmions,
and torons}
\label{sm:sec:director-geometry}

An axisymmetric nematic director can be parameterized without imposing a
particular radial profile as
\begin{equation}
 \vect n(\rho,\varphi,z)=
 \sin\theta(\rho,z)\cos\tau(\rho,z)\,\vect e_\rho+
 \sin\theta(\rho,z)\sin\tau(\rho,z)\,\vect e_\varphi+
 \cos\theta(\rho,z)\,\vect e_z.
 \label{sm:eq:director-parameterization}
\end{equation}
Here \(\theta\) is the polar tilt and \(\tau\) is the director helicity,
defined as the angle of the transverse director relative to
\(\vect e_\rho\).  Where the transverse projection is nonzero, its
unoriented axis is defined modulo \(\pi\).  For the full three-dimensional
director, the nematic identification is
\((\theta,\tau)\sim(\pi-\theta,\tau+\pi)\).
Equation~\eqref{sm:eq:director-parameterization} remains exactly axisymmetric
even when \(\theta\) and \(\tau\) vary strongly in both \(\rho\) and \(z\).
Analytical localized solutions of the Frank free energy provide an earlier
two-dimensional example of this family
\cite{bogdanov2003nematic_skyrmions}.

A straight skyrmion tube may have a \(z\)-independent wall and constant
helicity.  It then approaches the local commuting limit: the projected
retarder axis does not rotate with depth, and the locally generated
antisymmetric zero-order amplitude vanishes.  A nonsingular baby skyrmion spans
the cell without point defects but may bend smoothly as finite anchoring
modulates \(\theta\) and \(\tau\) along \(z\).  Weak axial modulation is a
perturbation of the commuting-tube limit, not a new angular sector.  The modeled toron
instead closes its reversed core within the cell.  Its curved wall terminates
at two axial point defects; its radius \(R_{\rm w}(z)\), defined by
\(n_z=0\), varies through the cell; and its helicity rotates approximately
linearly between the substrates.  These geometric changes preserve
rotational covariance about \(z\) but make the local retarders at different
depths noncommuting.  They change the amplitudes and phases within the allowed
\(m=0,\pm2\) order set without opening any new angular order.

This hierarchy is geometric and perturbative rather than purely topological.
A straight Bloch-like tube can acquire \(b\) through marching-wave transverse
coupling, whereas
an additional meridional mirror in a radial N\'eel-like texture can forbid the
same amplitude.  The robust result common to the entire axisymmetric family is the
\(m=0,\pm2\) selection rule; proximity to the commuting limit determines how
weakly \(b\) is populated before diffraction.  Published
polarizing micrographs show axisymmetric baby-skyrmion configurations
\cite{ackerman_boyle_smalyukh2017squirming} and distinct accumulated-retardance
patterns for translationally invariant skyrmions and defect-terminated torons
\cite{tai2020surface_anchoring}.  Their lensing and beam-steering responses
also establish that propagation through these textures is an observable part
of the optical map \cite{hess2020control_light}.  These observations motivate
applying the present symmetry hierarchy beyond the toron used for the
complex-field test.

The wall geometry and helicity of the model toron are quantified below from
wall crossings and weighted angular statistics; complementary
three-dimensional and sectional views appear in
Sec.~\ref{sm:sec:director-model}.  A deliberately tilted boundary condition,
which removes material axisymmetry rather than merely changing the allowed
radial profiles, provides the positive control in
Sec.~\ref{sm:sec:tilted-control}.

The relaxed toron also has approximate substrate-exchange symmetry: a rotation
by \(\pi\) about an in-plane axis interchanges the two substrates and the
neighborhoods of the two point defects.  Together with reciprocity, this
geometric property gives the Jones-component and image-axis
constraints derived in Sec.~\ref{sm:sec:surface-exchange}.  It is stronger than
material axisymmetry but does not require \(b=0\).

\subsection{Independent-column and marching-wave propagation}
\label{sm:sec:radial-evolution}

The uniaxial material tensor is
\begin{equation}
 \varepsilon_{ij}=n_o^2\delta_{ij}
 +(n_e^2-n_o^2)n_i n_j.
 \label{sm:eq:dielectric}
\end{equation}
At each fixed transverse position, independent-column propagation obeys
\begin{equation}
 \partial_z\Psi=\ii G\Psi,\qquad
 J_{\rm loc}=\mathcal P\exp\left(\ii\int_0^L G\dd z\right),
 \label{sm:eq:local-jones}
\end{equation}
where \(\Psi=(E_x,E_y)^{\mathsf T}\), \(G\) is the local transverse
generator, \(\mathcal P\) orders the exponential along \(z\), and \(L\) is the
cell thickness.  In this local-retarder model, the generator is
Hermitian in the same transverse basis, \(G^\dagger=G\).  Consequently, the
path-ordered propagator satisfies
\(J_{\rm loc}^\dagger J_{\rm loc}=I_2\).  The path-ordered product retains
noncommuting retarders along \(z\) but excludes coupling between neighboring
columns.

The marching-wave model is a one-way vector beam-propagation method that
retains the forward Maxwell branch and alternates anisotropic evolution with
homogeneous Fourier-space propagation
\cite{feit_fleck1978,van_roey1981,poy_zumer2020bpm,
petelin2023pom_bpm_preprint}.  For the axisymmetric director in
Eq.~\eqref{sm:eq:director-parameterization}, the reciprocal local generator is
\begin{equation}
 G=g_0I_2+g_1\Sigma_1+g_2\Sigma_2,\qquad
 (g_1,g_2)=g(\cos2\tau,\sin2\tau).
 \label{sm:eq:local-generator}
\end{equation}
In the electric-only paraxial reduction these quantities are explicit.  With
\begin{equation}
 h=\frac{\Delta\varepsilon n_o^2\sin^2\theta}{\varepsilon_{zz}},
 \qquad
 \bar\varepsilon_\perp=
 \left(n_o^2+\frac h2\right)I_2+
 \frac h2(\cos2\tau\,\Sigma_1+\sin2\tau\,\Sigma_2),
 \label{sm:eq:local-schur-axis}
\end{equation}
the zeroth-derivative part of
Eq.~\eqref{sm:eq:electric-only-radial-system} is
\begin{equation}
 g_0=\frac{k_0^2(n_o^2+h/2)-k_b^2}{2k_b},\qquad
 g=\frac{k_0^2h}{4k_b}.
 \label{sm:eq:explicit-local-generator}
\end{equation}
For physical interpretation, the leading transverse-derivative expansion of
the forward branch can be written as
\begin{equation}
 \partial_z\vect q=
 \ii
 \begin{pmatrix}
 g_0&0&g_1&g_2\\
 0&g_0&g_2&-g_1\\
 g_1&g_2&g_0&0\\
 g_2&-g_1&0&g_0
 \end{pmatrix}\vect q
 +
 \frac{\ii}{2k_b}
 \begin{pmatrix}\mathcal L_0a\\\mathcal L_0b\\
                 \mathcal L_2c\\\mathcal L_2d\end{pmatrix}
 +\mathcal W_{\rm vec}[\theta,\tau]\vect q,
 \label{sm:eq:radial-system}
\end{equation}
where
\begin{equation}
 \mathcal L_m=\partial_\rho^2+\rho^{-1}\partial_\rho-\frac{m^2}{\rho^2},
 \qquad \vect q=(a,b,c,d)^{\mathsf T}.
 \label{sm:eq:radial-laplacian}
\end{equation}
For a uniform incident field, the entrance condition is
\(\vect q=(1,0,0,0)^{\mathsf T}\).  Equation~\eqref{sm:eq:radial-system} provides
an analytical decomposition of the forward branch.  The numerical solver,
however, advances the tangential electric--magnetic state in
Eqs.~\eqref{sm:eq:exact-radial-e} and \eqref{sm:eq:exact-radial-h}; the forward
TE/TM projector supplies the nonlocal relation between \(\vect q_H\) and
\(\vect q_E\).

Equation~\eqref{sm:eq:radial-system} separates the forward evolution into three
contributions: the ordered local-retarder stack, scalar diffraction, and
polarization-dependent vector corrections.  Scalar diffraction changes
radial amplitudes and phases without changing angular order.  The final term,
\(\mathcal W_{\rm vec}\), collects the derivative couplings absent from both
the local-retarder matrix and the scalar Laplacians.  Within the electric-only
one-way approximation, it is a closed operator on the four electric
coefficients.  After removing the background carrier as in
Eq.~\eqref{sm:eq:electric-only-radial-system} and suppressing tildes, define
\(\vect v=\mathcal D(\vect q)\) and split
\begin{equation}
 \vect\ell=\vect\ell_0+\vect\ell_1,\qquad
 \vect\ell_0^{\mathsf T}=-\vect\eta^{\mathsf T}\widehat J,\qquad
 \vect\ell_1^{\mathsf T}=
 \frac{\ii k_b}{k_0^2\varepsilon_{zz}}\vect v^{\mathsf T}.
 \label{sm:eq:radial-vector-ell}
\end{equation}
The vector term in Eq.~\eqref{sm:eq:radial-system} is explicitly
\begin{equation}
 \mathcal W_{\rm vec}\vect q=
 \frac{1}{2\ii k_b}\left\{
 -k_0^2\Pi_4(\vect g\vect\ell_1^{\mathsf T})
 +\mathcal G(\vect v)
 +\ii k_b\mathcal G(\vect\ell_0+\vect\ell_1)
 \right\}.
 \label{sm:eq:explicit-vector-operator}
\end{equation}
Every quantity in Eq.~\eqref{sm:eq:explicit-vector-operator} is defined
componentwise in Eqs.~\eqref{sm:eq:radial-matrix-projection},
\eqref{sm:eq:radial-divergence-map}, \eqref{sm:eq:radial-gradient-map}, and
\eqref{sm:eq:radial-material-blocks}.  Thus \(\mathcal W_{\rm vec}\) is
shorthand for an explicit combination of radial derivatives, not an
additional phenomenological term.  The \(E\)--\(H\) equations integrated by
\texttt{topovec} are
Eqs.~\eqref{sm:eq:exact-radial-e}--\eqref{sm:eq:exact-radial-h}.

The antisymmetric amplitude \(b\) obeys
\begin{equation}
 \left(\partial_z-\ii g_0-\frac{\ii}{2k_b}\mathcal L_0\right)b
 =\ii(g_2c-g_1d)+(\mathcal W_{\rm vec}\vect q)_b.
 \label{sm:eq:b-evolution}
\end{equation}
The first source is the noncommutation of the instantaneous retarder axis with
the accumulated local propagator; the second is the
polarization-dependent transverse-coupling term
\(\mathcal W_{\rm vec}\).  Both mechanisms populate the same symmetry-allowed
order-zero amplitude \(b\), but depend differently on the director helicity.  In particular,
Eq.~\eqref{sm:eq:explicit-vector-operator} gives
\begin{equation}
 \begin{split}
 (\mathcal W_{\rm vec}\vect q)_b={}&\frac{1}{2\ii k_b}
 \bigg\{-\frac{k_0^2}{2}
 \left(g_\varphi\ell_{1,\rho}-g_\rho\ell_{1,\varphi}\right)
 -\frac12\left(\partial_\rho+\frac1\rho\right)v_\varphi\\
 &\hspace{5em}
 -\frac{\ii k_b}{2}
 \left(\partial_\rho+\frac1\rho\right)
 (\ell_{0,\varphi}+\ell_{1,\varphi})\bigg\},\\
 v_\varphi=-\partial_\rho b+\partial_\rho d+\frac{2d}{\rho},\\
 \ell_{0,\varphi}&=-\eta_\rho(-b+d)-\eta_\varphi(a-c),\qquad
 \ell_{1,\varphi}=
 \frac{\ii k_b}{k_0^2\varepsilon_{zz}}v_\varphi .
 \end{split}
 \label{sm:eq:explicit-b-vector-source}
\end{equation}

\subsection{Substrate exchange and mirror constraints}
\label{sm:sec:surface-exchange}

Consider a texture with exact substrate-exchange symmetry: a half turn about
an in-plane axis interchanges two identical substrates and maps the director
field onto itself.  Lorentz reciprocity then relates the forward transmission
kernel to the transpose of the spatially reflected reverse kernel.  In the
spatially local limit, choosing the in-plane axis as \(\varphi=0\) gives
\begin{equation}
 d(\rho)=0,\qquad B_+(\rho)=B_-(\rho).
 \label{sm:eq:surface-exchange}
\end{equation}
This exact local relation equates the two opposite conversion amplitudes but
does not require \(b=0\).  The relaxed model toron has only approximate
substrate-exchange symmetry, so Eq.~\eqref{sm:eq:surface-exchange} is a limiting
diagnostic rather than an exact constraint on that field.  Scalar diffraction
preserves the equality if it holds at the exit plane.  Under vectorial,
spatially nonlocal propagation, reciprocity constrains the complete pair of
forward and reverse kernels rather than the reduced forward response; the
latter may therefore have \(B_+\ne B_-\).

The same symmetry explains why the image axes may remain fixed even when
their contrast changes substantially.  Align the crossed input and analyzer
axes with the in-plane half-turn axis.  Then
\begin{align}
 J_{yx}&=b+c\sin2\varphi,\nonumber\\
 I_\times&=|b|^2+\frac{|c|^2}{2}
 -\frac{|c|^2}{2}\cos4\varphi
 +2\operatorname{Re}(b^*c)\sin2\varphi,\nonumber\\
 H_2&=-\ii\operatorname{Re}(b^*c),\nonumber\\
 H_4&=-\frac14|c|^2.
 \label{sm:eq:surface-exchange-axis-locking}
\end{align}
Thus, wherever the harmonics are nonzero,
\(\arg H_2=\pm\pi/2\) and \(\arg H_4=\pi\): substrate exchange
and reciprocity lock the mirror axes to the external linear polarizers rather
than allowing them to rotate continuously with \(\rho\) or \(z\).  A
zero of \(\operatorname{Re}(b^*c)\) reverses the sign of \(H_2\) and
interchanges its bright and dark directions by \(\pi/2\), while leaving the
geometric mirror family unchanged.  At that zero the response is \(C_4\) if
\(c\ne0\); it is circular only if the fourth harmonic also vanishes.

A cylindrically symmetric scalar propagator and a centered circular pupil do
not mix the sine and cosine partners of a fixed angular order.  They therefore
preserve Eq.~\eqref{sm:eq:surface-exchange-axis-locking} during homogeneous
defocus, although the propagated radial phases can create repeated sign
reversals of \(\operatorname{Re}(b^*c)\).  Marching-wave propagation inside
the sample need not preserve the reduced pointwise condition \(d=0\), so the
locking is exact for substrate-symmetric independent-column propagation
and only approximate for marching-wave propagation and for the experiment.

The additional vertical mirror symmetry of an exactly radial texture
\(\tau=0\pmod\pi\) imposes
\begin{equation}
 J(\rho,-\varphi)=M J(\rho,\varphi)M^{-1},
 \qquad M=\operatorname{diag}(1,-1),
 \label{sm:eq:mirror}
\end{equation}
and hence \(b=d=0\), independently of diffraction strength.  A constant
nonzero Bloch-like helicity lacks this mirror and permits
marching-wave generation of \(b\).  If the director itself loses
rotational covariance, odd angular orders become allowed; this constitutes a
qualitatively different route to optical asymmetry.

\subsection{Helicity: local commutators and transverse coupling}
\label{sm:sec:helicity}

For the rank-one projector onto a local retarder axis,
\begin{equation}
 \Pi(\alpha)=
 \begin{pmatrix}
 \cos^2\alpha&\sin\alpha\cos\alpha\\
 \sin\alpha\cos\alpha&\sin^2\alpha
 \end{pmatrix},
\end{equation}
two layers satisfy
\begin{equation}
 [\Pi(\alpha_1),\Pi(\alpha_2)]
 =\tfrac12\sin[2(\alpha_1-\alpha_2)]K.
 \label{sm:eq:retarder-commutator}
\end{equation}
A fixed projected axis gives a commuting stack and no local \(b\).
Longitudinal variation of \(\tau\) generates \(b\) through the first
nonvanishing Magnus commutator.  Rescaling a weak variation
\(\delta\tau\) by a dimensionless amplitude \(s\),
\(\delta\tau\mapsto s\,\delta\tau\), therefore gives \(b=O(s)\).  This is a
perturbative forward prediction, not a unique inverse
relation between \(b\) and helicity.  At finite variation, contributions from
different depths interfere coherently, and vectorial diffraction adds a
second complex contribution to \(b\).  The controls in
Sec.~\ref{sm:sec:controls} distinguish the population of \(b\) from its
contribution to observable twofold interference.

At constant helicity, the local propagator belongs to the algebra
\begin{equation}
 U_0=a_0(\rho)I_2+c_0(\rho)\Sigma_\tau(\varphi),\qquad
 \Sigma_\tau=\cos2\tau\,\Sigma_1+\sin2\tau\,\Sigma_2.
 \label{sm:eq:constant-helicity-algebra}
\end{equation}
The scalar Laplacian preserves this algebra,
\begin{equation}
 \nabla_\perp^2[fI_2+g\Sigma_\tau]
 =\mathcal L_0f\,I_2+\mathcal L_2g\,\Sigma_\tau,
 \label{sm:eq:scalar-closure}
\end{equation}
so scalar diffraction cannot generate \(K\) at any order.  The vectorial
derivative operator does not obey this closure.  Its mirror-odd permittivity
components include
\begin{equation}
 \varepsilon_{z\varphi}=\Delta\varepsilon\sin\theta\cos\theta\sin\tau,
 \qquad
 \varepsilon_{\rho\varphi}
 =\tfrac12\Delta\varepsilon\sin^2\theta\sin2\tau .
 \label{sm:eq:mirror-odd}
\end{equation}
Writing the vector-diffractive contribution to the antisymmetric amplitude as
\(b_{\rm v}\), symmetry requires
\begin{equation}
 b_{\rm v}(\rho;-\tau)=-b_{\rm v}(\rho;\tau),\qquad
 b_{\rm v}(\rho;0)=0.
 \label{sm:eq:b-parity}
\end{equation}

For a family with constant helicity \(\tau\), at fixed polar-tilt profile
\(\theta(\rho,z)\), geometry, and optical parameters, this odd dependence can
be written as
\begin{equation}
 b_{\rm v}(\rho;\tau)=\mathcal B_1(\rho)\sin\tau+
 \mathcal B_2(\rho)\sin2\tau+\cdots.
 \label{sm:eq:b-vector}
\end{equation}
Here \(\mathcal B_j(\rho)\) are complex radial response coefficients.

The dimensionless scalar diffraction strength is
\(\mu_{\rm diff}=L/(2k_bw_{\rm eff}^2)\), where \(w_{\rm eff}\) is the
effective transverse scale of the active wall or optical fringe.
The scalar Duhamel correction about the local propagator is
\begin{equation}
 U(L,0)=U_0(L,0)+\frac{\ii}{2k_b}
 \int_0^L U_0(L,z)\nabla_\perp^2U_0(z,0)\dd z+\cdots .
 \label{sm:eq:duhamel}
\end{equation}
When the local stack has already generated \(b\), this correction changes its
phase relation to the order-\(\pm2\) amplitudes and generically produces
\(H_2=O(\mu_{\rm diff})\), even though the local value is zero.  At constant
helicity, scalar diffraction alone remains confined to
Eq.~\eqref{sm:eq:constant-helicity-algebra}; the vectorial terms in
Eq.~\eqref{sm:eq:explicit-vector-operator} are then required to generate \(b\).
Increasing the sample thickness or decreasing the active wall or fringe width
increases the nonlocal correction.  Wavelength changes both
\(\mu_{\rm diff}\) and the local retardance, so no universal monotonic
wavelength dependence is expected.

Population of \(b\) must therefore be distinguished from visible symmetry
breaking.  Local lossless evolution can generate a large \(b\) while retaining
\(\operatorname{Re}(b^*c)=\operatorname{Re}(b^*d)=0\), and hence
\(H_2=0\).  Vectorial diffraction may add only a smaller contribution
to \(b\), yet alter the relative phase so that \(H_2\) becomes
nonzero.  The controls in Sec.~\ref{sm:sec:controls} quantify these two aspects
through the \(b\) power and the phase-sensitive interference efficiency.

\subsection{Homogeneous propagation and a coherent pupil}
\label{sm:sec:propagation}

Let \(\zeta\) denote the homogeneous propagation distance from an exit or
reconstruction plane; it is distinct from the material-depth coordinate
\(z\).  For a radial profile \(f(\rho)\) of azimuthal order \(m\), define its
order-\(m\) Hankel transform by
\begin{equation}
 \widetilde f_m(\kappa)=
 \int_0^\infty f(\rho)J_m(\kappa\rho)\rho\dd\rho,
 \label{sm:eq:hankel}
\end{equation}
and the corresponding pupil-filtered propagation operator by
\begin{equation}
 \mathcal U_m^{(P)}(\zeta)f=
 \int_0^\infty\widetilde f_m(\kappa)P(\kappa)
 \ee^{\ii\zeta[\sqrt{k^2-\kappa^2}-k]}
 J_m(\kappa\rho)\kappa\dd\kappa.
 \label{sm:eq:hankel-propagator}
\end{equation}
Here \(\kappa\) is the radial spatial angular frequency, \(J_m\) is the
Bessel function of the first kind, \(k\) is the wavenumber in the homogeneous
medium, and \(P(\kappa)\) is the complex coherent pupil function.  With
\(\Theta\) denoting the unit-step function, an ideal centered circular aperture
has \(P=\Theta(k_0\mathrm{NA}-\kappa)\).  The four radial coefficients then
evolve as
\begin{equation}
 \begin{pmatrix}a\\b\\c\\d\end{pmatrix}(\rho,\zeta)=
 \operatorname{diag}
 (\mathcal U_0^{(P)},\mathcal U_0^{(P)},
  \mathcal U_2^{(P)},\mathcal U_2^{(P)})
 \begin{pmatrix}a_{\rm e}\\b_{\rm e}\\c_{\rm e}\\d_{\rm e}\end{pmatrix}.
 \label{sm:eq:focus-law}
\end{equation}
The subscript \({\rm e}\) denotes the coefficients at the exit or
reconstruction plane.  Equation~\eqref{sm:eq:focus-law} is diagonal in azimuthal
order: a centered scalar pupil and homogeneous defocus neither mix angular
orders nor generate \(b\) when \(b_{\rm e}=0\).  They nevertheless act
differently on the radial spectra of the \(m=0\) and \(|m|=2\) sectors,
thereby changing their local amplitudes and relative phases.  The bilinear
observables \(H_2\) and \(H_4\) can consequently be suppressed, restored, or
reoriented as functions of \(\rho\) and \(\zeta\) without any change in the
material configuration.

For \(P=1\), Eq.~\eqref{sm:eq:focus-law} describes free homogeneous propagation;
for \(\zeta=0\), it describes coherent-pupil filtering without an additional
propagation phase.  The diagonal form requires a centered, rotationally
invariant, polarization-independent pupil.  An off-center or elliptical pupil
generally mixes angular sectors, whereas a rotationally covariant,
polarization-dependent matrix pupil may preserve the covariant sector while
altering its polarization quadratures and pointwise normality.  These are
distinct instrumental mechanisms.  Aperture-induced symmetry anomalies in
polarizing microscopy have been reported previously
\cite{inoue_kubota1958diffraction_anomaly}.

\begin{figure}[p]
 \centering
 \includegraphics[width=0.96\linewidth]
 {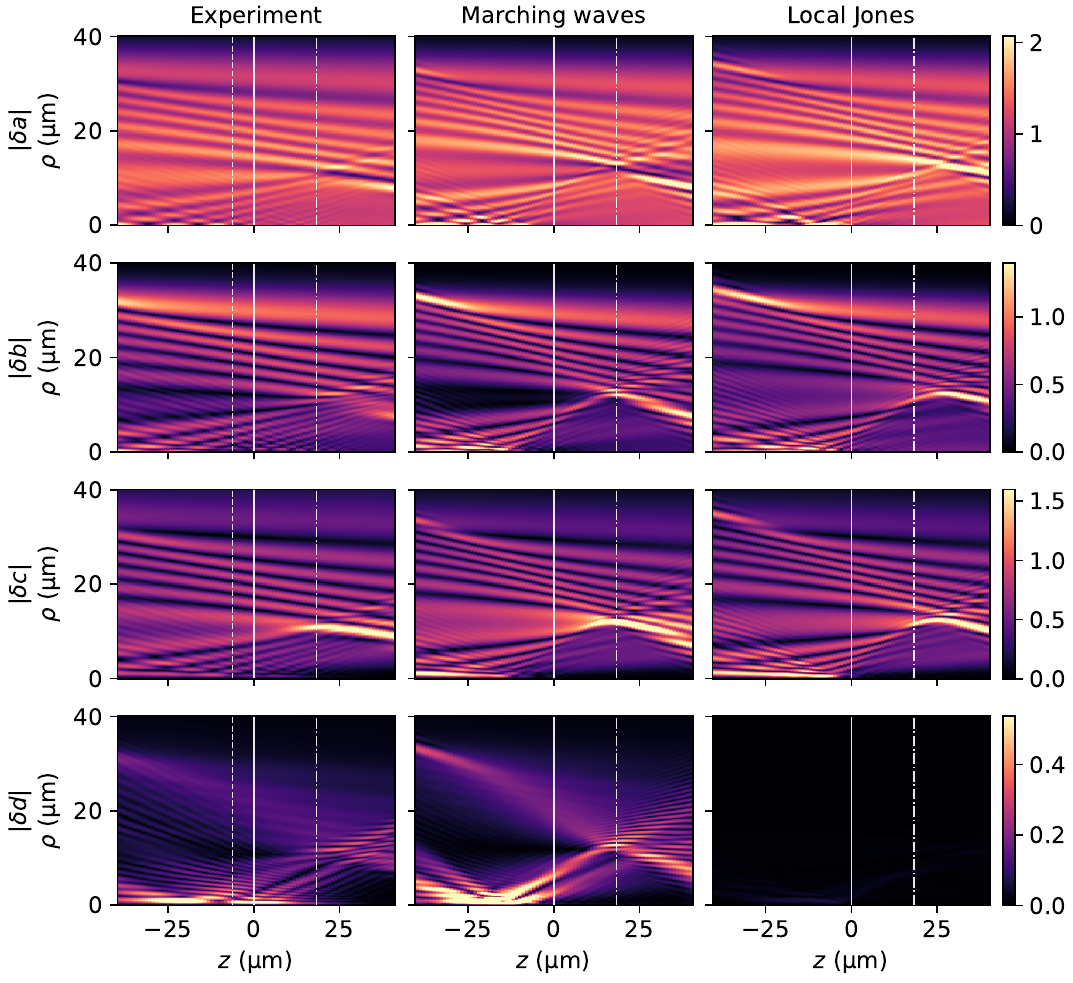}
 \caption{Homogeneous propagation of the radial response coefficients at
 \(\lambda=520\,\mathrm{nm}\).  Rows show
 \(\lvert a-a_\infty\rvert\), \(\lvert b-b_\infty\rvert\),
 \(\lvert c-c_\infty\rvert\), and \(\lvert d-d_\infty\rvert\); columns show
 the reconstructed experiment, pupil-matched marching-wave propagation, and
 pupil-matched independent-column Jones propagation.  Each row shares one
 linear color scale across all three columns.  The abscissa is the common
 model-referenced coordinate \(z\).  Solid lines mark the calculated exit
 surfaces and the corresponding experimental estimate at \(z=0\);
 dash-dotted lines mark the common annular-focus coordinate
 \(z_f=18.30\,\unitmu\mathrm m\) used for experimental--model registration.  The
 dashed line in the experimental column marks the holographic reconstruction
 plane at
 \(z=-6.23\,\unitmu\mathrm m\).}
 \label{sm:fig:coefficient-evolution}
\end{figure}

\begin{figure}[p]
 \centering
 \includegraphics[width=0.96\linewidth]
 {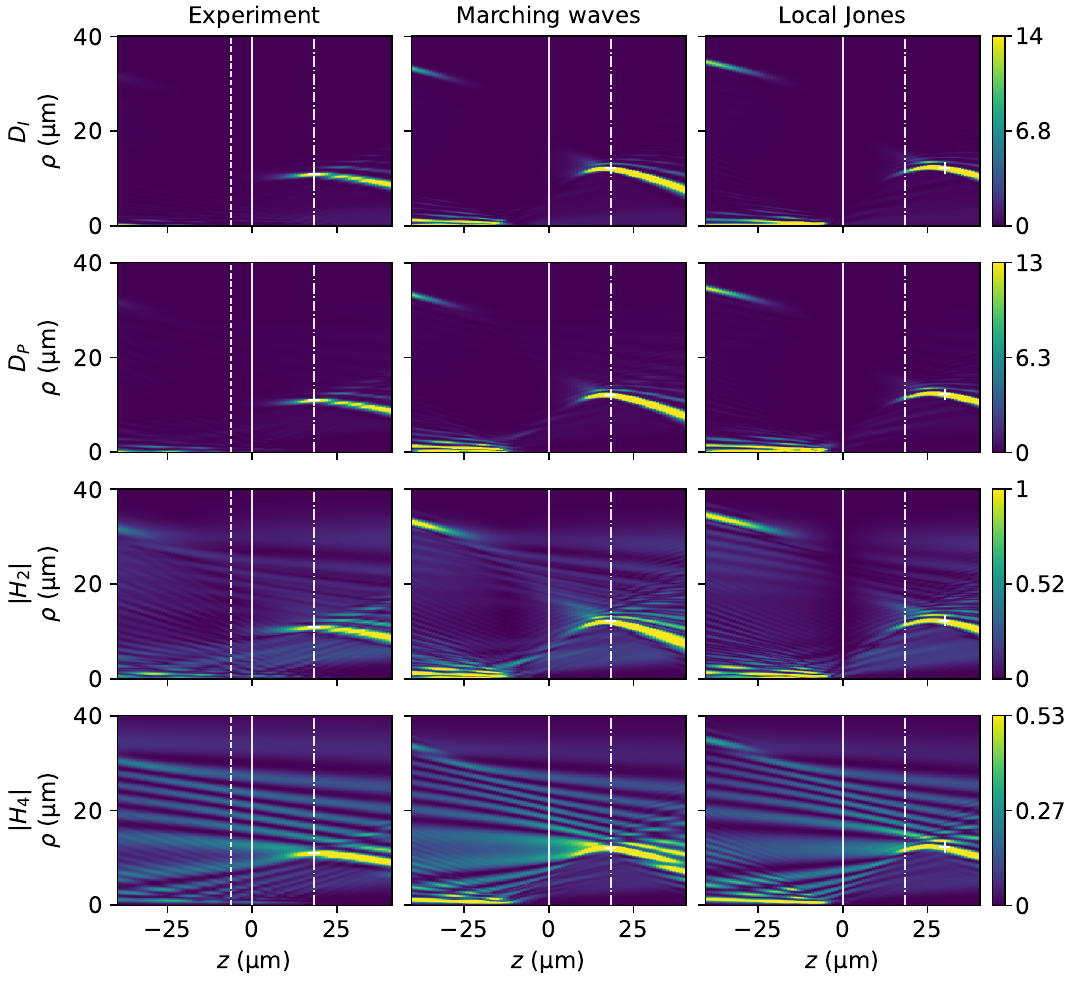}
 \caption{Axial evolution of \(D_I\), \(D_P\), and the crossed-polarizer
 harmonics \(|H_2|\) and
 \(|H_4|\) under homogeneous propagation.  Columns show the
 experiment, pupil-matched marching-wave propagation, and pupil-matched
 independent-column Jones propagation.  Each row shares one linear color
 scale; its upper limit is the joint 99.5th percentile.  In each column, the
 cross marks the two-dimensional maximum of
 \(I_{\rm av}(z,\rho)=\tfrac12\operatorname{tr}(T^\dagger T)\), repeated in
 all four rows.  After focus registration, the experimental and marching-wave
 maxima share \(z_f=18.30\,\unitmu\mathrm m\); the independent-column maximum is
 shown at its separately predicted coordinate.  Solid lines mark the
 calculated exit surfaces and the model-based experimental estimate \(z=0\),
 dash-dotted lines mark \(z_f\), and the dashed line in the experimental
 column marks the holographic reconstruction plane at
 \(z=-6.23\,\unitmu\mathrm m\).}
 \label{sm:fig:observable-evolution}
\end{figure}

\begin{figure}[p]
 \centering
 \includegraphics[width=0.96\linewidth]
 {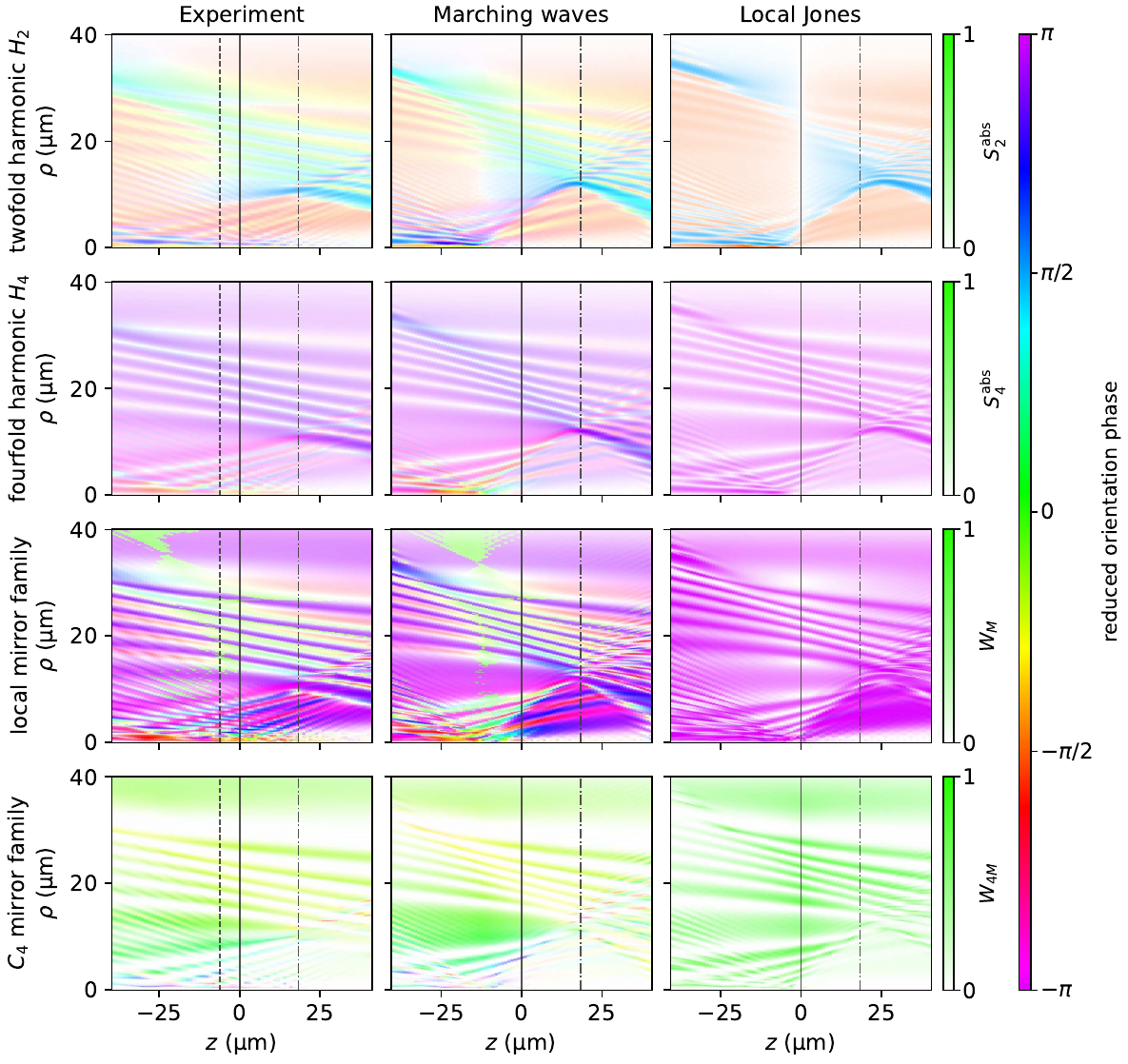}
 \caption{Harmonic orientations and resolved mirror-axis families at
 \(\lambda=520\,\mathrm{nm}\).  Columns show the experiment, pupil-matched
 marching-wave propagation, and pupil-matched independent-column Jones
 propagation.  The first two rows show the phases of \(H_2\) and \(H_4\),
 with saturation \(S_m^{\rm abs}\) from
 Eq.~\eqref{sm:eq:absolute-axis-saturation}.  The third row shows the best
 resolved mirror family \(W_M\), and the fourth its nested
 \(C_4\)-compatible subset \(W_{4M}\).  The common cyclic color scale encodes
 the reduced family phase; white denotes either unresolved angular modulation
 or low mirror fidelity, as distinguished by Eq.~\eqref{sm:eq:axis-visibility}.
 Vertical lines use the conventions of Fig.~\ref{sm:fig:coefficient-evolution}.}
 \label{sm:fig:symmetry-axis-evolution}
\end{figure}

\begin{figure*}[t]
 \centering
 \includegraphics[width=\textwidth]{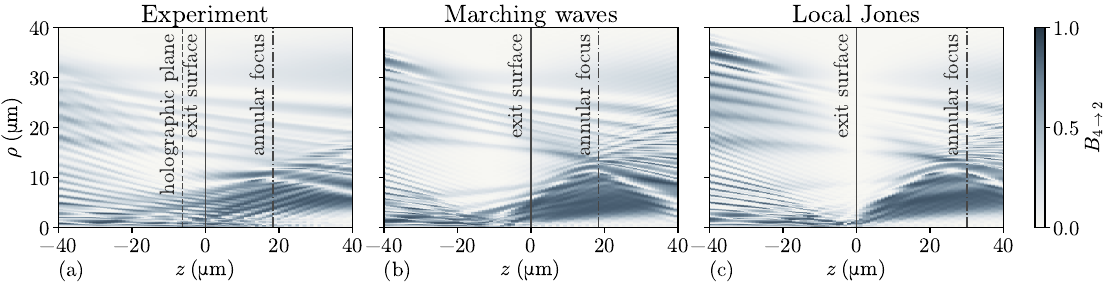}
 \caption{Focus-resolved visible reduction from \(C_4\) to \(C_2\) at
 \(\lambda=520\,\mathrm{nm}\).  Gray scale gives
 \(B_{4\to2}(\rho,z)\) for (a) experiment, (b) pupil-matched marching-wave
 propagation, and (c) pupil-matched independent-column Jones
 propagation.  Solid lines mark the calculated exit surfaces and the
 corresponding model-referenced experimental estimate, dash-dotted lines
 the annular foci, and the dashed line in (a) the holographic reconstruction
 plane.}
 \label{sm:fig:visible-c4-focus}
\end{figure*}

\begin{figure}[t]
 \centering
 \includegraphics[width=0.76\linewidth]
 {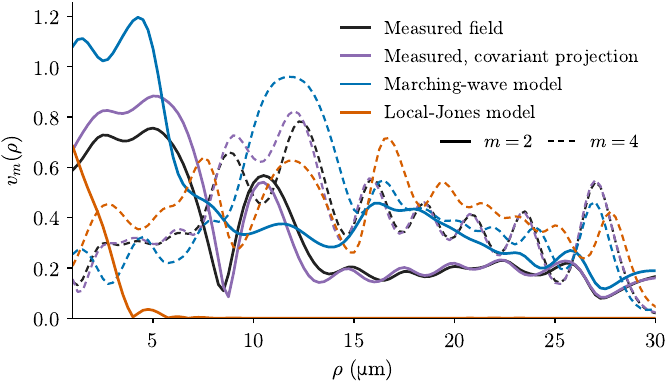}
 \caption{Radial coexistence of twofold and fourfold crossed-polarizer
 intensity at the model-referenced exit plane and \(520\,\mathrm{nm}\).  Solid
 and dashed curves show \(v_2(\rho)\) and
 \(v_4(\rho)\), respectively, for the experiment, its exactly
 rotationally covariant projection, marching-wave propagation, and
 independent-column Jones propagation.  The profiles are radially smoothed
 with a Gaussian kernel of standard deviation \(0.75\,\unitmu\mathrm m\).}
\label{sm:fig:radial-crossover}
\end{figure}

All propagation plots below (Figs.~\ref{sm:fig:coefficient-evolution}--\ref{sm:fig:radial-crossover})
use a common model-referenced observation
coordinate, again denoted by \(z\).  For both calculations, \(z=0\) is the
known liquid-crystal exit surface.  For the experiment, the
reconstruction-plane coordinate \(\zeta\) in Eq.~\eqref{sm:eq:focus-forward} is
translated by the single focus-registration parameter,
\(z=\zeta-\zeta_{\rm MW}\).  Experimental \(z=0\) is therefore the
model-based estimate of the exit plane obtained from the marching-wave
calculation, not an independently measured interface position.  The
experimental field is propagated over
\(\zeta\in[-33,47]\,\unitmu\mathrm m\), and both calculated fields over
\(z\in[-40,40]\,\unitmu\mathrm m\), using a \(1\,\unitmu\mathrm m\) step.  The
figures show the complete translated ranges.  The subscript \(\infty\) denotes
the corresponding coefficient of the uniform outer background.

In the upper two rows of Fig.~\ref{sm:fig:symmetry-axis-evolution}, color
saturation represents the amplitude of each harmonic relative to its maximum
over all three data sets and the full displayed \((z,\rho)\) domain:
\begin{equation}
 H_{m,\max}=\max_{\nu,z,\rho}|H_m^{(\nu)}|,\qquad
 S_m^{\rm abs}=\sqrt{\min(1,|H_m|/H_{m,\max})}.
 \label{sm:eq:absolute-axis-saturation}
\end{equation}
Here \(\nu\) labels the experiment, marching-wave propagation, and
independent-column Jones propagation.  The third and fourth rows use
\(W_M=VF_M\) and \(W_{4M}=VF_{4M}\) from
Eq.~\eqref{sm:eq:axis-visibility}, with
\(\epsilon=0.01Q_{99}(H_0)\), where \(Q_{99}\) denotes the joint 99th
percentile over the plotted data.  Their hues represent the geometric family
phases \(\Phi_M=4\psi_M\) and \(\Phi_{4M}=8\psi_4\), reduced to \([ -\pi,\pi)\).  A full hue cycle
therefore traverses one family of equivalent mirror lines, without choosing
an arbitrary bright or dark representative.  Since \(W_{4M}\leq W_M\), the
fourth row is visibly nested inside the third.

The nearly constant hues in Fig.~\ref{sm:fig:symmetry-axis-evolution} are
consistent with the axis locking in
Eq.~\eqref{sm:eq:surface-exchange-axis-locking}.  For the independent-column
field, \(H_4\) remains on the negative real axis, whereas
\(H_2\) lies on either branch of the imaginary axis.  The abrupt color
changes in the \(H_2\) row occur when \(\operatorname{Re}(b^*c)\) changes
sign: the bright and dark twofold directions are interchanged by \(\pi/2\),
but the geometric mirror-axis family shown in the third row does not rotate.
If \(H_4\) remains nonzero, the angular profile at this crossing is
fourfold rather than circular.  The plotted coordinate describes homogeneous
propagation after the sample.  A centered circular scalar propagator
preserves the sine--cosine parity fixed by
Eq.~\eqref{sm:eq:surface-exchange-axis-locking} while changing radial amplitudes
and phases.  The modest continuous color drift in the marching-wave and experimental
columns reflects departures from the reduced pointwise condition \(d=0\), as
well as experimental imperfections; rotational covariance alone does not
require such a drift.

White in either mirror-family row indicates that the corresponding product in
Eq.~\eqref{sm:eq:axis-visibility} is small.  This can have two physically
distinct origins, distinguished pointwise by the separately evaluated factors
\(V\) and \(F_M\).  If
\(V\simeq0\), the image is locally almost circular, every line is equivalent,
and no preferred mirror line can be reported; the two upper rows then also
show weak absolute harmonic amplitudes in the usual case.  If \(V\) is
appreciable but \(F_M\simeq0\), the angular pattern is visible while the
phases of \(H_2\) and \(H_4\) are incompatible, so no mirror line survives;
rotational \(C_2\) nevertheless remains exact.  Finally, a colored third row
with a white fourth row denotes a resolved mirror-symmetric \(C_2\) pattern
whose quarter-turn symmetry is broken.  Thus loss of color in the mirror rows
does not by itself mean loss of rotational \(C_2\).  Each side color bar shows
the fixed-hue saturation map used for that row, whereas the common cyclic
color bar represents orientation only.  The separate fields \(V\),
\(\varepsilon_M\), and \(\varepsilon_4\) provide an unambiguous pointwise
diagnosis.  Solid vertical lines mark \(z=0\),
dash-dotted lines the common marching-wave registration focus, and the dashed
line in the experimental column the holographic plane.

To suppress orientation information and isolate the visible reduction of
rotational symmetry from \(C_4\) to \(C_2\), we use
\begin{equation}
 B_{4\to2}=V(1-\Gamma_4)
 =V\frac{|H_2|^2}{|H_2|^2+|H_4|^2}.
 \label{sm:eq:visible-c4-breaking}
\end{equation}
This quantity vanishes both for a purely fourfold pattern and where the total
angular modulation is unresolved.  When \(H_2\) dominates it approaches the
resolved modulation \(V\), and hence approaches unity only when that
modulation is itself near unity.  It measures the visible contribution
forbidden by a quarter turn; mirror fidelity remains a separate observable.

\clearpage

Figure~\ref{sm:fig:visible-c4-focus} shows that at the model-referenced exit
plane, the experimental and marching-wave fields contain pronounced twofold
regions, whereas the independent-column field is nearly fourfold.
Homogeneous post-sample propagation nevertheless creates twofold bands even
for the independent-column field.  This is consistent with
Eq.~\eqref{sm:eq:focus-law}: propagation preserves the allowed azimuthal orders
but changes their local amplitudes and relative phases.

The complementary radial section at the model-referenced exit plane is shown
in Fig.~\ref{sm:fig:radial-crossover}.  It uses the pointwise crossed-intensity
visibility and the radial-\(L^2\) amplitude defined in
Eqs.~\eqref{sm:eq:local-harmonic-visibility} and
\eqref{sm:eq:integrated-visibility}.  The one-dimensional profiles are smoothed
with a radial Gaussian kernel of standard deviation \(0.75\,\unitmu\mathrm m\) to
suppress pixel noise; the mirror and quarter-turn tests themselves are
evaluated pointwise, without radial smoothing.

In experiment,
\((\mathcal V_2,
\mathcal V_4)\) changes from \((0.520,0.467)\) over
\(1\leq\rho\leq10\,\unitmu\mathrm m\) to \((0.218,0.398)\) over
\(10\leq\rho\leq28\,\unitmu\mathrm m\).  Marching-wave propagation gives
\((0.550,0.441)\) and \((0.313,0.484)\), respectively, whereas the
independent-column Jones field remains fourfold dominated.  The rotationally
covariant projection of the experiment gives \((0.604,0.511)\) and
\((0.210,0.401)\), preserving the same radial ordering.  Both the maps and the
radial profiles therefore show a predominantly twofold central response that
becomes more nearly fourfold toward the outer wall.

At the native \(520\,\mathrm{nm}\) holographic plane, the exact covariant
projector is applied in the circular basis: \(J_{++}\) and \(J_{--}\) retain
only \(m=0\), \(J_{-+}\) only \(m=+2\), and \(J_{+-}\) only \(m=-2\).
This circular-polarization--order locking is stronger than retaining the set
\(m=0,\pm2\) somewhere in the laboratory-basis matrix.  On the
full-resolution native holographic grid,
\(\epsilon_{\rm cov}=11.208\%\).  The projection bounds only content outside
the covariant representation; structural perturbations within it remain in
the projected field.  As a numerical consistency check,
direct compensation of the angular winding and
Eq.~\eqref{sm:eq:coefficient-extraction} agree to an rms difference below
\(1.9\times10^{-15}\).

The annular maximum of
\(I_{\rm av}=\tfrac12\operatorname{tr}(T^\dagger T)\) occurs at
\((\zeta_f,\rho_f)=(24.52,10.98)\,\unitmu\mathrm m\) in the raw experimental
propagation and at \((z_f,\rho_f)=(18.30,12.15)\,\unitmu\mathrm m\) in the
marching-wave calculation.  Aligning this optical landmark gives
\(\zeta_{\rm MW}=\zeta_f-z_f=6.23\,\unitmu\mathrm m\), placing the holographic
reconstruction plane at \(z=-6.23\,\unitmu\mathrm m\) and the experimental
annular focus at \(z=18.30\,\unitmu\mathrm m\).  This is a model-referenced axial
registration, not an independent interface measurement.  Full-map radial
residuals remain after the shift, and the inferred origin depends on the
forward model: independent-column propagation predicts its annular maximum at
\((z_f,\rho_f)=(30.08,12.24)\,\unitmu\mathrm m\).  The two models
thus differ in axial focus position by \(11.78\,\unitmu\mathrm m\)
and in focus radius by only
\(0.08\,\unitmu\mathrm m\), or \(0.7\%\).  For the selected director-model
parameters, the focus match adjusts only the axial offset.  The focus radius,
fringe slopes and spacing, Jones-component amplitudes, and full axial
evolution are compared without further adjustment.  These comparisons remain
conditional on the preceding manual parameter selection described in
Sec.~\ref{sm:sec:director-model}.

As a direct exit-plane axisymmetry control, we propagate the complete
two-dimensional experimental Jones field homogeneously by
\(\zeta_{\rm MW}=6.23\,\unitmu\mathrm m\), apply the same covariant projector in
the circular basis, and resynthesize the corresponding strictly covariant
field.  At this model-referenced exit plane, evaluated on the factor-two
low-pass-decimated propagation grid,
\(\epsilon_{\rm cov}=11.142\%\), compared with \(11.208\%\) on the
full-resolution native grid.
Nevertheless, over \(4\leq\rho\leq28\,\unitmu\mathrm m\), the integrated
crossed-polarizer amplitudes \((m=2,m=4)\) change only from
\((0.2512,0.4051)\) to \((0.2530,0.4116)\).  Because intensity is
nonlinear, this near equality could conceal cancellation.  We therefore
resolve its origin using the exact decomposition below.
For the projected field over the same radial interval,
\[
 \frac{\int |b|^2\rho\,\dd\rho}{\int H_0\rho\,\dd\rho}=0.578,
 \qquad
 \frac{\int (|c|^2+|d|^2)\rho\,\dd\rho/2}
 {\int H_0\rho\,\dd\rho}=0.422 .
\]
The fractional modulations use Gaussian-annulus intensity coefficients with
\(\sigma_\rho=0.75\,\unitmu\mathrm m\), whereas the \(H_0\) partition is
integrated directly from the projected radial amplitudes; both use the
\(\rho\,\dd\rho\) measure.  The measured field already contains the
experimental detection response, so no separate instrument-matching pupil is
applied.  Before homogeneous propagation, its twofold amplitude on the same
factor-two low-pass-decimated grid is
\(\mathcal V_2=0.205\); the full-resolution value is
\(0.206\).
With
\(u_{\rm cov}=\vect e_y^\dagger J_{\rm cov}\vect e_x\),
\(u_{\rm res}=\vect e_y^\dagger(J-J_{\rm cov})\vect e_x\), and
\(u=u_{\rm cov}+u_{\rm res}\), we therefore use the exact decomposition
\begin{equation}
 H_m=H_m^{cc}+H_m^{cr}+H_m^{rr},
 \qquad
 I^{cc}=|u_{\rm cov}|^2,\quad
 I^{cr}=2\operatorname{Re}(u_{\rm cov}u_{\rm res}^*),\quad
 I^{rr}=|u_{\rm res}|^2 .
 \label{sm:eq:covariant-residual-intensity}
\end{equation}
For \(m=2\), the rms norms of these three terms relative to the full harmonic
are \(1.020\), \(0.121\), and \(0.213\).  For
\(\alpha\in\{cc,cr,rr\}\), define the signed projection coefficient
\[
 s_\alpha=
 \frac{\operatorname{Re}\int H_2^*(\rho)H_2^\alpha(\rho)\,
 \rho\,\dd\rho}
 {\int|H_2(\rho)|^2\rho\,\dd\rho}.
\]
The corresponding values are \(1.002\), \(-0.0720\), and \(+0.0702\).
The residual-interference and residual-intensity terms are therefore nonzero
but cancel almost exactly in projection onto the measured \(H_2\), whereas the
covariant contribution has a projection coefficient close to unity.  The complete
complex radial profiles differ by
\(\lVert H_2^{cc}-H_2\rVert/\lVert H_2\rVert=0.189\).
The same calculation for the nine individual acquisitions
gives \(s_{cc}=0.877\)--\(1.057\).  Thus the measured noncovariant remainder is
not the dominant source of the observed twofold image.

The sloping ridges in Figs.~\ref{sm:fig:coefficient-evolution} and
\ref{sm:fig:observable-evolution} are propagation fringes.  For two nearby
radial spectral components, let \(\Delta\kappa\) and \(\Delta k_z\) be their
transverse and longitudinal wavenumber differences, let
\(\bar\kappa,\bar k_z\) be the corresponding means, and let \(\sigma=\pm1\)
select the radial phase branch.  With constant reference phase
\(\Delta\phi_0\), their relative phase is
\[
 \Delta\Phi=\Delta\phi_0+\sigma\Delta\kappa\,\rho+\Delta k_z\zeta,
\]
so a constant-phase ridge has
\begin{equation}
 \frac{\dd\rho}{\dd\zeta}
 =-\sigma\frac{\Delta k_z}{\Delta\kappa}
 \simeq\sigma\frac{\bar\kappa}{\bar k_z}.
 \label{sm:eq:ridge-slope}
\end{equation}
The measured and calculated \(H_2,H_4\) ridges have slopes of the same sign
and similar magnitude, but neither field is uniformly steeper: the
experimental \(H_2\) ridges are steeper, whereas the calculated \(H_4\)
ridges are steeper.  We therefore use the ridges only as a qualitative check
of the propagated radial phase.

\subsection{Source-wavelength, discrete-rotation, and normality audits}
\label{sm:sec:experimental-audits}

The two nominally \(520\,\mathrm{nm}\) input columns were measured with
different diode-laser modules.  To quantify the resulting propagation
systematic, we repeated the homogeneous continuation with the horizontal and
vertical columns propagated at \(518.2\) and \(522\,\mathrm{nm}\),
respectively, before any change of polarization basis.  At the
model-referenced exit plane, this changes the background-subtracted complex
Jones field by \(0.261\%\) in relative \(L^2\) norm.  The unprojected
\(\mathcal V_2\) changes from \(0.251179\) to \(0.250922\),
and the projected value from \(0.252982\) to \(0.252938\).  The complete
complex projected \(H_2\) profile changes by \(0.399\%\), while
\(\epsilon_{\rm cov}\) changes from \(11.142\%\) to \(11.224\%\).  Over all
five fixed planes used in the through-focus analysis, the largest relative
changes are \(1.37\%\) for the background-subtracted Jones field and
\(1.32\%\) for the projected complex \(H_2\) profile.  Using a common
representative wavelength for homogeneous continuation therefore has a
negligible effect on the exit-plane harmonic comparison and only a small
effect throughout the reported focus range.

We next test covariance directly on the centered Cartesian grid, without
polar interpolation.  Let \(S=J-J_\infty\) and
\[
 \Delta_\beta(\vect r)=
 S(\Rmat_\beta\vect r)-\Rmat_\beta S(\vect r)\Rmat_{-\beta}.
\]
After factor-two low-pass decimation, the field occupies an odd
\(705\times705\) grid centered exactly on pixel \((352,352)\), so rotations
by \(90^\circ\) and \(180^\circ\) are exact index permutations.  We normalize
\(\lVert\Delta_\beta\rVert_2\) by the rms norm of its two terms, using the
same Gaussian-annulus estimator and \(4\leq\rho\leq28\,\unitmu\mathrm m\)
interval as for the harmonics.  The coherently averaged exit-plane field
gives
\[
 \epsilon_{90}=0.2115,\qquad \epsilon_{180}=0.1624 .
\]
The half turn does not mix the two input columns and therefore provides a
within-column baseline, whereas the quarter turn also tests their mutual
rotational consistency.  Across the nine individual acquisitions the robust
median and scatter are \(0.251\pm0.018\) and \(0.203\pm0.025\),
respectively.  The direct-grid residuals include noise, physical
nonaxisymmetry, and inter-column calibration or wavelength mismatch; they are
not assigned to any one source.  Their reduction under coherent averaging
quantifies the improvement in rotational consistency.  The split-column
result separately shows that using \(518.2\) and \(522\,\mathrm{nm}\)
for the two input columns in the homogeneous continuation changes the projected complex \(H_2\) profile by
only \(0.399\%\); it does not resolve the other contributions.

Finally, projection onto a linear covariant subspace need not preserve the
nonlinear condition of pointwise normality.  We therefore evaluate
\[
 N[J]=J^\dagger J-JJ^\dagger
\]
pixel by pixel both before and after projection.  With
\(I_J=\operatorname{tr}(J^\dagger J)/2\), define the normalized rms
defect
\[
 \delta_N=
 \left[
 \frac{\int_{4}^{28}
  \left\langle\lVert N\rVert_F^2/2\right\rangle_{\sigma_\rho}
  \rho\,\dd\rho}
 {\int_{4}^{28}
  \left\langle I_J^2\right\rangle_{\sigma_\rho}
  \rho\,\dd\rho}
 \right]^{1/2},
 \qquad \sigma_\rho=0.75\,\unitmu\mathrm m .
\]
For the averaged field, \(\delta_N=0.4167\) before projection and
\(0.3483\) after projection.  Across the nine acquisitions the corresponding
values are \(0.4475\pm0.0116\) and \(0.3541\pm0.0037\).  Thus projection
reduces rather than creates the full normality defect.  For the projected
field, the directly evaluated order-two contraction
\(\operatorname{tr}[N(\Sigma_1+\ii\Sigma_2)]/8\) agrees with the independently
calculated complex \(H_2\) profile to \(1.6\times10^{-6}\) in relative
radial-\(L^2\) norm.  Before projection the same two profiles differ by
\(10.9\%\), as expected for a field that is not exactly covariant, while the
order-two normality profiles before and after projection differ by
\(17.2\%\).  The measured field is therefore already non-normal on the scale
relevant to \(H_2\); the covariant projection does not manufacture the
failure of the null test.

\subsection{Controlled helicity and diffraction tests}
\label{sm:sec:controls}

These controls are derived from the relaxed three-dimensional director but do
not use it in unmodified form.  We construct an exactly axisymmetric synthetic
texture from its radial polar-tilt profile \(\theta(\rho,z)\), imposing in each
layer either the layer-resolved mean helicity \(\bar\tau(z)\) extracted from
the relaxed director or its constant midplane value
\(\bar\tau(L/2)=-\pi/2\).  The construction preserves the radial geometry
while separating longitudinal helicity variation from transverse
propagation.

The scalar control values quoted below are integrated over a disk of radius
\(R=28\,\unitmu\mathrm m\) centered on the toron axis; the background is estimated
outside \(31\,\unitmu\mathrm m\).  In the cumulative curves, \(R\) is the variable
disk radius.  The incident-normalized power carried by the antisymmetric
amplitude is
\begin{equation}
 P_b(<R)=2\int_{\rho\leq R}|b-b_\infty|^2\dd^2r.
 \label{sm:eq:b-power}
\end{equation}
Because the Jones fields are normalized by the incident background amplitude,
\(P_b\) has units of area and requires no control-specific normalization.  In
the discrete calculation, each pixel carries its physical area.

To separate the power in \(b\) from its contribution to the measured second
harmonic, define the phase-sensitive interference efficiency
\begin{equation}
 \eta_2^2(<R)=
 \frac{\int_{\rho\leq R}\{[\operatorname{Re}(b^*c)]^2+
 [\operatorname{Re}(b^*d)]^2\}\dd^2r}
 {\int_{\rho\leq R}|b|^2(|c|^2+|d|^2)\dd^2r}.
 \label{sm:eq:phase-unlocking}
\end{equation}
For an exactly covariant field, its numerator is
\(\int|H_2|^2\dd^2r\).  Thus \(\eta_2\)
measures, in an rms sense, how efficiently the relative phases of \(b,c,d\)
generate the allowed interference term.  It is undefined, rather than zero,
when the denominator vanishes.  The observable rms \(m=2\) modulation is the
already defined \(\mathcal V_2[0,R]\) of
Eq.~\eqref{sm:eq:integrated-visibility}.  The triplet
\((P_b,\eta_2,\mathcal V_2)\) therefore separates mode
power, phase-sensitive interference efficiency, and observable intensity
modulation.  Replacing \(R\) by a variable upper limit gives the cumulative
curves.  Disk integration uses \(2\pi\rho\dd\rho\), and all nonlinear products
entering \(P_b\) and \(\eta_2\) are evaluated before integration.  The radial
intensity harmonics use the same soft-annulus estimator as the other harmonic
results, with a \(0.75\,\unitmu\mathrm m\) Gaussian width.

All four controls use \(\lambda=520\,\mathrm{nm}\), the primary
comparison wavelength.  Control \(\mathrm{0}\) fixes
\(\bar\tau=-\pi/2\) and uses independent-column propagation.  Control
\(\mathrm H\) uses the extracted layer-resolved \(\bar\tau(z)\) but remains
independent-column.  Control \(\mathrm D\) introduces marching-wave transverse coupling at
fixed helicity, and the synthetic \(\mathrm{HD}\) control combines both effects:
\begin{equation}
\begin{array}{c|ccc}
 &P_b\;(\unitmu\mathrm m^2)&\eta_2&
 \mathcal V_2[0,R]\\ \hline
 \mathrm{0}&6.87\times10^{-6}&2.00\times10^{-3}&7.51\times10^{-7}\\
 \mathrm H&2129&6.83\times10^{-6}&1.09\times10^{-5}\\
 \mathrm D&93.7&0.755&0.342\\
 \mathrm{HD}&2143&0.381&0.332
\end{array}.
\label{sm:eq:hd-table}
\end{equation}
The power ratios are \(P_b(\mathrm H)/P_b(\mathrm{HD})=99.4\%\) and
\(P_b(\mathrm D)/P_b(\mathrm H)=4.4\%\).  Despite its much smaller
\(b\) power, \(\mathrm D\) produces an \(m=2\) modulation 3.0\% larger than
\(\mathrm{HD}\).  The controls are not additive because the helicity- and
diffraction-generated fields combine as coherent complex amplitudes.  Within
this synthetic continuation, marching-wave transverse coupling at fixed
helicity is therefore sufficient to produce the observed scale of the
twofold response, whereas longitudinal helicity variation in the
independent-column model produces a large but almost perfectly phase-locked
\(b\) amplitude.

For the complete \(520\,\mathrm{nm}\) fields at the model-referenced exit
plane, both calculated fields use the same centered circular pupil with the
measured cutoff, whereas the measured field already contains the experimental
detection response and receives no separate instrument-matching pupil.  Over
\(4\leq\rho\leq28\,\unitmu\mathrm m\), the crossed-polarizer harmonics are
\[
\begin{array}{c|cc}
 &\mathcal V_2&
 \mathcal V_4\\ \hline
\text{experiment}&0.2512&0.4051\\
\text{covariant projection}&0.2530&0.4116\\
\text{marching-wave}&0.3387&0.4802\\
\text{independent-column}&0.00472&0.4364
\end{array}.
\]
For the same director and pupil, the marching-wave model gives a
strong twofold modulation.  It includes in-layer transverse propagation
and anisotropic longitudinal-field coupling.  The models also differ in
their treatment of local retardance
(Sec.~\ref{sm:sec:optical-numerics}).  The second harmonic contains both
\(-\ii\operatorname{Re}(b^*c)\) and \(\operatorname{Re}(b^*d)\); the full
complex calculation retains their coherent sum.

\section{Physical model and numerical implementation}
\label{sm:sec:methods}

\subsection{Liquid-crystal free energy and relaxed toron state}
\label{sm:sec:director-model}

The director is obtained by minimizing the sum of the bulk Frank--Oseen
energy and the surface anchoring term given below.  The bulk contribution is
\begin{equation}
 \begin{aligned}
 F=\frac12\int\big[&
 K_1(\nabla\cdot\vect n)^2
 +K_2(\vect n\cdot\nabla\times\vect n+q_0)^2\\
 &+K_3|\vect n\times\nabla\times\vect n|^2\big]\dd^3\vect r.
 \end{aligned}
 \label{sm:eq:frank}
\end{equation}
Here \(K_1\), \(K_2\), and \(K_3\) are the splay, twist, and bend elastic
constants, and \(q_0\) is the preferred chiral wavenumber in the sign
convention of Eq.~\eqref{sm:eq:frank}.  The accompanying source-data record
\path{data_src/directors.toml} specifies that the director was relaxed for
\(1000\,\mathrm{s}\) of simulated time with \texttt{lcsim}
\cite{lcsim_git}, using the \texttt{cayley-fixed} integrator with automatic
safe-step selection.  The archived state,
\path{data_src/directors/toron_27um_5cb_soft.npz}, contains a
\(200\times200\times51\) grid of unit vectors in \((x,y,z)\).  The
accompanying record gives an isotropic node spacing of \(0.548\,\unitmu\mathrm m\)
and a cell thickness of \(27.4\,\unitmu\mathrm m\).  At all four lateral
boundaries the director is fixed along the cell normal,
\(\vect n=\vect e_z\).
Finite homeotropic anchoring at the two substrates contributes
\begin{equation}
 F_{\rm s}=\frac{W_{\rm s}}{2}\int_{\partial\Omega_{\rm sub}}
 \left[1-(\vect n\cdot\vect e_z)^2\right]\dd A,
 \qquad W_{\rm s}=5.48\times10^{-6}\,\mathrm{J\,m^{-2}}.
 \label{sm:eq:soft-anchoring}
\end{equation}
Here \(\partial\Omega_{\rm sub}\) denotes the two substrate surfaces.  The
archive records this \texttt{SoftAxis} condition separately for the top and
bottom boundaries; Eq.~\eqref{sm:eq:soft-anchoring} specifies the meaning of the
stored strength.  The elastic and chiral parameters are
\begin{equation}
 K_1=6.4,\qquad K_2=3.0,\qquad K_3=9.1\quad\mathrm{pN},
 \qquad q_0=-0.2285\,\unitmu\mathrm m^{-1}.
 \label{sm:eq:director-parameters}
\end{equation}
With the convention in Eq.~\eqref{sm:eq:frank}, the preferred bulk twist is
\(\vect n\cdot\nabla\times\vect n=-q_0\), corresponding to
\(2\pi/|q_0|=27.5\,\unitmu\mathrm m\).  The rotational viscosity used in the
relaxational solver was \(\gamma_1=203.6\,\mathrm{mPa\,s}\); it sets the
relaxation time scale but not the stationary points of
Eqs.~\eqref{sm:eq:frank} and~\eqref{sm:eq:soft-anchoring}.

The elastic constants and anchoring strength were selected manually to
reproduce experimental observations; they are not independent measurements
for this mixture and interface.  This selection did not constitute a
systematic multiparameter fit or an estimate of parameter uncertainties.
The resulting director and refractive indices are held fixed in the
comparison of local-Jones and marching-wave propagation.

The archive contains the final director, grid, boundary conditions, material
blocks, chirality, viscosity, and integrator identifier.  This file provides
the reproducible starting point for the primary optical calculations reported
here.  It does not contain the preceding relaxation trajectory, the
discretized initial seed, or a recorded terminal residual, so the optical
results can be reproduced from the archived metastable state, but the
relaxation itself cannot be reproduced exactly from this file alone.  The
comparisons are conditional on this director: they establish neither its uniqueness
as a free-energy minimum nor an inverse reconstruction of the experimental
texture.  Director-model uncertainty is therefore distinct from numerical
convergence of the optical solver.

\begin{figure*}[t]
 \centering
 \includegraphics[width=\textwidth]{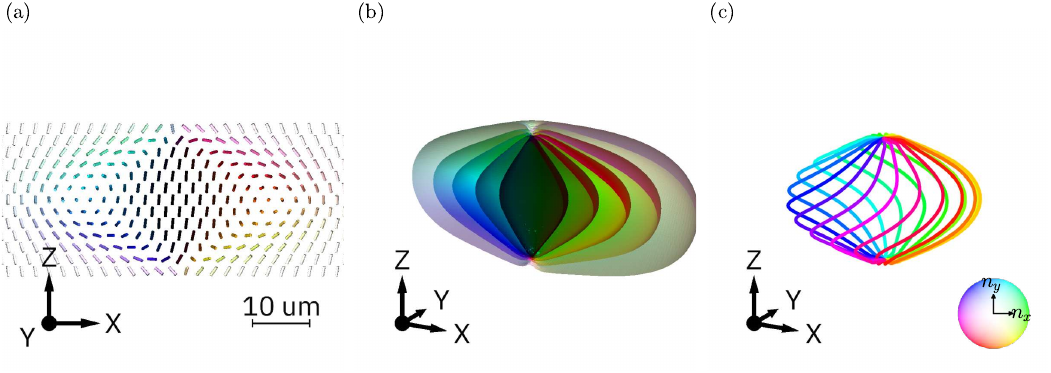}
 \caption{Three-dimensional views of the relaxed toron director used in the
 primary optical calculations.  (a) Sparse headless director glyphs in the
 central meridional section.  (b) Nested \(n_z\) isosurfaces show the closed
 wall and its two axial terminations.  (c) Director curves drawn on the
 \(n_z=0\) wall show the helicity rotation through the cell.}
\label{sm:fig:director-views}
\end{figure*}

The relaxed toron comprises a reversed central region enclosed by a curved
domain wall and terminated by two axial point defects
(Fig.~\ref{sm:fig:director-views}).  We define the wall
center by \(n_z=0\), and its inner and outer half-widths by the crossings
\(n_z=-1/2\) and \(n_z=+1/2\).  The local in-plane helicity is
\begin{equation}
 \tau_{\rm loc}=\frac12\arg
 [(n_x+\ii n_y)^2\ee^{-2\ii\varphi}],
 \label{sm:eq:director-helicity}
\end{equation}
which is invariant under the nematic identification \(\vect n\sim-\vect n\).

To quantify the depth-dependent helicity, we define geometric and optically
weighted complex nematic moments,
\begin{align}
 M_{\rm g}(z)&=\sum_{x,y}(n_x+\ii n_y)^2\ee^{-2\ii\varphi},\nonumber\\
 M_{\rm o}(z)&=\sum_{x,y}\gamma(x,y,z)
 \frac{(n_x+\ii n_y)^2\ee^{-2\ii\varphi}}{n_x^2+n_y^2},
 \label{sm:eq:helicity-moments}\\
 \gamma&=k_0\max(n_{\rm eff}-n_o,0),\qquad
 n_{\rm eff}=\frac{n_o}{
 \sqrt{n_z^2+(n_o/n_e)^2(n_x^2+n_y^2)}}.\nonumber
\end{align}
Terms with vanishing transverse director are omitted from \(M_{\rm o}\).  For
either moment, \(\arg M(z)\) is unwrapped continuously along \(z\), and the
mean helicity is \(\bar\tau(z)=\tfrac12\arg M(z)\).  Linear fits are weighted
by \(|M(z)|\).  The geometric and optically weighted slopes are
\(-0.0991\,\unitmu\mathrm m^{-1}\) and \(-0.0986\,\unitmu\mathrm m^{-1}\), with
weighted rms residuals \(0.49^\circ\) and \(0.45^\circ\), respectively.
Their ratios to \(q_0\) are \(0.434\) and \(0.432\).  The small residuals
quantify the nearly linear helicity rotation of the archived director; the
optical weight suppresses regions of weak local retardance, especially near the point
defects.

\begin{figure*}[t]
 \centering
 \includegraphics[width=\textwidth]{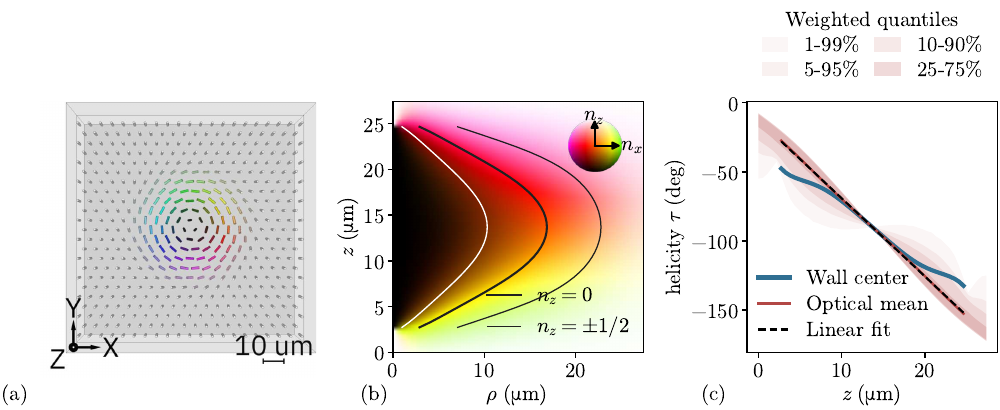}
 \caption{Quantitative characterization of the relaxed untilted toron with
 axisymmetric boundary conditions used in the optical calculations.
 (a) Headless director glyphs in the
 central \(xy\) plane.  (b) Meridional section; hue gives the in-plane
 direction, saturation gives the tilt, and lines mark \(n_z=0,\pm1/2\).
 (c) Helicity at the wall center, the optically weighted layer mean, weighted
 1--99\%, 5--95\%, 10--90\%, and 25--75\% quantile bands, and the weighted
 linear fit.  The wall center is defined by \(n_z=0\).}
 \label{sm:fig:director-quantitative}
\end{figure*}

Figure~\ref{sm:fig:director-quantitative} summarizes the central-plane texture,
wall geometry, and depth-dependent helicity of the same relaxed state.

\subsection{Optical material model}
\label{sm:sec:optical-model}

The optical calculation uses the lossless uniaxial tensor in
Eq.~\eqref{sm:eq:dielectric}.  Pure-5CB dispersion at
\(25\,{}^\circ\mathrm C\) \cite{li2005refractive} gives
\begin{equation}
\begin{array}{c|ccc}
\lambda\ (\mathrm{nm})&453&520&662\\ \hline
n_o&1.5582&1.5441&1.5299\\
n_e&1.7673&1.7336&1.7018
\end{array}.
\label{sm:eq:indices}
\end{equation}
The \(520\,\mathrm{nm}\) field is used for the primary quantitative analysis.
The \(453\)- and \(662\)-nm calculations provide qualitative checks against
the nominal \(450\)- and \(660\)-nm measurements shown in
Sec.~\ref{sm:sec:linear-comparisons}.
Dispersion data for the precise 5CB--S811 mixture were unavailable, so its
real refractive indices are approximated by the pure-5CB values in
Eq.~\eqref{sm:eq:indices}.  Spatially uniform attenuation cancels under
normalization by the outer response.  Polarization-dependent loss combined
with noncommuting retardance, which could provide a local non-normal
alternative to the experimental null test, is not included.  The forward
model treats the liquid crystal as a lossless uniaxial layer and neglects
substrate reflections and backward waves.
Centered isotropic planar interfaces preserve the \(m=0,\pm2\) selection rule,
although a reflection-aware Berreman calculation could shift the radial
phases.

\subsection{Independent-column and marching-wave propagation}
\label{sm:sec:optical-numerics}

The independent-column calculation multiplies the lossless layer matrices in
propagation order.  It retains the full path-ordered local retardance,
including helicity-induced commutators, but transfers no field between
neighboring transverse columns.  The marching-wave calculation in
\texttt{topovec} \cite{topovec_git} advances the forward electric--magnetic
state through the same director field.  Each step applies the exact homogeneous
transverse Fourier propagator, projects onto the forward TE/TM branches
of the isotropic reference medium, and
integrates the anisotropic residual.  Even in a uniform anisotropic
layer, this projection can change the relative phase of the two
eigenpolarizations compared with the local-Jones result.  The difference
persists in the limit of vanishing axial step.  The associated longitudinal-field
coupling is written explicitly in Eqs.~\eqref{sm:eq:exact-radial-e}--
\eqref{sm:eq:exact-radial-h}; \(\mathcal W_{\rm vec}\) denotes the same terms only
within the electric-only analytical reduction of Eq.~\eqref{sm:eq:radial-system}.

The director grid is interpolated transversely by a factor of two, producing
\(400\times400\) output maps with \(0.274\,\unitmu\mathrm m\) sampling.
The two input basis polarizations are propagated independently and stored as
the columns of \(J\).  Calculations use single-precision arithmetic.  For the
marching-wave calculations, the maximum internal phase increment is
\(0.025\) rad at 453 and 520 nm and \(0.0125\) rad at 662 nm.  The coherent
imaging pupil is not applied during in-layer propagation; it is imposed once
afterward for instrument-matched comparisons.
Propagation through a uniform homeotropic director supplies the common complex
phase reference.
The calculated fields are stored in
\begin{equation}
\begin{array}{l}
\texttt{data/simulated/toron\_27um\_5cb\_soft/}\\
\qquad\texttt{\{450,520,660\}nm\_\{spectral-jones,marching-wave\}.npz}.
\end{array}
\label{sm:eq:production-paths}
\end{equation}
The filenames use the nominal experimental wavelengths; each manifest records
the numerical wavelength actually used.  Every compressed NumPy archive
contains \(J_{xx},J_{xy},J_{yx},J_{yy}\), the physical pixel pitch, and a
simulation manifest that records the director source, component and axis
order, wavelength, refractive indices, subdivision, phase increment,
precision, and propagation model.  For instrument-matched
comparisons, an ideal centered circular pupil with cutoff inferred from the
holographic Jones spectra is applied to both calculated exit fields after
propagation through the cell.

\section{Forward-model validation and reproducibility}
\label{sm:sec:validation}

The following controls test distinct parts of the forward model with the
selected parameters held fixed and no additional fit to the experimental
harmonic amplitudes.  Total-intensity redistribution
probes transverse energy transport; a deliberately tilted boundary condition
tests the angular-order classification after material axisymmetry is broken;
and phase-step and symmetry-reconstruction tests quantify selected numerical
sensitivities.

\subsection{Lensing under homogeneous propagation}
\label{sm:sec:lensing}

Lensing is most directly visible in
\(S_0=|E_x|^2+|E_y|^2\), which sums both transverse output components and
requires no analyzer.  For unit input, a lossless independent-column map
leaves \(S_0\) unchanged at the exit surface; spatial redistribution appears
only after subsequent homogeneous propagation or coherent-pupil filtering.
Marching-wave propagation additionally permits transverse energy flow within
the liquid crystal, producing annular focusing, radial transport, and an
angular-order-two anisotropy of the local Gram matrix.

The basis-invariant quantity
\(I_{\rm av}=\tfrac12\operatorname{tr}Q\) provides the annular optical landmark
used for the one-parameter axial registration in
Figs.~\ref{sm:fig:coefficient-evolution} and
\ref{sm:fig:observable-evolution}.  Its position depends on the forward model and
therefore cannot by itself locate the material interface.  The focus match is
used only as the explicitly stated axial registration parameter, not as a
general validation of the full propagated field.

\subsection{Controlled breaking of axisymmetry}
\label{sm:sec:tilted-control}

\begin{figure}[p]
 \centering
 \includegraphics[width=0.98\linewidth]{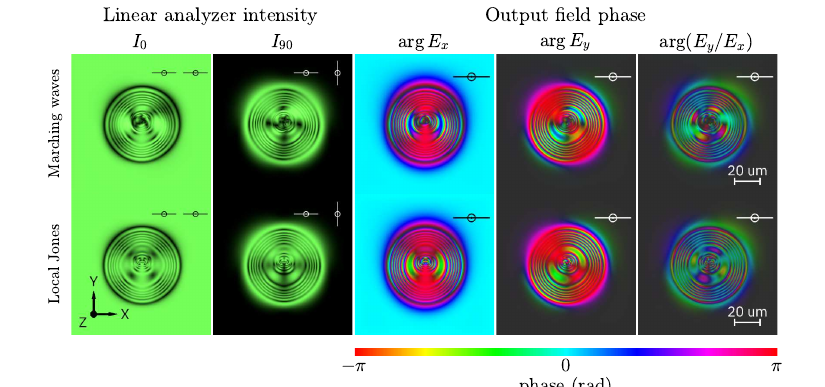}
 \caption{Positive control with both substrate anchoring directions tilted by
 \(1.718^\circ\) at \(520\,\mathrm{nm}\).  Rows show marching-wave and
 independent-column Jones propagation.  Columns show the two linear-analyzer
 intensities and the phases of \(E_x\), \(E_y\), and \(E_y/E_x\) for unit
 \(x\)-polarized input.  The tilt breaks rotational covariance about \(z\)
 and opens odd angular orders in both optical models.}
 \label{sm:fig:tilted-control}
\end{figure}

To distinguish propagation-selected symmetry within the covariant sector from
explicitly broken material axisymmetry, we use a positive control
(Fig.~\ref{sm:fig:tilted-control}) with both
substrates anchored along
\begin{equation}
 \vect n_{\rm s}=\frac{(0.03,0,1)}{\sqrt{1+0.03^2}},
 \qquad \beta_{\rm s}=\arctan(0.03)=1.718^\circ.
 \label{sm:eq:tilted-anchoring}
\end{equation}
No nontrivial rotation about \(z\) leaves this boundary condition invariant,
so Eq.~\eqref{sm:eq:jones-covariance} and the restriction to
\(m=0,\pm2\) no longer apply.  Between the first and last resolved planes
containing the reversed \(n_z<0\) core, its area centroid shifts transversely
by \(3.61\,\unitmu\mathrm m\) over an axial span of
\(21.92\,\unitmu\mathrm m\).  We fix the inclined state's structural origin
independently of the optical fields as the volume centroid of the transverse
director weight \(n_x^2+n_y^2\).  At this origin, the out-of-set
angular-order leakage amplitudes are \(27.2\%\) for independent-column Jones
propagation and \(27.7\%\) for marching-wave propagation.  The stronger
full covariance residuals are \(29.0\%\) and \(29.7\%\), respectively.  Of the
out-of-set angular-order leakage power, \(77.8\%\) and \(76.4\%\) lie in
the leading odd orders \(m=\pm1,\pm3\).  Optimizing a separate optical center
for each tilted field lowers the leakage amplitudes to \(22.3\%\) and \(23.7\%\);
these are reported only as lower bounds because the fit removes part of the
odd-order signal being tested.  Even a weak anchoring tilt therefore opens
orders that are absent from the ideal axisymmetric model.  By contrast, the
twofold crossed-polarizer response of the untilted toron survives projection
onto the covariant sector.

\begin{figure*}[p]
 \centering
 \includegraphics[width=\textwidth]{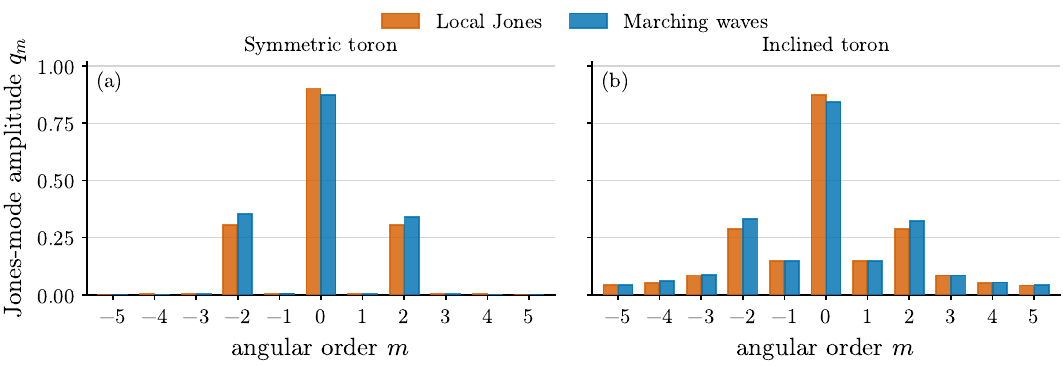}
 \caption{Amplitudes \(q_m\) of the signed angular orders \(m\) in the complete
 background-subtracted Jones response at \(520\,\mathrm{nm}\), normalized
 separately for each field according to
 Eq.~\eqref{sm:eq:jones-mode-amplitude}.  Only the liquid-crystal exit surface
 \(z=0\) is shown: (a) the symmetric toron and (b) the inclined toron.
 Orange and blue bars denote independent-column Jones and marching-wave
 propagation, respectively.  Both panels use a common linear scale,
 \(R=28\,\unitmu\mathrm m\), the fixed source-director structural origin, and no
 imaging pupil.  For the inclined state that origin is the volume centroid of
 \(n_x^2+n_y^2\), not a fit to either optical field.}
 \label{sm:fig:tilted-mode-power}
\end{figure*}

To resolve the residual by signed angular order,
Fig.~\ref{sm:fig:tilted-mode-power} compares the complete-matrix amplitudes
\(q_m\) from Eq.~\eqref{sm:eq:jones-mode-amplitude} for
\(-5\leq m\leq5\) within \(R=28\,\unitmu\mathrm m\).  The symmetric toron is
concentrated in \(m=0,\pm2\), whereas the inclined anchoring populates large
odd-order components in both optical models.  The plotted spectra use the
director-defined structural axis stored with each simulated field.  An
independent fit that minimizes odd optical orders is retained only as a
sensitivity bound: for the inclined state it removes a substantial part of
the signal being tested and therefore is not used for the primary bars.

The archived control uses the same cell geometry, material parameters, and
finite anchoring strength \(W_{\rm s}\) as the primary director; only the two
stored \texttt{SoftAxis} directions are tilted as in
Eq.~\eqref{sm:eq:tilted-anchoring}.  Its \(520\)-nm optical fields use the same
material indices and numerical tolerances as the primary comparison.  The
separately normalized spectra in Fig.~\ref{sm:fig:tilted-mode-power} compare
relative modal amplitudes rather than the absolute response amplitudes of the
two relaxed states.

\clearpage

\subsection{Numerical accuracy checks}
\label{sm:sec:convergence}

The director relaxation was continued to \(t_{\rm rel}=1000\,\mathrm{s}\).
For the cell thickness \(L=27.4\,\unitmu\mathrm m\), the rough elastic time scales
\(\gamma_1L^2/K_i\) span \(17\)--\(51\,\mathrm{s}\), so the run covers
approximately \(20\)--\(60\) such thickness-scale intervals.  This comparison
supports the chosen relaxation duration but is not a residual-based
convergence bound: the archived final state contains neither the relaxation
trajectory nor a terminal torque or energy-change metric.

Both primary finite-anchoring optical calculations use a transverse
interpolation factor of two.  To quantify phase-step sensitivity, we repeat
the primary \(520\)-nm marching-wave calculation with the maximum internal
phase increment reduced from \(0.025\) to \(0.0125\) rad.  This targeted
refinement does not constitute a multiparameter convergence bound.

For the same archived finite-anchoring director, this reduction decreases the
normalized amplitude of \(b\) by \(0.246\%\) and the integrated
crossed-intensity twofold amplitude by \(3.09\%\).  After one optimal global
complex scaling, the background-subtracted Jones fields differ by
\(0.677\%\) rms within \(\rho\leq28\,\unitmu\mathrm m\); each background is
measured over \(41\leq\rho\leq45\,\unitmu\mathrm m\).  These differences
characterize phase-step sensitivity for the primary field.

For the factor-two-decimated experimental field at its native holographic
plane, replacing the cubic complex-field interpolation in the covariant
projector with quintic interpolation changes \(\epsilon_{\rm cov}\) from
\(11.14048\%\) to \(11.14085\%\), the projected \(\mathcal V_2\) from
\(0.2188471\) to \(0.2188458\), and the projected
\(\mathcal V_4\) from \(0.4061427\) to \(0.4061413\).
There is no propagation in this check; its center, background, radial grid,
and estimators are fixed.

No numerical bounds are assigned to the neglected backward waves, the
one-way Maxwell approximation, director discretization, or the finite
transverse domain.

\subsection{Reproducibility map}
\label{sm:sec:reproducibility}

OpenAI Codex and ChatGPT (GPT-5.4 and GPT-6 Astra) assisted with
calculation, analysis, and figure-generation code under the authors'
direction.  The authors reviewed the generated code and verified the results.
Figures were plotted from measured and calculated data using the accompanying
scripts.

The publication repository \cite{fourfold_repository} contains the
calculation, analysis, and figure code, data manifests, and build wrappers.
The matching code and data release is archived on Zenodo.
Experimental inputs are distributed separately from calculated results and
caches.  The experimental-data artifact contains the 27 reconstructed
single-acquisition Jones fields (nine at each wavelength) and the five
direct-POM states at 12 and 42 ms together with their dark and reference
measurements.  The calculated-data artifact includes the two saved director
states, retaining their existing \texttt{data\_src/directors/} paths, the
derived products in \texttt{data/}, and the final figures in \texttt{figs/}.
Optical calculations start from the saved directors.  The camera holograms from which the calibrated
Jones fields were reconstructed lie outside this reproducibility boundary,
and no external
experimental-data directory is required by the published workflow.

The primary data-to-result chain is:
\begin{itemize}
 \item the 27 source Jones fields are the files \path{1.npz}--\path{9.npz}
 in each of the \path{450/}, \path{520/}, and \path{660/} subdirectories of
 \path{data_src/holography/}; the consistency panels are cropped directly
 from these fields using centers determined by the registration procedure;
 \item their registered complex averages are
 \path{data/experimental/results_toron25um_50X_averaged/}, in files
 \path{450nm_average.npz}, \path{520nm_average.npz}, and
 \path{660nm_average.npz};
 \item the measured pupil cutoff and three-wavelength angular-order checks are
 recorded in \path{data/analysis/averaged_pupil_spectrum/report.json} and
 \path{data/analysis/experimental_angular_modes/report.json};
 \item the two-exposure direct-POM inputs are in
 \path{data_src/direct_pom/520nm/}; HDR combination and axial matching produce
 \path{data/experimental/results_toron25um_50X_focus_hdr/} and
 \path{data/analysis/results_toron25um_50X_focus_recovery/};
 the registered \(I_{90}\) comparison is recorded in
 \path{data/analysis/results_toron25um_50X_focus_recovery_soft/focus_i90_comparison_report.json};
 \item the director used for the primary optical calculations is
 \path{data_src/directors/toron_27um_5cb_soft.npz};
 \item the inclined positive control starts from
 \path{data_src/directors/toron_27um_5cb_soft_inclined_0.03.npz};
 \item the six calculated Jones fields are those in
 Eq.~\eqref{sm:eq:production-paths};
 \item the half-phase-step \(520\)-nm check is
 \path{data/simulated/toron_27um_5cb_soft/520nm_marching-wave_phase-step-0p0125.npz};
 its three convergence metrics are recorded in
 \path{data/analysis/phase_step_convergence_520_soft/report.json};
 \item the four synthetic helicity/diffraction fields and their report are in
 \path{data/analysis/constant_helicity_diffraction_control_520_soft/};
 \item the TopoVec render cache for the experimental source fields and their
 coherent averages is in
 \path{data/analysis/results_toron25um_50X_averaging/topovec_panels/};
 \item the primary 520-nm harmonics and control metrics are recorded in
 \path{data/analysis/main_symmetry_520_soft/};
 \item the saved \(a,b,c,d\) profiles and homogeneous-propagation data are in
 \path{data/analysis/axisymmetric_jones_evolution_soft/};
 \item the covariant-residual decomposition is in
 \path{data/analysis/covariant_residual_520_soft/report.json};
 \item the director-helicity diagnostics are in
 \path{data/analysis/director_helicity_soft/director_helicity_analysis.json};
 \item center/background and interpolation checks are in
 \path{data/analysis/numerical_validation/};
 \item the tilted positive-control fields are in
 \path{data/simulated/toron_27um_5cb_soft_inclined_0.03/}, and their angular
 spectra and reversed-core geometry are recorded in
 \path{data/analysis/tilted_angular_modes_520_soft/}.
\end{itemize}
Every calculated-field NPZ file contains its physical pixel pitch together
with embedded provenance metadata.  Production simulations use a simulation
manifest, while the synthetic controls record equivalent metadata.
The average manifests list all nine source fields and centers and specify that
no global phase correction was fitted.  Simulation manifests contain the
director source, axis and component order, optical constants, numerical
parameters, precision, and model.  Analysis reports store the radial
intervals, pupil cutoffs, focus offsets, and principal numerical diagnostics
consumed by the publication figures.

The Python environment is reproduced from the root-level \texttt{uv.lock} with
\texttt{uv sync --locked}.  CPU analysis and figure generation use the
default environment; optical simulations use the additional
\texttt{simulation} dependencies.  All Python entry points are then run from the repository
root with \texttt{uv run}, without a separate project-selection option.  The
calculations reported here use the published \texttt{topovec} 0.1.17
package and \texttt{lcsim} 0.1.12 at revision
\texttt{4ba0759ac688}.  Data unpacking,
environment setup, every experimental reduction, optical calculation,
analysis, and figure command are given in \texttt{REPRODUCTION.md}.  Numerical
stages populate \texttt{data/}; the intentionally separate
\texttt{make -f reproduce.mk figures} target then renders the publication PDFs from those
derived products without rerunning the expensive calculations.  This
two-stage boundary allows figures to be rebuilt repeatedly while preserving a
complete path from \texttt{data\_src/} to every reported result.

%% file: paper.bbl
%
%
%
%
%
%
%
%
\begin{thebibliography}{33}%
\makeatletter
\providecommand \@ifxundefined [1]{%
 \@ifx{#1\undefined}
}%
\providecommand \@ifnum [1]{%
 \ifnum #1\expandafter \@firstoftwo
 \else \expandafter \@secondoftwo
 \fi
}%
\providecommand \@ifx [1]{%
 \ifx #1\expandafter \@firstoftwo
 \else \expandafter \@secondoftwo
 \fi
}%
\providecommand \natexlab [1]{#1}%
\providecommand \enquote  [1]{``#1''}%
\providecommand \bibnamefont  [1]{#1}%
\providecommand \bibfnamefont [1]{#1}%
\providecommand \citenamefont [1]{#1}%
\providecommand \href@noop [0]{\@secondoftwo}%
\providecommand \href [0]{\begingroup \@sanitize@url \@href}%
\providecommand \@href[1]{\@@startlink{#1}\@@href}%
\providecommand \@@href[1]{\endgroup#1\@@endlink}%
\providecommand \@sanitize@url [0]{\catcode `\\12\catcode `\$12\catcode
  `\&12\catcode `\#12\catcode `\^12\catcode `\_12\catcode `\%12\relax}%
\providecommand \@@startlink[1]{}%
\providecommand \@@endlink[0]{}%
\providecommand \url  [0]{\begingroup\@sanitize@url \@url }%
\providecommand \@url [1]{\endgroup\@href {#1}{\urlprefix }}%
\providecommand \urlprefix  [0]{URL }%
\providecommand \Eprint [0]{\href }%
\providecommand \doibase [0]{https://doi.org/}%
\providecommand \selectlanguage [0]{\@gobble}%
\providecommand \bibinfo  [0]{\@secondoftwo}%
\providecommand \bibfield  [0]{\@secondoftwo}%
\providecommand \translation [1]{[#1]}%
\providecommand \BibitemOpen [0]{}%
\providecommand \bibitemStop [0]{}%
\providecommand \bibitemNoStop [0]{.\EOS\space}%
\providecommand \EOS [0]{\spacefactor3000\relax}%
\providecommand \BibitemShut  [1]{\csname bibitem#1\endcsname}%
\let\auto@bib@innerbib\@empty
%
\bibitem [{\citenamefont {Smalyukh}\ \emph {et~al.}(2010)\citenamefont
  {Smalyukh}, \citenamefont {Lansac}, \citenamefont {Clark},\ and\
  \citenamefont {Trivedi}}]{smalyukh2010torons}%
  \BibitemOpen
  \bibfield  {author} {\bibinfo {author} {\bibfnamefont {I.~I.}\ \bibnamefont
  {Smalyukh}}, \bibinfo {author} {\bibfnamefont {Y.}~\bibnamefont {Lansac}},
  \bibinfo {author} {\bibfnamefont {N.~A.}\ \bibnamefont {Clark}},\ and\
  \bibinfo {author} {\bibfnamefont {R.~P.}\ \bibnamefont {Trivedi}},\
  }\bibfield  {title} {\bibinfo {title} {Three-dimensional structure and
  multistable optical switching of triple-twisted particle-like excitations in
  anisotropic fluids},\ }\href {https://doi.org/10.1038/nmat2592} {\bibfield
  {journal} {\bibinfo  {journal} {Nat. Mater.}\ }\textbf {\bibinfo {volume}
  {9}},\ \bibinfo {pages} {139} (\bibinfo {year} {2010})}\BibitemShut {NoStop}%
\bibitem [{\citenamefont {Ackerman}\ and\ \citenamefont
  {Smalyukh}(2017)}]{ackerman_smalyukh2017knot}%
  \BibitemOpen
  \bibfield  {author} {\bibinfo {author} {\bibfnamefont {P.~J.}\ \bibnamefont
  {Ackerman}}\ and\ \bibinfo {author} {\bibfnamefont {I.~I.}\ \bibnamefont
  {Smalyukh}},\ }\bibfield  {title} {\bibinfo {title} {Diversity of knot
  solitons in liquid crystals manifested by linking of preimages in torons and
  hopfions},\ }\href {https://doi.org/10.1103/PhysRevX.7.011006} {\bibfield
  {journal} {\bibinfo  {journal} {Phys. Rev. X}\ }\textbf {\bibinfo {volume}
  {7}},\ \bibinfo {pages} {011006} (\bibinfo {year} {2017})}\BibitemShut
  {NoStop}%
\bibitem [{\citenamefont {Ackerman}\ \emph {et~al.}(2017)\citenamefont
  {Ackerman}, \citenamefont {Boyle},\ and\ \citenamefont
  {Smalyukh}}]{ackerman_boyle_smalyukh2017squirming}%
  \BibitemOpen
  \bibfield  {author} {\bibinfo {author} {\bibfnamefont {P.~J.}\ \bibnamefont
  {Ackerman}}, \bibinfo {author} {\bibfnamefont {T.}~\bibnamefont {Boyle}},\
  and\ \bibinfo {author} {\bibfnamefont {I.~I.}\ \bibnamefont {Smalyukh}},\
  }\bibfield  {title} {\bibinfo {title} {Squirming motion of baby skyrmions in
  nematic fluids},\ }\href {https://doi.org/10.1038/s41467-017-00659-5}
  {\bibfield  {journal} {\bibinfo  {journal} {Nat. Commun.}\ }\textbf {\bibinfo
  {volume} {8}},\ \bibinfo {pages} {673} (\bibinfo {year} {2017})}\BibitemShut
  {NoStop}%
\bibitem [{\citenamefont {Lobanov}\ \emph {et~al.}(2022)\citenamefont
  {Lobanov}, \citenamefont {Aksenova}, \citenamefont {Orlova}, \citenamefont
  {Darmoroz}, \citenamefont {Uzdin},\ and\ \citenamefont
  {Kiselev}}]{lobanov2022optical_imaging}%
  \BibitemOpen
  \bibfield  {author} {\bibinfo {author} {\bibfnamefont {I.}~\bibnamefont
  {Lobanov}}, \bibinfo {author} {\bibfnamefont {E.}~\bibnamefont {Aksenova}},
  \bibinfo {author} {\bibfnamefont {T.}~\bibnamefont {Orlova}}, \bibinfo
  {author} {\bibfnamefont {D.}~\bibnamefont {Darmoroz}}, \bibinfo {author}
  {\bibfnamefont {V.}~\bibnamefont {Uzdin}},\ and\ \bibinfo {author}
  {\bibfnamefont {A.~D.}\ \bibnamefont {Kiselev}},\ }\bibfield  {title}
  {\bibinfo {title} {Optical imaging and analytical design of localized
  topological structures in chiral liquid crystals},\ }\href
  {https://doi.org/10.3390/sym14122476} {\bibfield  {journal} {\bibinfo
  {journal} {Symmetry}\ }\textbf {\bibinfo {volume} {14}},\ \bibinfo {pages}
  {2476} (\bibinfo {year} {2022})}\BibitemShut {NoStop}%
\bibitem [{\citenamefont {Varanytsia}\ \emph {et~al.}(2017)\citenamefont
  {Varanytsia}, \citenamefont {Posnjak}, \citenamefont {Mur}, \citenamefont
  {Joshi}, \citenamefont {Darrah}, \citenamefont {Mu{\v{s}}evi{\v{c}}},
  \citenamefont {{\v{C}}opar},\ and\ \citenamefont
  {Chien}}]{varanytsia2017topology}%
  \BibitemOpen
  \bibfield  {author} {\bibinfo {author} {\bibfnamefont {A.}~\bibnamefont
  {Varanytsia}}, \bibinfo {author} {\bibfnamefont {G.}~\bibnamefont {Posnjak}},
  \bibinfo {author} {\bibfnamefont {U.}~\bibnamefont {Mur}}, \bibinfo {author}
  {\bibfnamefont {V.}~\bibnamefont {Joshi}}, \bibinfo {author} {\bibfnamefont
  {K.}~\bibnamefont {Darrah}}, \bibinfo {author} {\bibfnamefont
  {I.}~\bibnamefont {Mu{\v{s}}evi{\v{c}}}}, \bibinfo {author} {\bibfnamefont
  {S.}~\bibnamefont {{\v{C}}opar}},\ and\ \bibinfo {author} {\bibfnamefont
  {L.-C.}\ \bibnamefont {Chien}},\ }\bibfield  {title} {\bibinfo {title}
  {Topology-commanded optical properties of bistable electric-field-induced
  torons in cholesteric bubble domains},\ }\href
  {https://doi.org/10.1038/s41598-017-16241-4} {\bibfield  {journal} {\bibinfo
  {journal} {Sci. Rep.}\ }\textbf {\bibinfo {volume} {7}},\ \bibinfo {pages}
  {16149} (\bibinfo {year} {2017})}\BibitemShut {NoStop}%
\bibitem [{\citenamefont {Marrucci}\ \emph {et~al.}(2006)\citenamefont
  {Marrucci}, \citenamefont {Manzo},\ and\ \citenamefont
  {Paparo}}]{marrucci2006}%
  \BibitemOpen
  \bibfield  {author} {\bibinfo {author} {\bibfnamefont {L.}~\bibnamefont
  {Marrucci}}, \bibinfo {author} {\bibfnamefont {C.}~\bibnamefont {Manzo}},\
  and\ \bibinfo {author} {\bibfnamefont {D.}~\bibnamefont {Paparo}},\
  }\bibfield  {title} {\bibinfo {title} {Optical spin-to-orbital angular
  momentum conversion in inhomogeneous anisotropic media},\ }\href
  {https://doi.org/10.1103/PhysRevLett.96.163905} {\bibfield  {journal}
  {\bibinfo  {journal} {Phys. Rev. Lett.}\ }\textbf {\bibinfo {volume} {96}},\
  \bibinfo {pages} {163905} (\bibinfo {year} {2006})}\BibitemShut {NoStop}%
\bibitem [{\citenamefont {Rubano}\ \emph {et~al.}(2019)\citenamefont {Rubano},
  \citenamefont {Cardano}, \citenamefont {Piccirillo},\ and\ \citenamefont
  {Marrucci}}]{rubano2019qplate_review}%
  \BibitemOpen
  \bibfield  {author} {\bibinfo {author} {\bibfnamefont {A.}~\bibnamefont
  {Rubano}}, \bibinfo {author} {\bibfnamefont {F.}~\bibnamefont {Cardano}},
  \bibinfo {author} {\bibfnamefont {B.}~\bibnamefont {Piccirillo}},\ and\
  \bibinfo {author} {\bibfnamefont {L.}~\bibnamefont {Marrucci}},\ }\bibfield
  {title} {\bibinfo {title} {Q-plate technology: a progress review},\ }\href
  {https://doi.org/10.1364/JOSAB.36.000D70} {\bibfield  {journal} {\bibinfo
  {journal} {J. Opt. Soc. Am. B}\ }\textbf {\bibinfo {volume} {36}},\ \bibinfo
  {pages} {D70} (\bibinfo {year} {2019})}\BibitemShut {NoStop}%
\bibitem [{\citenamefont {Karimi}\ \emph {et~al.}(2009)\citenamefont {Karimi},
  \citenamefont {Piccirillo}, \citenamefont {Marrucci},\ and\ \citenamefont
  {Santamato}}]{karimi2009qplate_propagation}%
  \BibitemOpen
  \bibfield  {author} {\bibinfo {author} {\bibfnamefont {E.}~\bibnamefont
  {Karimi}}, \bibinfo {author} {\bibfnamefont {B.}~\bibnamefont {Piccirillo}},
  \bibinfo {author} {\bibfnamefont {L.}~\bibnamefont {Marrucci}},\ and\
  \bibinfo {author} {\bibfnamefont {E.}~\bibnamefont {Santamato}},\ }\bibfield
  {title} {\bibinfo {title} {Light propagation in a birefringent plate with
  topological charge},\ }\href {https://doi.org/10.1364/OL.34.001225}
  {\bibfield  {journal} {\bibinfo  {journal} {Opt. Lett.}\ }\textbf {\bibinfo
  {volume} {34}},\ \bibinfo {pages} {1225} (\bibinfo {year}
  {2009})}\BibitemShut {NoStop}%
\bibitem [{\citenamefont {Jisha}\ \emph {et~al.}(2017)\citenamefont {Jisha},
  \citenamefont {Alberucci}, \citenamefont {Marrucci},\ and\ \citenamefont
  {Assanto}}]{jisha2017diffraction_pb}%
  \BibitemOpen
  \bibfield  {author} {\bibinfo {author} {\bibfnamefont {C.~P.}\ \bibnamefont
  {Jisha}}, \bibinfo {author} {\bibfnamefont {A.}~\bibnamefont {Alberucci}},
  \bibinfo {author} {\bibfnamefont {L.}~\bibnamefont {Marrucci}},\ and\
  \bibinfo {author} {\bibfnamefont {G.}~\bibnamefont {Assanto}},\ }\bibfield
  {title} {\bibinfo {title} {Interplay between diffraction and the
  {Pancharatnam--Berry} phase in inhomogeneously twisted anisotropic media},\
  }\href {https://doi.org/10.1103/PhysRevA.95.023823} {\bibfield  {journal}
  {\bibinfo  {journal} {Phys. Rev. A}\ }\textbf {\bibinfo {volume} {95}},\
  \bibinfo {pages} {023823} (\bibinfo {year} {2017})}\BibitemShut {NoStop}%
\bibitem [{\citenamefont {Inou{\'e}}\ and\ \citenamefont
  {Kubota}(1958)}]{inoue_kubota1958diffraction_anomaly}%
  \BibitemOpen
  \bibfield  {author} {\bibinfo {author} {\bibfnamefont {S.}~\bibnamefont
  {Inou{\'e}}}\ and\ \bibinfo {author} {\bibfnamefont {H.}~\bibnamefont
  {Kubota}},\ }\bibfield  {title} {\bibinfo {title} {Diffraction anomaly in
  polarizing microscopes},\ }\href {https://doi.org/10.1038/1821725a0}
  {\bibfield  {journal} {\bibinfo  {journal} {Nature}\ }\textbf {\bibinfo
  {volume} {182}},\ \bibinfo {pages} {1725} (\bibinfo {year}
  {1958})}\BibitemShut {NoStop}%
\bibitem [{\citenamefont {Arbabi}\ \emph {et~al.}(2015)\citenamefont {Arbabi},
  \citenamefont {Horie}, \citenamefont {Bagheri},\ and\ \citenamefont
  {Faraon}}]{arbabi2015dielectric_metasurfaces}%
  \BibitemOpen
  \bibfield  {author} {\bibinfo {author} {\bibfnamefont {A.}~\bibnamefont
  {Arbabi}}, \bibinfo {author} {\bibfnamefont {Y.}~\bibnamefont {Horie}},
  \bibinfo {author} {\bibfnamefont {M.}~\bibnamefont {Bagheri}},\ and\ \bibinfo
  {author} {\bibfnamefont {A.}~\bibnamefont {Faraon}},\ }\bibfield  {title}
  {\bibinfo {title} {Dielectric metasurfaces for complete control of phase and
  polarization with subwavelength spatial resolution and high transmission},\
  }\href {https://doi.org/10.1038/nnano.2015.186} {\bibfield  {journal}
  {\bibinfo  {journal} {Nat. Nanotechnol.}\ }\textbf {\bibinfo {volume} {10}},\
  \bibinfo {pages} {937} (\bibinfo {year} {2015})}\BibitemShut {NoStop}%
\bibitem [{\citenamefont {Devlin}\ \emph {et~al.}(2017)\citenamefont {Devlin},
  \citenamefont {Ambrosio}, \citenamefont {Rubin}, \citenamefont {Mueller},\
  and\ \citenamefont {Capasso}}]{devlin2017arbitrary_oam}%
  \BibitemOpen
  \bibfield  {author} {\bibinfo {author} {\bibfnamefont {R.~C.}\ \bibnamefont
  {Devlin}}, \bibinfo {author} {\bibfnamefont {A.}~\bibnamefont {Ambrosio}},
  \bibinfo {author} {\bibfnamefont {N.~A.}\ \bibnamefont {Rubin}}, \bibinfo
  {author} {\bibfnamefont {J.~P.~B.}\ \bibnamefont {Mueller}},\ and\ \bibinfo
  {author} {\bibfnamefont {F.}~\bibnamefont {Capasso}},\ }\bibfield  {title}
  {\bibinfo {title} {Arbitrary spin-to-orbital angular momentum conversion of
  light},\ }\href {https://doi.org/10.1126/science.aao5392} {\bibfield
  {journal} {\bibinfo  {journal} {Science}\ }\textbf {\bibinfo {volume}
  {358}},\ \bibinfo {pages} {896} (\bibinfo {year} {2017})}\BibitemShut
  {NoStop}%
\bibitem [{\citenamefont {Kwon}\ \emph {et~al.}(2018)\citenamefont {Kwon},
  \citenamefont {Sounas}, \citenamefont {Cordaro}, \citenamefont {Polman},\
  and\ \citenamefont {Al{\`u}}}]{kwon2018nonlocal_metasurfaces}%
  \BibitemOpen
  \bibfield  {author} {\bibinfo {author} {\bibfnamefont {H.}~\bibnamefont
  {Kwon}}, \bibinfo {author} {\bibfnamefont {D.}~\bibnamefont {Sounas}},
  \bibinfo {author} {\bibfnamefont {A.}~\bibnamefont {Cordaro}}, \bibinfo
  {author} {\bibfnamefont {A.}~\bibnamefont {Polman}},\ and\ \bibinfo {author}
  {\bibfnamefont {A.}~\bibnamefont {Al{\`u}}},\ }\bibfield  {title} {\bibinfo
  {title} {Nonlocal metasurfaces for optical signal processing},\ }\href
  {https://doi.org/10.1103/PhysRevLett.121.173004} {\bibfield  {journal}
  {\bibinfo  {journal} {Phys. Rev. Lett.}\ }\textbf {\bibinfo {volume} {121}},\
  \bibinfo {pages} {173004} (\bibinfo {year} {2018})}\BibitemShut {NoStop}%
\bibitem [{\citenamefont {Song}\ \emph {et~al.}(2021)\citenamefont {Song},
  \citenamefont {van~de Groep}, \citenamefont {Kim},\ and\ \citenamefont
  {Brongersma}}]{song2021nonlocal_metasurfaces}%
  \BibitemOpen
  \bibfield  {author} {\bibinfo {author} {\bibfnamefont {J.-H.}\ \bibnamefont
  {Song}}, \bibinfo {author} {\bibfnamefont {J.}~\bibnamefont {van~de Groep}},
  \bibinfo {author} {\bibfnamefont {S.~J.}\ \bibnamefont {Kim}},\ and\ \bibinfo
  {author} {\bibfnamefont {M.~L.}\ \bibnamefont {Brongersma}},\ }\bibfield
  {title} {\bibinfo {title} {Non-local metasurfaces for spectrally decoupled
  wavefront manipulation and eye tracking},\ }\href
  {https://doi.org/10.1038/s41565-021-00967-4} {\bibfield  {journal} {\bibinfo
  {journal} {Nat. Nanotechnol.}\ }\textbf {\bibinfo {volume} {16}},\ \bibinfo
  {pages} {1224} (\bibinfo {year} {2021})}\BibitemShut {NoStop}%
\bibitem [{\citenamefont {Fukuda}\ and\ \citenamefont {{\v
  Z}umer}(2011)}]{fukuda_zumer2011skyrmions}%
  \BibitemOpen
  \bibfield  {author} {\bibinfo {author} {\bibfnamefont {J.-i.}\ \bibnamefont
  {Fukuda}}\ and\ \bibinfo {author} {\bibfnamefont {S.}~\bibnamefont {{\v
  Z}umer}},\ }\bibfield  {title} {\bibinfo {title} {Quasi-two-dimensional
  skyrmion lattices in a chiral nematic liquid crystal},\ }\href
  {https://doi.org/10.1038/ncomms1250} {\bibfield  {journal} {\bibinfo
  {journal} {Nat. Commun.}\ }\textbf {\bibinfo {volume} {2}},\ \bibinfo {pages}
  {246} (\bibinfo {year} {2011})}\BibitemShut {NoStop}%
\bibitem [{\citenamefont {Tai}\ and\ \citenamefont
  {Smalyukh}(2020)}]{tai2020surface_anchoring}%
  \BibitemOpen
  \bibfield  {author} {\bibinfo {author} {\bibfnamefont {J.-S.~B.}\
  \bibnamefont {Tai}}\ and\ \bibinfo {author} {\bibfnamefont {I.~I.}\
  \bibnamefont {Smalyukh}},\ }\bibfield  {title} {\bibinfo {title} {Surface
  anchoring as a control parameter for stabilizing torons, skyrmions, twisted
  walls, fingers, and their hybrids in chiral nematics},\ }\href
  {https://doi.org/10.1103/PhysRevE.101.042702} {\bibfield  {journal} {\bibinfo
   {journal} {Phys. Rev. E}\ }\textbf {\bibinfo {volume} {101}},\ \bibinfo
  {pages} {042702} (\bibinfo {year} {2020})}\BibitemShut {NoStop}%
\bibitem [{sup()}]{supplemental_material}%
  \BibitemOpen
  \href@noop {} {}\bibinfo {note} {See Supplemental Material included in this
  preprint (p.~\pageref{arxiv:supplement}) for experimental and numerical
  methods, complete symmetry derivations, definitions of the reported
  observables, convergence tests, and additional wavelength and control data,
  which includes
  Refs.~\cite{berreman1972,yeh1980,yeh_gu1999,feit_fleck1978,van_roey1981,petelin2023pom_bpm_preprint,armitage2014jones_symmetry,bogdanov2003nematic_skyrmions,hess2020control_light}}\BibitemShut
  {NoStop}%
\bibitem [{\citenamefont {Reztsov}\ \emph {et~al.}(2025)\citenamefont
  {Reztsov}, \citenamefont {Chernykh}, \citenamefont {Orlova},\ and\
  \citenamefont {Petrov}}]{reztsov2025toron_phm}%
  \BibitemOpen
  \bibfield  {author} {\bibinfo {author} {\bibfnamefont {T.~V.}\ \bibnamefont
  {Reztsov}}, \bibinfo {author} {\bibfnamefont {A.~V.}\ \bibnamefont
  {Chernykh}}, \bibinfo {author} {\bibfnamefont {T.}~\bibnamefont {Orlova}},\
  and\ \bibinfo {author} {\bibfnamefont {N.}~\bibnamefont {Petrov}},\
  }\bibfield  {title} {\bibinfo {title} {A dynamic analysis of toron formation
  in chiral nematic liquid crystals using a polarization holographic
  microscope},\ }\href {https://doi.org/10.3390/polym17131849} {\bibfield
  {journal} {\bibinfo  {journal} {Polymers}\ }\textbf {\bibinfo {volume}
  {17}},\ \bibinfo {pages} {1849} (\bibinfo {year} {2025})}\BibitemShut
  {NoStop}%
\bibitem [{\citenamefont {Poy}\ and\ \citenamefont {{\v
  Z}umer}(2020)}]{poy_zumer2020bpm}%
  \BibitemOpen
  \bibfield  {author} {\bibinfo {author} {\bibfnamefont {G.}~\bibnamefont
  {Poy}}\ and\ \bibinfo {author} {\bibfnamefont {S.}~\bibnamefont {{\v
  Z}umer}},\ }\bibfield  {title} {\bibinfo {title} {Physics-based multistep
  beam propagation in inhomogeneous birefringent media},\ }\href
  {https://doi.org/10.1364/OE.400984} {\bibfield  {journal} {\bibinfo
  {journal} {Opt. Express}\ }\textbf {\bibinfo {volume} {28}},\ \bibinfo
  {pages} {24327} (\bibinfo {year} {2020})}\BibitemShut {NoStop}%
\bibitem [{\citenamefont {Shalev}\ \emph {et~al.}(2025)\citenamefont {Shalev},
  \citenamefont {Misura}, \citenamefont {Georgieva}, \citenamefont {Chernykh},
  \citenamefont {Petrov}, \citenamefont {Orlova}, \citenamefont {Lobanov},
  \citenamefont {Aksenova}, \citenamefont {Uzdin},\ and\ \citenamefont
  {Kiselev}}]{shalev2025oam_spherulite}%
  \BibitemOpen
  \bibfield  {author} {\bibinfo {author} {\bibfnamefont {A.~N.}\ \bibnamefont
  {Shalev}}, \bibinfo {author} {\bibfnamefont {A.~A.}\ \bibnamefont {Misura}},
  \bibinfo {author} {\bibfnamefont {A.~O.}\ \bibnamefont {Georgieva}}, \bibinfo
  {author} {\bibfnamefont {A.~V.}\ \bibnamefont {Chernykh}}, \bibinfo {author}
  {\bibfnamefont {N.~V.}\ \bibnamefont {Petrov}}, \bibinfo {author}
  {\bibfnamefont {T.}~\bibnamefont {Orlova}}, \bibinfo {author} {\bibfnamefont
  {I.~S.}\ \bibnamefont {Lobanov}}, \bibinfo {author} {\bibfnamefont {E.~V.}\
  \bibnamefont {Aksenova}}, \bibinfo {author} {\bibfnamefont {V.~M.}\
  \bibnamefont {Uzdin}},\ and\ \bibinfo {author} {\bibfnamefont {A.~D.}\
  \bibnamefont {Kiselev}},\ }\bibfield  {title} {\bibinfo {title}
  {Polarization-controlled orbital angular momentum of light passing through a
  cholesteric spherulite},\ }\href {https://doi.org/10.1364/OL.561669}
  {\bibfield  {journal} {\bibinfo  {journal} {Opt. Lett.}\ }\textbf {\bibinfo
  {volume} {50}},\ \bibinfo {pages} {4866} (\bibinfo {year}
  {2025})}\BibitemShut {NoStop}%
\bibitem [{\citenamefont {Lobanov}\ \emph {et~al.}(2026)\citenamefont
  {Lobanov}, \citenamefont {Reztsov}, \citenamefont {Shalev}, \citenamefont
  {Orlova}, \citenamefont {Aksenova}, \citenamefont {Uzdin}, \citenamefont
  {Petrov},\ and\ \citenamefont {Chernykh}}]{fourfold_repository}%
  \BibitemOpen
  \bibfield  {author} {\bibinfo {author} {\bibfnamefont {I.~S.}\ \bibnamefont
  {Lobanov}}, \bibinfo {author} {\bibfnamefont {T.~V.}\ \bibnamefont
  {Reztsov}}, \bibinfo {author} {\bibfnamefont {A.~N.}\ \bibnamefont {Shalev}},
  \bibinfo {author} {\bibfnamefont {T.}~\bibnamefont {Orlova}}, \bibinfo
  {author} {\bibfnamefont {E.~V.}\ \bibnamefont {Aksenova}}, \bibinfo {author}
  {\bibfnamefont {V.~M.}\ \bibnamefont {Uzdin}}, \bibinfo {author}
  {\bibfnamefont {N.~V.}\ \bibnamefont {Petrov}},\ and\ \bibinfo {author}
  {\bibfnamefont {A.~V.}\ \bibnamefont {Chernykh}},\ }\href
  {https://doi.org/10.5281/zenodo.24323744} {\bibinfo {title}
  {{symmetry-in-pom}: source code and data for emergent fourfold symmetry in
  polarization images}},\ \bibinfo {howpublished}
  {\url{https://github.com/alepoydes/symmetry-in-pom}} (\bibinfo {year}
  {2026}),\ \bibinfo {note} {version 0.1.0}\BibitemShut {NoStop}%
\bibitem [{\citenamefont {Lobanov}(2026{\natexlab{a}})}]{lcsim_git}%
  \BibitemOpen
  \bibfield  {author} {\bibinfo {author} {\bibfnamefont {I.~S.}\ \bibnamefont
  {Lobanov}},\ }\href@noop {} {\bibinfo {title} {{lcsim}: liquid-crystal
  director-field simulation code}},\ \bibinfo {howpublished}
  {\url{https://gitlab.com/alepoydes/lcsim}} (\bibinfo {year}
  {2026}{\natexlab{a}}),\ \bibinfo {note} {git repository, commit
  4ba0759ac6889d903309f559d7a15201a0cafc21; accessed 2026-07-24}\BibitemShut
  {NoStop}%
\bibitem [{\citenamefont {Li}\ \emph {et~al.}(2005)\citenamefont {Li},
  \citenamefont {Wen}, \citenamefont {Gauza}, \citenamefont {Lu},\ and\
  \citenamefont {Wu}}]{li2005refractive}%
  \BibitemOpen
  \bibfield  {author} {\bibinfo {author} {\bibfnamefont {J.}~\bibnamefont
  {Li}}, \bibinfo {author} {\bibfnamefont {C.-H.}\ \bibnamefont {Wen}},
  \bibinfo {author} {\bibfnamefont {S.}~\bibnamefont {Gauza}}, \bibinfo
  {author} {\bibfnamefont {R.}~\bibnamefont {Lu}},\ and\ \bibinfo {author}
  {\bibfnamefont {S.-T.}\ \bibnamefont {Wu}},\ }\bibfield  {title} {\bibinfo
  {title} {Refractive indices of liquid crystals for display applications},\
  }\href {https://doi.org/10.1109/JDT.2005.853357} {\bibfield  {journal}
  {\bibinfo  {journal} {J. Disp. Technol.}\ }\textbf {\bibinfo {volume} {1}},\
  \bibinfo {pages} {51} (\bibinfo {year} {2005})}\BibitemShut {NoStop}%
\bibitem [{\citenamefont {Lobanov}(2026{\natexlab{b}})}]{topovec_git}%
  \BibitemOpen
  \bibfield  {author} {\bibinfo {author} {\bibfnamefont {I.~S.}\ \bibnamefont
  {Lobanov}},\ }\href@noop {} {\bibinfo {title} {{topovec}: director-field and
  optical rendering tools for liquid-crystal topological textures}},\ \bibinfo
  {howpublished} {\url{https://pypi.org/project/topovec/0.1.17/}} (\bibinfo
  {year} {2026}{\natexlab{b}}),\ \bibinfo {note} {published package, version
  0.1.17; accessed 2026-07-26}\BibitemShut {NoStop}%
\bibitem [{\citenamefont {Berreman}(1972)}]{berreman1972}%
  \BibitemOpen
  \bibfield  {author} {\bibinfo {author} {\bibfnamefont {D.~W.}\ \bibnamefont
  {Berreman}},\ }\bibfield  {title} {\bibinfo {title} {Optics in stratified and
  anisotropic media: 4 by 4-matrix formulation},\ }\href
  {https://doi.org/10.1364/JOSA.62.000502} {\bibfield  {journal} {\bibinfo
  {journal} {J. Opt. Soc. Am.}\ }\textbf {\bibinfo {volume} {62}},\ \bibinfo
  {pages} {502} (\bibinfo {year} {1972})}\BibitemShut {NoStop}%
\bibitem [{\citenamefont {Yeh}(1980)}]{yeh1980}%
  \BibitemOpen
  \bibfield  {author} {\bibinfo {author} {\bibfnamefont {P.}~\bibnamefont
  {Yeh}},\ }\bibfield  {title} {\bibinfo {title} {Optics of anisotropic layered
  media: A new 4 by 4 matrix algebra},\ }\href@noop {} {\bibfield  {journal}
  {\bibinfo  {journal} {Surf. Sci.}\ }\textbf {\bibinfo {volume} {96}},\
  \bibinfo {pages} {41} (\bibinfo {year} {1980})}\BibitemShut {NoStop}%
\bibitem [{\citenamefont {Yeh}\ and\ \citenamefont {Gu}(1999)}]{yeh_gu1999}%
  \BibitemOpen
  \bibfield  {author} {\bibinfo {author} {\bibfnamefont {P.}~\bibnamefont
  {Yeh}}\ and\ \bibinfo {author} {\bibfnamefont {C.}~\bibnamefont {Gu}},\
  }\href@noop {} {\emph {\bibinfo {title} {Optics of Liquid Crystal
  Displays}}}\ (\bibinfo  {publisher} {Wiley},\ \bibinfo {address} {New York},\
  \bibinfo {year} {1999})\BibitemShut {NoStop}%
\bibitem [{\citenamefont {Feit}\ and\ \citenamefont
  {Fleck}(1978)}]{feit_fleck1978}%
  \BibitemOpen
  \bibfield  {author} {\bibinfo {author} {\bibfnamefont {M.~D.}\ \bibnamefont
  {Feit}}\ and\ \bibinfo {author} {\bibfnamefont {J.~A.}\ \bibnamefont
  {Fleck}},\ }\bibfield  {title} {\bibinfo {title} {Light propagation in
  graded-index optical fibers},\ }\href {https://doi.org/10.1364/AO.17.003990}
  {\bibfield  {journal} {\bibinfo  {journal} {Appl. Opt.}\ }\textbf {\bibinfo
  {volume} {17}},\ \bibinfo {pages} {3990} (\bibinfo {year}
  {1978})}\BibitemShut {NoStop}%
\bibitem [{\citenamefont {Van~Roey}\ \emph {et~al.}(1981)\citenamefont
  {Van~Roey}, \citenamefont {van~der Donk},\ and\ \citenamefont
  {Lagasse}}]{van_roey1981}%
  \BibitemOpen
  \bibfield  {author} {\bibinfo {author} {\bibfnamefont {J.}~\bibnamefont
  {Van~Roey}}, \bibinfo {author} {\bibfnamefont {J.}~\bibnamefont {van~der
  Donk}},\ and\ \bibinfo {author} {\bibfnamefont {P.~E.}\ \bibnamefont
  {Lagasse}},\ }\bibfield  {title} {\bibinfo {title} {Beam-propagation method:
  analysis and assessment},\ }\href {https://doi.org/10.1364/JOSA.71.000803}
  {\bibfield  {journal} {\bibinfo  {journal} {J. Opt. Soc. Am.}\ }\textbf
  {\bibinfo {volume} {71}},\ \bibinfo {pages} {803} (\bibinfo {year}
  {1981})}\BibitemShut {NoStop}%
\bibitem [{\citenamefont {Petelin}\ \emph {et~al.}(2023)\citenamefont
  {Petelin}, \citenamefont {Nys},\ and\ \citenamefont
  {Neyts}}]{petelin2023pom_bpm_preprint}%
  \BibitemOpen
  \bibfield  {author} {\bibinfo {author} {\bibfnamefont {A.}~\bibnamefont
  {Petelin}}, \bibinfo {author} {\bibfnamefont {I.}~\bibnamefont {Nys}},\ and\
  \bibinfo {author} {\bibfnamefont {K.}~\bibnamefont {Neyts}},\ }\href
  {https://doi.org/10.1364/opticaopen.22434013.v1} {\bibinfo {title}
  {Polarization microscope simulations using an efficient transfer matrix-based
  beam propagation method}},\ \bibinfo {howpublished} {Optica Open preprint}
  (\bibinfo {year} {2023})\BibitemShut {NoStop}%
\bibitem [{\citenamefont {Armitage}(2014)}]{armitage2014jones_symmetry}%
  \BibitemOpen
  \bibfield  {author} {\bibinfo {author} {\bibfnamefont {N.~P.}\ \bibnamefont
  {Armitage}},\ }\bibfield  {title} {\bibinfo {title} {Constraints on {Jones}
  transmission matrices from time-reversal invariance and discrete spatial
  symmetries},\ }\href {https://doi.org/10.1103/PhysRevB.90.035135} {\bibfield
  {journal} {\bibinfo  {journal} {Phys. Rev. B}\ }\textbf {\bibinfo {volume}
  {90}},\ \bibinfo {pages} {035135} (\bibinfo {year} {2014})}\BibitemShut
  {NoStop}%
\bibitem [{\citenamefont {Bogdanov}\ \emph {et~al.}(2003)\citenamefont
  {Bogdanov}, \citenamefont {R{\"o}{\ss}ler},\ and\ \citenamefont
  {Shestakov}}]{bogdanov2003nematic_skyrmions}%
  \BibitemOpen
  \bibfield  {author} {\bibinfo {author} {\bibfnamefont {A.~N.}\ \bibnamefont
  {Bogdanov}}, \bibinfo {author} {\bibfnamefont {U.~K.}\ \bibnamefont
  {R{\"o}{\ss}ler}},\ and\ \bibinfo {author} {\bibfnamefont {A.~A.}\
  \bibnamefont {Shestakov}},\ }\bibfield  {title} {\bibinfo {title} {Skyrmions
  in nematic liquid crystals},\ }\href
  {https://doi.org/10.1103/PhysRevE.67.016602} {\bibfield  {journal} {\bibinfo
  {journal} {Phys. Rev. E}\ }\textbf {\bibinfo {volume} {67}},\ \bibinfo
  {pages} {016602} (\bibinfo {year} {2003})}\BibitemShut {NoStop}%
\bibitem [{\citenamefont {Hess}\ \emph {et~al.}(2020)\citenamefont {Hess},
  \citenamefont {Poy}, \citenamefont {Tai}, \citenamefont {{\v Z}umer},\ and\
  \citenamefont {Smalyukh}}]{hess2020control_light}%
  \BibitemOpen
  \bibfield  {author} {\bibinfo {author} {\bibfnamefont {A.~J.}\ \bibnamefont
  {Hess}}, \bibinfo {author} {\bibfnamefont {G.}~\bibnamefont {Poy}}, \bibinfo
  {author} {\bibfnamefont {J.-S.~B.}\ \bibnamefont {Tai}}, \bibinfo {author}
  {\bibfnamefont {S.}~\bibnamefont {{\v Z}umer}},\ and\ \bibinfo {author}
  {\bibfnamefont {I.~I.}\ \bibnamefont {Smalyukh}},\ }\bibfield  {title}
  {\bibinfo {title} {Control of light by topological solitons in soft chiral
  birefringent media},\ }\href {https://doi.org/10.1103/PhysRevX.10.031042}
  {\bibfield  {journal} {\bibinfo  {journal} {Phys. Rev. X}\ }\textbf {\bibinfo
  {volume} {10}},\ \bibinfo {pages} {031042} (\bibinfo {year}
  {2020})}\BibitemShut {NoStop}%
\end{thebibliography}%
